\documentclass[a4paper]{article}

\newif\ifsubmission
\submissiontrue
\DeclareFontShape{OT1}{cmr}{m}{scit}{<->ssub*cmr/m/sc}{}
\usepackage{slantsc}

\usepackage[margin=1in]{geometry}
\usepackage{amsmath}
\usepackage{amsthm}
\usepackage{aliascnt}
\usepackage{amssymb}
\usepackage{graphicx}
\graphicspath{{figs/}}
\usepackage{booktabs}
\usepackage{longtable}
\usepackage{multirow}
\usepackage[dvipsnames]{xcolor}
\usepackage{mathrsfs}
\usepackage[utf8]{inputenc}
\usepackage[T1]{fontenc}
\usepackage{lmodern}
\usepackage{upquote}
\usepackage{placeins}
\usepackage{adjustbox}
\usepackage{microtype}
\usepackage{enumerate}
\usepackage{verbatim}
\usepackage{tikz}
\usetikzlibrary{shapes,calc,math,backgrounds,matrix}

\usepackage[backend=biber,
bibstyle=numeric,
citestyle=numeric-comp,
maxcitenames=3,
maxbibnames=20,
sorting=ydnt
]{biblatex}
\usepackage{hyperref}
\hypersetup{
	colorlinks,
	allcolors=blue,
	unicode,
	bookmarksopen,
	bookmarksdepth=2,
	pdftitle={An Article Template},
	pdfauthor={Anonymous},
	pdfkeywords={article template, placeholder text, LaTeX}
}

\usepackage[capitalise,noabbrev]{cleveref}
\makeatletter
\newcommand*{\kpvcr@thm@refstepcounter}[2][]{\let\kpvcr@saved@refstepcounter\cref@old@refstepcounter
  \let\cref@old@refstepcounter\Hy@theorem@refstepcounter
  \refstepcounter[#1]{#2}\let\cref@old@refstepcounter\kpvcr@saved@refstepcounter
}
\patchcmd{\cref@thmnoarg}{\refstepcounter}{\Hy@theorem@refstepcounter}{}{\PackageError{kpvcr}{Cannot patch the untyped theorem anchor}{Recheck the installed amsthm, hyperref and cleveref versions.}}
\patchcmd{\cref@thmoptarg}{\refstepcounter}{\kpvcr@thm@refstepcounter}{}{\PackageError{kpvcr}{Cannot patch the typed theorem anchor}{Recheck the installed amsthm, hyperref and cleveref versions.}}
\makeatother
\crefname{cond}{Condition}{Conditions}
\creflabelformat{cond}{(#2C#1#3)}
\crefname{eitem}{}{}
\creflabelformat{eitem}{(#2#1#3)}
\crefname{step}{\textbf{Step}}{\textbf{Steps}}
\creflabelformat{step}{{\bf #2#1#3}}

\theoremstyle{plain}
\newtheorem{theorem}{Theorem}[section]
\newaliascnt{corollary}{theorem}
\newtheorem{corollary}[corollary]{Corollary}
\aliascntresetthe{corollary}
\newaliascnt{lemma}{theorem}
\newtheorem{lemma}[lemma]{Lemma}
\aliascntresetthe{lemma}

\newaliascnt{axiom}{theorem}

\aliascntresetthe{axiom}
\newaliascnt{conjecture}{theorem}

\aliascntresetthe{conjecture}
\newaliascnt{fact}{theorem}

\aliascntresetthe{fact}
\newaliascnt{hypothesis}{theorem}

\aliascntresetthe{hypothesis}
\newaliascnt{assumption}{theorem}

\aliascntresetthe{assumption}
\newaliascnt{proposition}{theorem}

\aliascntresetthe{proposition}
\newaliascnt{criterion}{theorem}

\aliascntresetthe{criterion}
\theoremstyle{definition}
\newaliascnt{definition}{theorem}
\newtheorem{definition}[definition]{Definition}
\aliascntresetthe{definition}
\newaliascnt{example}{theorem}
\newtheorem{example}[example]{Example}
\aliascntresetthe{example}
\newaliascnt{remark}{theorem}
\newtheorem{remark}[remark]{Remark}
\aliascntresetthe{remark}
\newaliascnt{problem}{theorem}

\aliascntresetthe{problem}
\newaliascnt{principle}{theorem}

\aliascntresetthe{principle}

\crefname{theorem}{Theorem}{Theorems}
\Crefname{theorem}{Theorem}{Theorems}
\crefname{corollary}{Corollary}{Corollaries}
\Crefname{corollary}{Corollary}{Corollaries}
\crefname{lemma}{Lemma}{Lemmas}
\Crefname{lemma}{Lemma}{Lemmas}
\crefname{claim}{Claim}{Claims}
\Crefname{claim}{Claim}{Claims}
\crefname{axiom}{Axiom}{Axioms}
\Crefname{axiom}{Axiom}{Axioms}
\crefname{conjecture}{Conjecture}{Conjectures}
\Crefname{conjecture}{Conjecture}{Conjectures}
\crefname{fact}{Fact}{Facts}
\Crefname{fact}{Fact}{Facts}
\crefname{hypothesis}{Hypothesis}{Hypotheses}
\Crefname{hypothesis}{Hypothesis}{Hypotheses}
\crefname{assumption}{Assumption}{Assumptions}
\Crefname{assumption}{Assumption}{Assumptions}
\crefname{proposition}{Proposition}{Propositions}
\Crefname{proposition}{Proposition}{Propositions}
\crefname{criterion}{Criterion}{Criteria}
\Crefname{criterion}{Criterion}{Criteria}
\crefname{definition}{Definition}{Definitions}
\Crefname{definition}{Definition}{Definitions}
\crefname{example}{Example}{Examples}
\Crefname{example}{Example}{Examples}
\crefname{remark}{Remark}{Remarks}
\Crefname{remark}{Remark}{Remarks}
\crefname{problem}{Problem}{Problems}
\Crefname{problem}{Problem}{Problems}
\crefname{principle}{Principle}{Principles}
\Crefname{principle}{Principle}{Principles}

\usepackage[linesnumbered,ruled,vlined,commentsnumbered]{algorithm2e}
\crefname{algocf}{Algorithm}{Algorithms}
\Crefname{algocf}{Algorithm}{Algorithms}
\usepackage{array}
\DeclareUnicodeCharacter{2194}{\ensuremath{\leftrightarrow}}
\DeclareUnicodeCharacter{2227}{\ensuremath{\wedge}}
\DeclareUnicodeCharacter{2200}{\ensuremath{\forall}}
\DeclareUnicodeCharacter{2208}{\ensuremath{\in}}
\DeclareUnicodeCharacter{2229}{\ensuremath{\cap}}

\makeatletter
\def\begin@lgo{\begin{minipage}{1in}\begin{tabbing}
        \quad\=\qquad\=\qquad\=\qquad\=\qquad\=\qquad\=\qquad\=\qquad\=\qquad\=\qquad\=\qquad\=\qquad\=\qquad\=\kill}
\def\end@lgo{\end{tabbing}\end{minipage}}

\newenvironment{algobox}
{\begin{tabular}{|l|}\hline\begin@lgo}
{\end@lgo\\\hline\end{tabular}}

\def\beginc@de{\noindent\begin{center}\begin{minipage}{1in}\ttfamily
        \begin{tabbing}~~~~\=~~~~\=~~~~\=~~~~\=~~~~\=~~~~\=~~~~\=~~~~\=\kill}
\def\endc@de{\end{tabbing}\end{minipage}\end{center}}

\makeatother

\def\mathsc#1{\text{\textsc{#1}}}

\usepackage[pagewise]{lineno}
\usepackage[colorinlistoftodos]{todonotes}

\definecolor{done}{rgb}{0.55, 0.71, 0.0}
\definecolor{highpriority}{rgb}{0.82, 0.1, 0.26}
\definecolor{lowpriority}{rgb}{1.0, 0.75, 0.0}
\definecolor{mediumpriority}{rgb}{1.0, 0.49, 0.0}

\newcommand{\Mcomment}[3]{\ifsubmission \else \unskip\ {\color{#1}\bfseries\sffamily(#2: #3)}\fi }
\newcommand{\TaskNewPerson}[3]{\expandafter\newcommand\csname #2\endcsname[1]{\Mcomment{#3}{#2}{##1}}\expandafter\newcommand\csname Task#1\endcsname[2][mediumpriority]{\expandafter\TaskPerson[backgroundcolor=##1]{#1}{##2}}}

\newcommand*{\TaskPerson}[3][]{\ifsubmission \else \todo[inline,#1]{\textbf{(#2)} #3}\fi }

\newcommand{\email}[1]{\href{mailto:#1}{\texttt{#1}}}

\newcommand{\dist}{\mathsf{dist}}

\newcommand{\onestep}[1]{{\overset{#1}{\leftrightarrow}}}
\newcommand{\sevstep}[1]{{\overset{#1}{\leftrightsquigarrow}}}

\IfFileExists{ccicons.sty}{\usepackage{ccicons}\newcommand{\ccbyicon}{\ccby\enspace}}{\newcommand{\ccbyicon}{}}
\newcommand{\ccbynotice}{\ccbyicon This work is licensed under a \href{https://creativecommons.org/licenses/by/4.0/}{Creative Commons Attribution 4.0 International License (CC BY 4.0)}.}
\makeatletter
\def\ps@ccbyfoot{\let\@mkboth\@gobbletwo
	\let\@oddhead\@empty
	\let\@evenhead\@empty
	\def\@oddfoot{\reset@font\vtop{\hsize\textwidth
		\centering\thepage\par\smallskip
		{\scriptsize\sffamily\ccbynotice\par}}}\let\@evenfoot\@oddfoot}
\newcommand{\ccbylicense}{\@ifstar
	{\pagestyle{ccbyfoot}\thispagestyle{ccbyfoot}}{\thispagestyle{ccbyfoot}}}
\makeatother

\newcommand{\templatelistoftodos}{\ifsubmission
	\else
		\listoftodos
	\fi
}

\hypersetup{
pdftitle={A Polynomial-Time Algorithm for k-Path Vertex Cover Reconfiguration on Trees under Token Sliding},
	pdfauthor={Duc A. Hoang},
pdfkeywords={combinatorial reconfiguration, token sliding, k-path vertex cover, tree, polynomial-time algorithm}
}

\newcommand{\PSPACE}{\mathsf{PSPACE}}
\newcommand{\NP}{\mathsf{NP}}
\newcommand{\kPVCR}{\textup{\textsc{\(k\)-PVCR}}}
\newcommand{\Rigid}{\mathcal{R}}
\newcommand{\NeverOcc}{\mathcal{H}}
\newcommand{\sub}{\operatorname{sub}}

\newcommand{\rmstep}[2]{{\overset{#1,\,#2}{\rightsquigarrow}}}

\title{\textbf{A Polynomial-Time Algorithm for $k$-Path Vertex Cover Reconfiguration on Trees under Token Sliding}}
\author{Duc A.\ Hoang\\
VNU University of Science, Vietnam National University\\
Hanoi, Vietnam\\
\email{hoanganhduc@hus.edu.vn}}
\date{}

\begin{document}
\maketitle

\begin{abstract}

A $k$-path vertex cover ($k$-PVC) of a graph is a set of vertices meeting every path on $k$ vertices.
In Token Sliding, one token occupies each vertex of a cover; a move slides one token along an edge to an unoccupied vertex, preserving the cover property.
We study whether two covers of a tree are joined by such moves.
The case $k=2$ admits a linear-time decision algorithm.
For every integer $k\geq2$, we give an $O(kn^2)$-time decision algorithm on trees of order $n$, with $k$ part of the input.
Two covers are reachable precisely when their rigid tokens, which never move, occupy the same set and every component remaining after their deletion has the same token count.
The proof transforms configurations of equal size without rigid tokens into a canonical configuration built on properly rooted subtrees, whose paths on $k$ vertices pass through their roots.
A rooted-subtree recurrence identifies movable candidate tokens.
We also give a direct recursive construction between the supplied endpoints, returning at most $8n^2\,3^{\lceil n/2\rceil}$ slides in $O(n^{10}3^{n/2})$ time.

\smallskip
\noindent
ChatGPT 6 Pro and Claude Fable 5.1 were used to assist with developing and reviewing mathematical arguments and drafting and revising the paper.
The author takes full responsibility for the correctness and content of the paper.

\end{abstract}
\relax\space 

\templatelistoftodos

\section{Introduction}\label{sec:intro}

Recently, combinatorial reconfiguration has attracted attention from both theoretical and practical viewpoints.
A reconfiguration problem asks whether one feasible solution of a source problem can be transformed into another by a sequence of small changes, each of which keeps the solution feasible.
For an overview of this research area, we refer readers to the surveys of Nishimura~\cite{Nishimura2018} and van den Heuvel~\cite{vandenHeuvel2013}.

For an integer $k\geq 2$, a \emph{$k$-path vertex cover} (\emph{$k$-PVC}) of a graph $G$ is a set $S\subseteq V(G)$ such that every path on $k$ vertices of $G$ contains a vertex of $S$.
Brešar, Kardoš, Katrenič and Semanišin~\cite{Bresar2011} introduced the minimum $k$-PVC problem, proved that it is $\NP$-hard for every fixed $k\geq 2$, and gave a linear-time algorithm on trees.
A $2$-PVC is a vertex cover.
The complement of a $3$-PVC is a \emph{dissociation set}, that is, a set of vertices inducing a subgraph of maximum degree at most one~\cite{Bresar2011,TuZhangShi2021}.

Hoang, Suzuki and Yagita~\cite{HoangSuzukiYagita2022} initiated the study of reconfiguring $k$-PVCs.
In the \emph{Token Sliding} model, a token is placed on each vertex of a $k$-PVC.
A move slides one token along an edge to an unoccupied vertex so that the occupied vertices still form a $k$-PVC.
A sequence of such moves is a \emph{TS-sequence}.
The problem $k$-Path Vertex Cover Reconfiguration (\kPVCR{}) under Token Sliding asks, given a graph $G$ and two $k$-PVCs $S$ and $S'$ of $G$ of the same size, whether there is a TS-sequence from $S$ to $S'$.

For $k=2$, taking complements identifies the problem with Token Sliding for independent sets: sliding a vertex-cover token from $u$ to $v$ moves an independent-set token from $v$ to $u$~\cite{HoangSuzukiYagita2022}.
Token Sliding for independent sets is $\PSPACE$-complete on general graphs~\cite{HearnDemaine2005,Ito2011}, and its complexity on restricted graph classes has been studied extensively.
We refer readers to Kamiński, Medvedev and Milanič~\cite{Kaminski2012} and to Nishimura~\cite{Nishimura2018} for examples.
On trees, Demaine et al.~\cite{Demaine2015} gave a linear-time decision algorithm.
Their characterization uses the tokens that never move and compares token counts after deleting the vertices with those tokens together with their neighbors.

For general $k$, Hoang, Suzuki and Yagita~\cite{HoangSuzukiYagita2022} proved that \kPVCR{} is $\PSPACE$-complete under each of Token Sliding, Token Jumping and Token Addition/Removal even on planar graphs of maximum degree three and bounded bandwidth.
They gave polynomial-time algorithms under all three models for paths and cycles, and under Token Jumping and Token Addition/Removal for trees.
For each fixed $k\geq2$, Hoang, Suzuki and Yagita~\cite{HoangSuzukiYagita2022} also construct shortest TS-sequences on paths in $O(n^2)$ time and TS-sequences on cycles in $O(n^2)$ time whenever the target is reachable, where $n$ is the order of the input graph.
They left open the complexity under Token Sliding on trees.
For caterpillars and $k\geq 4$, Hoang~\cite{Hoang2023} gave a polynomial-time algorithm based on the vertices with rigid tokens, which never move, and the token counts after deleting those vertices.
For each fixed $k\geq4$, Hoang~\cite{Hoang2023} also constructs a TS-sequence in $O(n^4)$ time between equal-size $k$-PVCs of an $n$-vertex caterpillar when neither has a rigid token.
For $k=3$, the same paper points out that verifying one condition of its characterization of rigid tokens in polynomial time may require further ideas~\cite{Hoang2023}.
As far as we know, the complexity on trees has remained open for every $k\geq 3$.

\subsection{Our Results}\label{sec:results}

In this paper, we give a polynomial-time decision algorithm for every $k\geq 2$ and an algorithm for finding a TS-sequence for every yes-instance.
The first result treats $k$ as part of the input.

\begin{theorem}\label{thm:decision}
There is an algorithm that, given an integer $k\geq 2$, a tree $T$ of order $n$, and two $k$-PVCs $S$ and $S'$ of $T$, decides in $O(kn^2)$ time whether there is a TS-sequence from $S$ to $S'$.
\end{theorem}

When $k>n$, every vertex set is a $k$-PVC, and the answer is yes exactly when $|S|=|S'|$.
\Cref{sec:algorithm} proves \cref{thm:decision}.
For $k=3$, we apply the algorithm of \cref{thm:decision} to the complements to decide in $O(n^2)$ time whether one dissociation set of a tree can be transformed into another by sliding its tokens along edges, one at a time, so that the occupied vertices always form a dissociation set.

The decision algorithm has two main ingredients.
\begin{itemize}

\item We characterize reachability by rigid tokens and token counts (\cref{thm:characterization}).
For a $k$-PVC $S$ of a tree $T$, let $\Rigid(S)$ be the set of vertices occupied in every $k$-PVC reachable from $S$; the tokens on these vertices never move and are called \emph{rigid}.
Two $k$-PVCs $S$ and $S'$ of $T$ are reachable from one another if and only if $\Rigid(S)=\Rigid(S')$ and they have the same number of tokens in every component of $T-\Rigid(S)$.
The condition is necessary because no token can enter or leave a component of $T-\Rigid(S)$, and it is sufficient by the Key Lemma (\cref{lem:key}): two $k$-PVCs of equal size whose tokens are all movable are reachable from one another.
The characterization has the form of the result of Hoang~\cite{Hoang2023} for caterpillars and $k\geq 4$, and for $k=2$ it describes the same relation as that of Demaine et al.~\cite{Demaine2015} (\cref{rem:demaine-k2}).

The Key Lemma rests on a \emph{canonical} form: every $k$-PVC of a given size whose tokens are all movable can be transformed into one canonical $k$-PVC determined by the tree, its vertex order and the size.
It is built one part at a time on the properly rooted subtrees of Brešar et al.~\cite{Bresar2011} (\cref{sec:properly-rooted}), that is, subtrees $U$ attached to the rest of the tree by one edge at a root $u$ such that $T[U]$ contains a path on $k$ vertices and $T[U-u]$ does not.
While $u$ is occupied, the tokens in $U-u$ do not interact with the rest of the tree, so tokens moved through $u$ can put a prescribed set on $U$, after which the construction continues on the smaller tree.
For $k=2$, Demaine et al.\ cut off stars formed by a leaf, its neighbor and the other leaves there; already for $k=3$, such small pieces do not suffice (\cref{ex:path7}), and properly rooted subtrees take their place.

\item We compute $\Rigid(S)$ with \cref{alg:compute-rigid}, which follows the idea of the linear-time algorithm of Demaine et al.~\cite{Demaine2015} for independent sets.
It starts with all occupied vertices as candidates and repeatedly removes a candidate whose token can move while the other candidates stay occupied; the candidates that remain form $\Rigid(S)$ (\cref{thm:rigid-algorithm}).
What is new for general $k$ is the test whether the token of a candidate can move.
A slide from a candidate $v$ creates only unoccupied paths through $v$, each made of $v$ and at most two unoccupied paths starting at other neighbors of $v$, and the components of the tree minus the candidates can be rearranged independently.
Hence, the test needs, for each such component $P$ and its vertex $r$ next to a candidate, only the least possible number of vertices of a longest unoccupied path starting at $r$.
The integer $c(P,r)$ is the minimum number of tokens in a configuration obtainable from the fully occupied configuration $P$ by slides in $T[P]$ and legal removals at $r$.
Among the configurations with $c(P,r)$ tokens obtained in this way, $\tau(P,r)$ is the least number of vertices of a longest unoccupied path starting at $r$, with value zero when $r$ is occupied.
Since the tokens of these components stay movable, the number needed by the test is $\tau(P,r)$ if $P$ holds exactly $c(P,r)$ tokens and zero otherwise (\cref{cor:min-eps-formula}), where one recurrence on rooted subtrees computes $c(P,r)$ and $\tau(P,r)$ for every $k\geq 2$.
\end{itemize}

Our second result outputs a TS-sequence for every yes-instance.
For this construction, running times count array, integer, and slide-record operations after reading the input.
A binary encoding of $k$ adds $O(\log(k+1))$ for reading that parameter.

\begin{theorem}\label{thm:construction}
There is an algorithm that, given an integer $k\geq 2$, a tree $T$ of order $n\geq1$, and two $k$-PVCs $S$ and $S'$ of $T$ such that $S'$ is reachable from $S$, outputs a TS-sequence from $S$ to $S'$ with at most $8n^2\,3^{\lceil n/2\rceil}$ slides in $O(n^{10}3^{n/2})$ time.
\end{theorem}

The construction in \cref{sec:construction} connects the two supplied endpoints directly.
It fixes the same properly rooted side in both endpoints and recursively connects their remaining restrictions once.
Within that side, a single transfer loop maintains occupied roots in nonempty branches, and exchanges along one path adjust consecutive prescribed sets.
The resulting quadratic local cost and the potential $(\sqrt{3})^p-1$ give the stated bound on the complete returned list.
The running-time bound uses exact tree traversals, the rooted recurrence, and linear assembly of the returned lists, as specified in \cref{subsec:construction-time}.
The exponential bound is an upper bound; we do not claim that this procedure must return exponentially many slides.

The rest of this paper is organized as follows: \cref{sec:preliminaries} introduces notation and basic properties of TS-sequences; \cref{sec:characterization} proves the reachability characterization; \cref{sec:algorithm} gives the decision algorithm and proves \cref{thm:decision}; \cref{sec:construction} proves \cref{thm:construction}; and \cref{sec:conclusion} presents the remaining questions.
\relax\space \section{Preliminaries}\label{sec:preliminaries}

This section introduces the graph notation, token sliding on $k$-PVCs, and properly rooted subtrees used throughout the paper.
We then prove the basic properties of TS-sequences needed for the characterization in \cref{sec:characterization} and the decision algorithm in \cref{sec:algorithm}, including a property of restrictions to one side of an edge (\cref{lem:attachment-movable}).

\subsection{Graph Notation}\label{sec:notation}

We use $V(G)$ and $E(G)$ to denote the sets of vertices and edges of a (simple, undirected) graph $G$, respectively.
The neighborhood of a vertex $v$ in $G$, denoted by $N_G(v)$, is the set $\{w \in V(G) : vw \in E(G)\}$.
The closed neighborhood of $v$ in $G$, denoted by $N_G[v]$, is simply the set $N_G(v) \cup \{v\}$.
For two sets $A$ and $B$, we sometimes use $A-B$ and $A+B$ to indicate $A\setminus B$ and $A\cup B$, respectively, and we write $A-x$ and $A+x$ for $A-\{x\}$ and $A+\{x\}$.
For $X\subseteq V(G)$, the subgraph of $G$ induced by $X$ is $G[X]$, and $G-X$ is obtained by removing $X$ and all incident edges; we write $G-v$ for $G-\{v\}$.
A \emph{path on $k$ vertices} is a sequence of $k$ distinct vertices in which consecutive vertices are adjacent.

Throughout the paper, $k\geq 2$ is an integer and $T$ is a finite nonempty tree of order $n$.
A fixed linear order of $V(T)$ resolves all choices between vertices.
A set $W\subseteq V(T)$ is \emph{connected} if $T[W]$ is connected.
We state the results for subsets of $V(T)$ because the proofs apply them to subtrees; the reader may take $W=V(T)$ at first reading.
For adjacent vertices $x,y\in W$, the set $W^x_y$ is the vertex set of the component of $T[W]-x$ containing $y$.
When $W$ is connected, $W^x_y$ and $W^y_x$ partition $W$, and the edge $xy$ is the only edge joining these two sets.
The notation follows Demaine et al.~\cite{Demaine2015}, who use $T^x_y$ when $W=V(T)$.
A connected set $P$ together with a vertex $r\in P$ is a \emph{rooted tree} $(P,r)$, in which parents, children, ancestors and descendants are defined as usual.
For a connected set $P$ rooted at $r\in P$, the sets $P^r_y$, for $y\in N_{T[P]}(r)$, are its \emph{branches}, rooted at the corresponding vertices $y$.
The \emph{rooted height} of $(P,r)$ is the number of vertices of a longest path of $T[P]$ starting at $r$.
The \emph{depth} of $v\in P$ is the number $\dist(v,r)$ of edges of the path from $r$ to $v$, and $\sub(v)$ is the vertex set of the subtree of $v$, that is, $v$ together with its descendants; the rooted tree is always clear from the context when $\sub(v)$ is used.
We sometimes identify a subtree with its vertex set.

For a finite multiset $X$ of nonnegative integers, $\sigma(X)$ denotes the sum of its two largest members, with missing members counted as zero.
In particular, $\sigma(\emptyset)=0$ and $\sigma(\{a\})=a$.
Deleting or decreasing members of $X$ cannot increase $\sigma(X)$.
All lists inside $\sigma$ below are multisets, so equal entries are counted separately.

\subsection{Token Sliding on \texorpdfstring{$k$}{k}-PVCs}\label{sec:token-sliding}

Let $W\subseteq V(T)$.
Recall that a $k$-PVC of $T[W]$ is a set $S\subseteq W$ meeting every path on $k$ vertices of $T[W]$.
Equivalently, $T[W-S]$ contains no path on $k$ vertices.
Following Hoang, Suzuki and Yagita~\cite{HoangSuzukiYagita2022} and Hoang~\cite{Hoang2023}, we regard each vertex of $S$ as \emph{occupied} by a token, and each vertex of $W-S$ as \emph{unoccupied}.
Every superset of a $k$-PVC in $W$ is a $k$-PVC of $T[W]$.

For two $k$-PVCs $S,S'$ of $T[W]$, we write $S\onestep{W}S'$ if $S'=S-u+v$ for an edge $uv$ of $T[W]$ with $u\in S$ and $v\notin S$.
We say that $S'$ is obtained by \emph{sliding} the token on $u$ to $v$.
For a $k$-PVC $S$, a vertex $u\in S$ and an unoccupied neighbor $v$ of $u$, we denote the slide of the token on $u$ to $v$ by $u\to v$, and it is \emph{legal} at $S$ if $S-u+v$ is a $k$-PVC of $T[W]$.
A TS-sequence in $T[W]$ is a sequence $\langle S_0,\dots,S_\ell\rangle$ of $k$-PVCs with $\ell\geq 0$ and $S_{i-1}\onestep{W}S_i$ for every $i\in\{1,\dots,\ell\}$.
The number of \emph{slides} of such a TS-sequence is $\ell$.
We write $S\sevstep{W}S'$ when such a TS-sequence starts at $S$ and ends at $S'$, and we denote by $[S]_W$ the set of $k$-PVCs reachable from $S$.
The relation $\sevstep{W}$ is an equivalence relation, and every member of $[S]_W$ has $|S|$ tokens.
When $W=V(T)$, we omit $W$ from subscripts.

Following Demaine et al.~\cite{Demaine2015} and Hoang~\cite{Hoang2023}, the token on $r\in S$ is \emph{$(T[W],S)$-rigid} if $r$ is occupied in every member of $[S]_W$, and \emph{$(T[W],S)$-movable} otherwise.
We denote by $\Rigid_W(S)$ the vertices with rigid tokens and by $\NeverOcc_W(S)$ the vertices that are never occupied in $[S]_W$; thus,
\[
\Rigid_W(S)=\bigcap_{S'\in[S]_W}S',
\qquad
\NeverOcc_W(S)=W-\bigcup_{S'\in[S]_W}S'.
\]
Both sets depend only on $[S]_W$.
When the ambient set and the $k$-PVC are clear, we simply say rigid or movable.
The equality $\Rigid_W(S)=\emptyset$ means that all tokens are movable; it does not imply that $\NeverOcc_W(S)=\emptyset$.
For $r\in W$, we define $\varepsilon_W(S,r)=0$ when $r\in S$; otherwise, $\varepsilon_W(S,r)$ is the number of vertices of a longest path of $T[W-S]$ starting at $r$.
Since $S$ is a $k$-PVC, $0\leq\varepsilon_W(S,r)\leq k-1$.

\begin{figure}[ht]
\centering
\begin{adjustbox}{max width=\textwidth}
\begin{tikzpicture}[v/.style={circle,draw=black,fill=white,inner sep=0pt,minimum size=4.5mm},
tok/.style={v,fill=black},rig/.style={v,fill=black,draw=black!30,line width=1.8pt},
nev/.style={v,fill=black!20},line width=.45pt]
\node[rig] (a_u0) at (0.0000,0.0000) {};
\node[tok] (a_u1) at (0.0000,1.2500) {};
\node[tok] (a_u2) at (0.0000,2.5000) {};
\node[v] (a_u3) at (-1.1600,3.5625) {};
\node[v] (a_u4) at (1.1600,3.5625) {};
\node[nev] (a_u5) at (-2.6100,-0.8125) {};
\node[nev] (a_u6) at (-2.6100,-1.8750) {};
\node[nev] (a_u7) at (-2.6100,-2.9375) {};
\node[nev] (a_u8) at (2.6100,-0.8125) {};
\node[nev] (a_u9) at (2.6100,-1.8750) {};
\node[nev] (a_u10) at (2.6100,-2.9375) {};
\node[tok] (a_u11) at (0.0000,-1.6875) {};
\node[v] (a_u12) at (-0.7975,-4.0000) {};
\node[v] (a_u13) at (0.7975,-4.0000) {};
\draw (a_u0) -- (a_u1);
\draw (a_u1) -- (a_u2);
\draw (a_u2) -- (a_u3);
\draw (a_u2) -- (a_u4);
\draw (a_u0) -- (a_u5);
\draw (a_u5) -- (a_u6);
\draw (a_u6) -- (a_u7);
\draw[line width=1.6pt] (a_u0) -- (a_u8);
\draw[line width=1.6pt] (a_u8) -- (a_u9);
\draw[line width=1.6pt] (a_u9) -- (a_u10);
\draw (a_u0) -- (a_u11);
\draw (a_u11) -- (a_u12);
\draw (a_u11) -- (a_u13);
\node[font=\scriptsize,anchor=west,inner sep=1pt] at (0.2900,0.0000) {$u_{0}$};
\node[font=\scriptsize,anchor=east,inner sep=1pt] at (-0.2900,1.2500) {$u_{1}$};
\node[font=\scriptsize,anchor=east,inner sep=1pt] at (-0.2900,2.5000) {$u_{2}$};
\node[font=\scriptsize,anchor=east,inner sep=1pt] at (-1.4500,3.5625) {$u_{3}$};
\node[font=\scriptsize,anchor=west,inner sep=1pt] at (1.4500,3.5625) {$u_{4}$};
\node[font=\scriptsize,anchor=east,inner sep=1pt] at (0.8700,3.5625) {$1$};
\node[font=\scriptsize,anchor=east,inner sep=1pt] at (-2.9000,-0.8125) {$u_{5}$};
\node[font=\scriptsize,anchor=west,inner sep=1pt] at (-2.3200,-0.8125) {$3$};
\node[font=\scriptsize,anchor=east,inner sep=1pt] at (-2.9000,-1.8750) {$u_{6}$};
\node[font=\scriptsize,anchor=west,inner sep=1pt] at (-2.3200,-1.8750) {$2$};
\node[font=\scriptsize,anchor=east,inner sep=1pt] at (-2.9000,-2.9375) {$u_{7}$};
\node[font=\scriptsize,anchor=west,inner sep=1pt] at (-2.3200,-2.9375) {$3$};
\node[font=\scriptsize,anchor=west,inner sep=1pt] at (2.9000,-0.8125) {$u_{8}$};
\node[font=\scriptsize,anchor=east,inner sep=1pt] at (2.3200,-0.8125) {$3$};
\node[font=\scriptsize,anchor=west,inner sep=1pt] at (2.9000,-1.8750) {$u_{9}$};
\node[font=\scriptsize,anchor=east,inner sep=1pt] at (2.3200,-1.8750) {$2$};
\node[font=\scriptsize,anchor=west,inner sep=1pt] at (2.9000,-2.9375) {$u_{10}$};
\node[font=\scriptsize,anchor=east,inner sep=1pt] at (2.3200,-2.9375) {$3$};
\node[font=\scriptsize,anchor=west,inner sep=1pt] at (0.2900,-1.6875) {$u_{11}$};
\node[font=\scriptsize,anchor=east,inner sep=1pt] at (-1.0875,-4.0000) {$u_{12}$};
\node[font=\scriptsize,anchor=west,inner sep=1pt] at (-0.5075,-4.0000) {$1$};
\node[font=\scriptsize,anchor=west,inner sep=1pt] at (1.0875,-4.0000) {$u_{13}$};
\node[font=\scriptsize,anchor=east,inner sep=1pt] at (0.5075,-4.0000) {$1$};
\node[font=\scriptsize,anchor=south,inner sep=1pt] at (-1.1600,3.9225) {$1$};
\draw[->,thick,green!50!black] (0.0170,2.9183) -- (-0.7448,3.6161);
\draw[->,thick,green!50!black] (-0.3529,0.2253) -- (-2.4473,-0.4267);
\draw[thick,red!80!black] (-1.4801,-0.0207) -- (-1.3201,-0.1807);
\draw[thick,red!80!black] (-1.4801,-0.1807) -- (-1.3201,-0.0207);
\path[use as bounding box] (-3.5,-4.5) rectangle (3.5,4.3);
\end{tikzpicture}
\relax\space \end{adjustbox}
\caption{A $4$-PVC $S=\{u_{0},u_{1},u_{2},u_{11}\}$ of the tree $T$ of order $14$.
All figures use this tree, with the fixed vertex order $u_{0}<u_{1}<\dots<u_{13}$.
Black vertices carry tokens; a thick gray outline marks a rigid token, light gray vertices are never occupied, and white vertices are the other unoccupied vertices.
Here $\Rigid(S)=\{u_{0}\}$, $\NeverOcc(S)=\{u_{5},u_{6},u_{7},u_{8},u_{9},u_{10}\}$, and $[S]$ has $18$ members.
The numbers beside unoccupied vertices give $\varepsilon_{V(T)}(S,u_i)$.
The arrow $u_{2}\to u_{3}$ is a legal slide; the crossed arrow $u_{0}\to u_{5}$ would leave the thick path $u_{10}u_{9}u_{8}u_{0}$ unoccupied.}
\label{fig:f1-basic}
\end{figure}
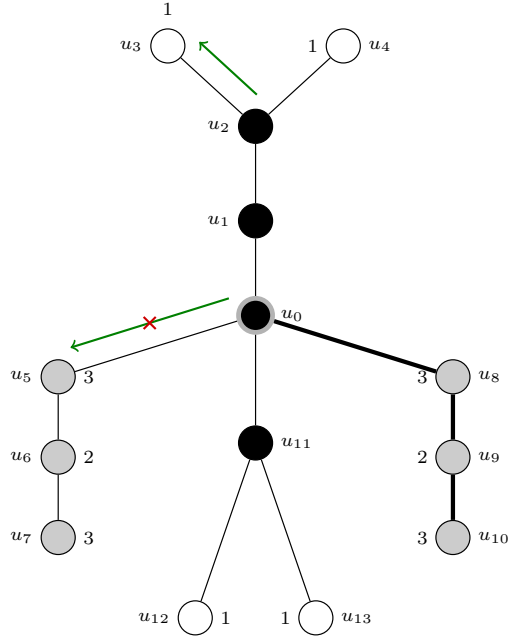

\cref{fig:f1-basic} illustrates rigid tokens, never-occupied vertices, and the values of $\varepsilon$.

Every newly uncovered path must contain the vertex vacated by the proposed slide.
The following test expresses that observation using $\sigma$.

\begin{lemma}\label{lem:slide-test}
Let $S$ be a $k$-PVC of $T[W]$, let $v\in S$, and let $z\in N_{T[W]}(v)-S$.
Then $S\onestep{W}S-v+z$ if and only if
\begin{equation}\label{eq:slide-test}
\sigma\!\left(\left\{\varepsilon_{W^v_y}(S\cap W^v_y,y):
 y\in N_{T[W]}(v)-z\right\}\right)\leq k-2.
\end{equation}
\end{lemma}

\begin{proof}
We determine the longest path of unoccupied vertices that contains $v$ after the proposed slide.
\cref{fig:f1-basic} illustrates the two outcomes of the slide test.
Every newly uncovered path must contain $v$, and it cannot use the occupied vertex $z$.
Such a path uses at most two branches at $v$.
The longest possible number of vertices on it is one plus the sum in \cref{eq:slide-test}, because paths in two distinct branches can be joined through $v$.
The new set is therefore a $k$-PVC exactly when that number is at most $k-1$.
\end{proof}

For an edge $ay$ of $T[W]$, the \emph{value} of $y$ at $a$ (at $S$) is $\varepsilon_{W^a_y}(S\cap W^a_y,y)$.
This value is zero when $y\in S$; otherwise, it is the number of vertices of a longest unoccupied path that starts at $y$ and avoids $a$.
A finite multiset $X$ of nonnegative integers \emph{fits} if $\sigma(X)\leq k-2$.
In these terms, \cref{lem:slide-test} says that the slide $v\to z$ is legal at $S$ if and only if the values at $v$ of the neighbors of $v$ in $T[W]$ other than $z$ fit.

We also use token removals at a specified vertex $r\in W$.
If $r\in Z$ and $Z-r$ is a $k$-PVC of $T[W]$, then $Z-r$ is obtained from $Z$ by \emph{removing the token on $r$}.
We write $Z\rmstep{W}{r}Z'$ if $Z'$ is obtained from $Z$ by zero or more slides in $T[W]$ and removals of tokens on $r$, with every intermediate set a $k$-PVC of $T[W]$.
In particular, slides may occur before, between, and after removals, and exactly $|Z|-|Z'|$ tokens are removed.

\subsection{Properly Rooted Subtrees}\label{sec:properly-rooted}

The proofs in \cref{sec:characterization,sec:algorithm} use the properly rooted subtrees of Brešar et al.~\cite{Bresar2011}.
A connected set $U$ rooted at $u$ is \emph{properly rooted} if $T[U]$ contains a path on $k$ vertices and $T[U-u]$ does not.
In particular, for an edge $uv$ of $T[W]$, the subtree $U=W^v_u$ is properly rooted precisely when it has these properties with root $u$.
The same definition applies when $U=W$ and $u$ is a specified root.
Every path on $k$ vertices of $T[U]$ contains $u$, so every subset of $U$ containing $u$ is a $k$-PVC of $T[U]$.

Let $P_1,\dots,P_D$ be the branches of a properly rooted $(U,u)$, with roots $w_1,\dots,w_D$ and rooted heights $a_1,\dots,a_D$.
We index the branches so that $a_1\leq\dots\leq a_D$.
Each branch contains no path on $k$ vertices, so $1\leq a_i\leq k-1$.
Let $m$ be the largest $j\in\{0,\dots,D\}$ such that $\sigma(\{a_1,\dots,a_j\})\leq k-2$, and let
\begin{equation}\label{eq:properly-rooted-parameters}
d=D-m,
\qquad
\gamma(U)=|U|-d.
\end{equation}
Among all submultisets of a given cardinality, the shortest heights minimize the two largest entries.
Consequently, $m$ is also the largest cardinality of a submultiset of the branch heights with $\sigma$ at most $k-2$.
Moreover, $d\geq 1$: otherwise, every path through $u$ would have at most $k-1$ vertices, and no branch contains a path on $k$ vertices.

For every $h\in\{0,\dots,\gamma(U)\}$, we define a \emph{prescribed set} $C_h\subseteq U$.
The set contains $u$ and $|U|-1-h$ further vertices.
For the latter vertices, the branch roots come first, in decreasing order of their rooted heights and with ties resolved by the fixed vertex order; all vertices of $U$ other than $u$ and the branch roots follow in the fixed vertex order.
The set $C_h$ consists of $u$ and the first $|U|-1-h$ vertices in this list.
In particular, when fewer than $D$ branch roots are occupied, the empty branches have the smallest heights.
The set $C_h$ is a $k$-PVC because it contains $u$.

\begin{figure}[ht]
\centering
\begin{adjustbox}{max width=\textwidth}
\begin{tikzpicture}[v/.style={circle,draw=black,fill=white,inner sep=0pt,minimum size=4.5mm},
tok/.style={v,fill=black},rig/.style={v,fill=black,draw=black!30,line width=1.8pt},
nev/.style={v,fill=black!20},line width=.45pt]
\node[tok] (a_u0) at (0.0000,0.0000) {};
\node[v] (a_u1) at (0.0000,1.0000) {};
\node[v] (a_u2) at (0.0000,2.0000) {};
\node[v] (a_u3) at (-0.8000,2.8500) {};
\node[v] (a_u4) at (0.8000,2.8500) {};
\node[tok] (a_u5) at (-1.8000,-0.6500) {};
\node[tok] (a_u6) at (-1.8000,-1.5000) {};
\node[tok] (a_u7) at (-1.8000,-2.3500) {};
\node[tok] (a_u8) at (1.8000,-0.6500) {};
\node[tok] (a_u9) at (1.8000,-1.5000) {};
\node[tok] (a_u10) at (1.8000,-2.3500) {};
\node[tok] (a_u11) at (0.0000,-1.3500) {};
\node[tok] (a_u12) at (-0.5500,-3.2000) {};
\node[v] (a_u13) at (0.5500,-3.2000) {};
\draw (a_u0) -- (a_u1);
\draw (a_u1) -- (a_u2);
\draw (a_u2) -- (a_u3);
\draw (a_u2) -- (a_u4);
\draw (a_u0) -- (a_u5);
\draw (a_u5) -- (a_u6);
\draw (a_u6) -- (a_u7);
\draw (a_u0) -- (a_u8);
\draw (a_u8) -- (a_u9);
\draw (a_u9) -- (a_u10);
\draw (a_u0) -- (a_u11);
\draw (a_u11) -- (a_u12);
\draw (a_u11) -- (a_u13);
\node[font=\small,anchor=east,inner sep=1pt] at (-0.3800,0.0000) {$u_{0}$};
\node[font=\small,anchor=east,inner sep=1pt] at (-0.3800,1.0000) {$u_{1}$};
\node[font=\small,anchor=east,inner sep=1pt] at (-0.3800,2.0000) {$u_{2}$};
\node[font=\small,anchor=east,inner sep=1pt] at (-1.1800,2.8500) {$u_{3}$};
\node[font=\small,anchor=west,inner sep=1pt] at (1.1800,2.8500) {$u_{4}$};
\node[font=\small,anchor=east,inner sep=1pt] at (-2.1800,-0.4500) {$u_{5}$};
\node[font=\small,anchor=east,inner sep=1pt] at (-2.1800,-0.8500) {$(1)$};
\node[font=\small,anchor=east,inner sep=1pt] at (-2.1800,-1.3000) {$u_{6}$};
\node[font=\small,anchor=east,inner sep=1pt] at (-2.1800,-1.7000) {$(4)$};
\node[font=\small,anchor=east,inner sep=1pt] at (-2.1800,-2.1500) {$u_{7}$};
\node[font=\small,anchor=east,inner sep=1pt] at (-2.1800,-2.5500) {$(5)$};
\node[font=\small,anchor=west,inner sep=1pt] at (2.1800,-0.4500) {$u_{8}$};
\node[font=\small,anchor=west,inner sep=1pt] at (2.1800,-0.8500) {$(2)$};
\node[font=\small,anchor=west,inner sep=1pt] at (2.1800,-1.3000) {$u_{9}$};
\node[font=\small,anchor=west,inner sep=1pt] at (2.1800,-1.7000) {$(6)$};
\node[font=\small,anchor=west,inner sep=1pt] at (2.1800,-2.1500) {$u_{10}$};
\node[font=\small,anchor=west,inner sep=1pt] at (2.1800,-2.5500) {$(7)$};
\node[font=\small,anchor=west,inner sep=1pt] at (0.3800,-1.1500) {$u_{11}$};
\node[font=\small,anchor=west,inner sep=1pt] at (0.3800,-1.5500) {$(3)$};
\node[font=\small,anchor=east,inner sep=1pt] at (-0.9300,-3.0000) {$u_{12}$};
\node[font=\small,anchor=east,inner sep=1pt] at (-0.9300,-3.4000) {$(8)$};
\node[font=\small,anchor=west,inner sep=1pt] at (0.9300,-3.0000) {$u_{13}$};
\node[font=\small,anchor=west,inner sep=1pt] at (0.9300,-3.4000) {$(9)$};
\draw[dotted,rounded corners=3pt] (-2.9000,-3.8500) rectangle (2.9000,0.4500);
\node[font=\small,anchor=center,inner sep=1pt] at (-2.6000,0.8200) {$U$};
\node[font=\small,anchor=center,inner sep=1pt] at (1.2000,1.5000) {$W'$};
\node[font=\small,anchor=center,inner sep=1pt] at (0.0000,-4.3000) {(a) $C_{1}$};
\node[tok] (b_u0) at (5.9000,0.0000) {};
\node[v] (b_u1) at (5.9000,1.0000) {};
\node[v] (b_u2) at (5.9000,2.0000) {};
\node[v] (b_u3) at (5.1000,2.8500) {};
\node[v] (b_u4) at (6.7000,2.8500) {};
\node[tok] (b_u5) at (4.1000,-0.6500) {};
\node[v] (b_u6) at (4.1000,-1.5000) {};
\node[v] (b_u7) at (4.1000,-2.3500) {};
\node[tok] (b_u8) at (7.7000,-0.6500) {};
\node[v] (b_u9) at (7.7000,-1.5000) {};
\node[v] (b_u10) at (7.7000,-2.3500) {};
\node[tok] (b_u11) at (5.9000,-1.3500) {};
\node[v] (b_u12) at (5.3500,-3.2000) {};
\node[v] (b_u13) at (6.4500,-3.2000) {};
\draw (b_u0) -- (b_u1);
\draw (b_u1) -- (b_u2);
\draw (b_u2) -- (b_u3);
\draw (b_u2) -- (b_u4);
\draw (b_u0) -- (b_u5);
\draw (b_u5) -- (b_u6);
\draw (b_u6) -- (b_u7);
\draw (b_u0) -- (b_u8);
\draw (b_u8) -- (b_u9);
\draw (b_u9) -- (b_u10);
\draw (b_u0) -- (b_u11);
\draw (b_u11) -- (b_u12);
\draw (b_u11) -- (b_u13);
\node[font=\small,anchor=east,inner sep=1pt] at (5.5200,0.0000) {$u_{0}$};
\node[font=\small,anchor=east,inner sep=1pt] at (5.5200,1.0000) {$u_{1}$};
\node[font=\small,anchor=east,inner sep=1pt] at (5.5200,2.0000) {$u_{2}$};
\node[font=\small,anchor=east,inner sep=1pt] at (4.7200,2.8500) {$u_{3}$};
\node[font=\small,anchor=west,inner sep=1pt] at (7.0800,2.8500) {$u_{4}$};
\node[font=\small,anchor=east,inner sep=1pt] at (3.7200,-0.4500) {$u_{5}$};
\node[font=\small,anchor=east,inner sep=1pt] at (3.7200,-0.8500) {$(1)$};
\node[font=\small,anchor=east,inner sep=1pt] at (3.7200,-1.3000) {$u_{6}$};
\node[font=\small,anchor=east,inner sep=1pt] at (3.7200,-1.7000) {$(4)$};
\node[font=\small,anchor=east,inner sep=1pt] at (3.7200,-2.1500) {$u_{7}$};
\node[font=\small,anchor=east,inner sep=1pt] at (3.7200,-2.5500) {$(5)$};
\node[font=\small,anchor=west,inner sep=1pt] at (8.0800,-0.4500) {$u_{8}$};
\node[font=\small,anchor=west,inner sep=1pt] at (8.0800,-0.8500) {$(2)$};
\node[font=\small,anchor=west,inner sep=1pt] at (8.0800,-1.3000) {$u_{9}$};
\node[font=\small,anchor=west,inner sep=1pt] at (8.0800,-1.7000) {$(6)$};
\node[font=\small,anchor=west,inner sep=1pt] at (8.0800,-2.1500) {$u_{10}$};
\node[font=\small,anchor=west,inner sep=1pt] at (8.0800,-2.5500) {$(7)$};
\node[font=\small,anchor=west,inner sep=1pt] at (6.2800,-1.1500) {$u_{11}$};
\node[font=\small,anchor=west,inner sep=1pt] at (6.2800,-1.5500) {$(3)$};
\node[font=\small,anchor=east,inner sep=1pt] at (4.9700,-3.0000) {$u_{12}$};
\node[font=\small,anchor=east,inner sep=1pt] at (4.9700,-3.4000) {$(8)$};
\node[font=\small,anchor=west,inner sep=1pt] at (6.8300,-3.0000) {$u_{13}$};
\node[font=\small,anchor=west,inner sep=1pt] at (6.8300,-3.4000) {$(9)$};
\draw[dotted,rounded corners=3pt] (3.0000,-3.8500) rectangle (8.8000,0.4500);
\node[font=\small,anchor=center,inner sep=1pt] at (3.3000,0.8200) {$U$};
\node[font=\small,anchor=center,inner sep=1pt] at (7.1000,1.5000) {$W'$};
\node[font=\small,anchor=center,inner sep=1pt] at (5.9000,-4.3000) {(b) $C_{6}$};
\node[tok] (c_u0) at (11.8000,0.0000) {};
\node[v] (c_u1) at (11.8000,1.0000) {};
\node[v] (c_u2) at (11.8000,2.0000) {};
\node[v] (c_u3) at (11.0000,2.8500) {};
\node[v] (c_u4) at (12.6000,2.8500) {};
\node[tok] (c_u5) at (10.0000,-0.6500) {};
\node[v] (c_u6) at (10.0000,-1.5000) {};
\node[v] (c_u7) at (10.0000,-2.3500) {};
\node[v] (c_u8) at (13.6000,-0.6500) {};
\node[v] (c_u9) at (13.6000,-1.5000) {};
\node[v] (c_u10) at (13.6000,-2.3500) {};
\node[v] (c_u11) at (11.8000,-1.3500) {};
\node[v] (c_u12) at (11.2500,-3.2000) {};
\node[v] (c_u13) at (12.3500,-3.2000) {};
\draw (c_u0) -- (c_u1);
\draw (c_u1) -- (c_u2);
\draw (c_u2) -- (c_u3);
\draw (c_u2) -- (c_u4);
\draw (c_u0) -- (c_u5);
\draw (c_u5) -- (c_u6);
\draw (c_u6) -- (c_u7);
\draw (c_u0) -- (c_u8);
\draw (c_u8) -- (c_u9);
\draw (c_u9) -- (c_u10);
\draw (c_u0) -- (c_u11);
\draw (c_u11) -- (c_u12);
\draw (c_u11) -- (c_u13);
\node[font=\small,anchor=east,inner sep=1pt] at (11.4200,0.0000) {$u_{0}$};
\node[font=\small,anchor=east,inner sep=1pt] at (11.4200,1.0000) {$u_{1}$};
\node[font=\small,anchor=east,inner sep=1pt] at (11.4200,2.0000) {$u_{2}$};
\node[font=\small,anchor=east,inner sep=1pt] at (10.6200,2.8500) {$u_{3}$};
\node[font=\small,anchor=west,inner sep=1pt] at (12.9800,2.8500) {$u_{4}$};
\node[font=\small,anchor=east,inner sep=1pt] at (9.6200,-0.4500) {$u_{5}$};
\node[font=\small,anchor=east,inner sep=1pt] at (9.6200,-0.8500) {$(1)$};
\node[font=\small,anchor=east,inner sep=1pt] at (9.6200,-1.3000) {$u_{6}$};
\node[font=\small,anchor=east,inner sep=1pt] at (9.6200,-1.7000) {$(4)$};
\node[font=\small,anchor=east,inner sep=1pt] at (9.6200,-2.1500) {$u_{7}$};
\node[font=\small,anchor=east,inner sep=1pt] at (9.6200,-2.5500) {$(5)$};
\node[font=\small,anchor=west,inner sep=1pt] at (13.9800,-0.4500) {$u_{8}$};
\node[font=\small,anchor=west,inner sep=1pt] at (13.9800,-0.8500) {$(2)$};
\node[font=\small,anchor=west,inner sep=1pt] at (13.9800,-1.3000) {$u_{9}$};
\node[font=\small,anchor=west,inner sep=1pt] at (13.9800,-1.7000) {$(6)$};
\node[font=\small,anchor=west,inner sep=1pt] at (13.9800,-2.1500) {$u_{10}$};
\node[font=\small,anchor=west,inner sep=1pt] at (13.9800,-2.5500) {$(7)$};
\node[font=\small,anchor=west,inner sep=1pt] at (12.1800,-1.1500) {$u_{11}$};
\node[font=\small,anchor=west,inner sep=1pt] at (12.1800,-1.5500) {$(3)$};
\node[font=\small,anchor=east,inner sep=1pt] at (10.8700,-3.0000) {$u_{12}$};
\node[font=\small,anchor=east,inner sep=1pt] at (10.8700,-3.4000) {$(8)$};
\node[font=\small,anchor=west,inner sep=1pt] at (12.7300,-3.0000) {$u_{13}$};
\node[font=\small,anchor=west,inner sep=1pt] at (12.7300,-3.4000) {$(9)$};
\draw[dotted,rounded corners=3pt] (8.9000,-3.8500) rectangle (14.7000,0.4500);
\node[font=\small,anchor=center,inner sep=1pt] at (9.2000,0.8200) {$U$};
\node[font=\small,anchor=center,inner sep=1pt] at (13.0000,1.5000) {$W'$};
\node[font=\small,anchor=center,inner sep=1pt] at (11.8000,-4.3000) {(c) $C_{8}$};
\path[use as bounding box] (-3.1,-4.7) rectangle (14.9,3.3);
\end{tikzpicture}
\relax\space \end{adjustbox}
\caption{The properly rooted subtree $U=T^{u_{1}}_{u_{0}}$ for $k=4$ is enclosed by a dotted outline, with $W'=\{u_{1},u_{2},u_{3},u_{4}\}$ outside it.
Panels (a), (b), and (c) show $C_{1}$, $C_{6}$, $C_{8}$, respectively; black vertices belong to the prescribed set and white vertices are unoccupied.
The numbers in parentheses beside vertex names give positions in the list defining $C_h$, with the root included separately.
The branches rooted at $u_{5}$, $u_{8}$, $u_{11}$ have rooted heights $3$, $3$, $2$, respectively, giving $D=3$, $m=1$, $d=2$, and $\gamma(U)=8$.
In particular, $C_{6}=\{u_{0},u_{5},u_{8},u_{11}\}$ and $C_{8}=\{u_{0},u_{5}\}$ illustrate \cref{lem:properly-rooted-count}.}
\label{fig:f3-prescribed}
\end{figure}
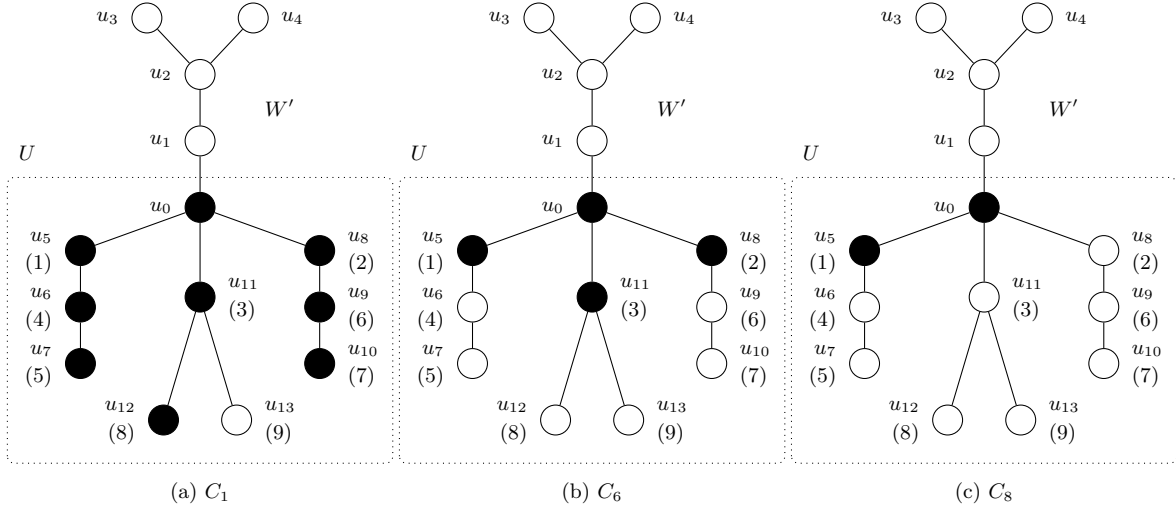

\cref{fig:f3-prescribed} illustrates a properly rooted subtree and its prescribed sets.

The next lemma gives the properly rooted subtrees used in the inductive proofs.

\begin{lemma}\label{lem:properly-rooted-existence}
Let $W\subseteq V(T)$ be connected.
\begin{enumerate}[(a)]
\item If $r\in W$ and $T[W-r]$ contains a path on $k$ vertices, then some edge $uv$ of $T[W]$ gives a properly rooted subtree $W^v_u$ with $r\notin W^v_u$.
\item If $T[W]$ contains a path on $k$ vertices and $|W|>k$, then some edge $uv$ of $T[W]$ gives a properly rooted subtree $W^v_u$.
\end{enumerate}
\end{lemma}

\begin{proof}
We choose a smallest edge side that contains a path on $k$ vertices and avoids $r$.
\begin{enumerate}[(a)]
\item The path lies in a component $W^r_y$ of $T[W-r]$, where $y$ is a neighbor of $r$, so some edge $uv$ of $T[W]$ has $W^v_u$ containing a path on $k$ vertices and avoiding $r$; among such edges, $uv$ is chosen with $|W^v_u|$ least.
For every neighbor $y\neq v$ of $u$, the set $W^u_y$ is contained in $W^v_u-u$, so it is smaller and avoids $r$.
By the choice of $uv$, it contains no path on $k$ vertices, so $W^v_u$ is properly rooted at $u$.
\item As $|W|>k$, some vertex $r\in W$ lies outside a given path on $k$ vertices, and $T[W-r]$ contains that path, so (a) applies.
\qedhere
\end{enumerate}
\end{proof}

\subsection{Basic Properties of TS-Sequences}\label{sec:basic}

We first show that fixing occupied vertices allows independent reconfiguration on the remaining components.
We then study adding and removing tokens at one vertex, and use these operations to relate rigid tokens to vertices that are never occupied.
The final consequences concern a maximum number of tokens on one side of an edge, the addition of a token to a $k$-PVC whose tokens are all movable, and the endpoint of an edge on one side of it.

\begin{lemma}\label{lem:fixing-occupied-set}
Let $S$ be a $k$-PVC of $T[W]$, let $X\subseteq S$, and let $Q$ be a union of vertex sets of components of $T[W]-X$.
\begin{enumerate}[(a)]
\item For every $k$-PVC $A$ of $T[Q]$, the set $(S-Q)+A$ is a $k$-PVC of $T[W]$.
\item Every TS-sequence in $T[Q]$ starting at $S\cap Q$ extends to a TS-sequence in $T[W]$ that agrees with $S$ outside $Q$.
\item Every TS-sequence in $T[W]$ whose members contain $X$ restricts, after repeated consecutive sets are omitted, to a TS-sequence in $T[Q]$.
\end{enumerate}
If $X\subseteq\Rigid_W(S)$, then $[S]_W$ consists exactly of the sets obtained by adjoining $X$ to independently chosen members of $[S\cap P]_P$ for the vertex sets $P$ of the components of $T[W]-X$.
\end{lemma}

\begin{proof}
We distinguish paths meeting $X$ from paths lying in a component of $T[W]-X$.
\begin{enumerate}[(a)]
\item A path meeting $X$ contains an occupied vertex.
A path avoiding $X$ lies in one component of $T[W]-X$, so a path on $k$ vertices in $Q$ meets $A$ and one outside $Q$ meets $S-Q$.
Thus, $(S-Q)+A$ is a $k$-PVC.
\item Each slide in $T[Q]$ uses the same occupied source and unoccupied target after $S-Q$ is adjoined.
Part (a) ensures that both endpoints of every slide are $k$-PVCs of $T[W]$.
\item No slide of the given sequence uses a vertex of $X$, since every such vertex remains occupied.
The endpoints of every slide therefore lie in one component of $T[W]-X$.
Its restriction to $Q$ is either a slide or an unchanged set, which proves the assertion.
\end{enumerate}
When $X\subseteq\Rigid_W(S)$, part (c) applies to every TS-sequence starting at $S$, and gives one inclusion in the final assertion.
Part (b), applied to the components one at a time, gives the reverse inclusion.
\end{proof}

We will often reconfigure within a tree where every subset of the required cardinality is a $k$-PVC.
The reachability assertion below is a special case of the connectivity result of Fabila-Monroy et al.~\cite{FabilaMonroy2012}; we include a direct proof.

\begin{lemma}\label{lem:unrestricted-sliding}
Let $W\subseteq V(T)$ be connected and let $0\leq t\leq |W|$.
If $T[W]$ contains no path on $k$ vertices or $|W|-t<k$, then every $t$-element subset of $W$ is a $k$-PVC, and every two such subsets are reachable from one another in $T[W]$.
\end{lemma}

\begin{proof}
We exchange one token with one unoccupied vertex at a time; every set below has $t$ elements, so it is a $k$-PVC: under either hypothesis of the lemma, there is no path on $k$ vertices at all, or there are too few unoccupied vertices to form one.
Let $A\subseteq W$ have $t$ elements, let $b\in A$ and $a\in W-A$, and let $a=v_0,v_1,\dots,v_L=b$ be the path of $T[W]$ joining them.
We show by induction on $L$ that $A$ reaches $A-b+a$.
For $L=1$, the exchange is the slide $b\to a$.
For $L\geq 2$, there are two cases.
\begin{itemize}
\item \textbf{Case 1: $v_{L-1}\in A$.}
By the induction hypothesis, $A$ reaches $A-v_{L-1}+a$.
The slide $b\to v_{L-1}$ then gives $A-b+a$.
\item \textbf{Case 2: $v_{L-1}\notin A$.}
The slide $b\to v_{L-1}$ first gives $A-b+v_{L-1}$.
By the induction hypothesis, this set reaches $A-b+a$.
\end{itemize}
Now let $A,B\subseteq W$ have $t$ elements.
If $A=B$, the zero-slide TS-sequence $\langle A\rangle$ joins them.
If $A\neq B$, then some $a\in B-A$ and $b\in A-B$ exist, $A$ reaches $A'=A-b+a$, and $|B-A'|=|B-A|-1$, so induction on $|B-A|$ joins $A$ to $B$.
\end{proof}

Adding a token at a fixed vertex preserves reachability when that vertex is unoccupied at both endpoints.

\begin{lemma}\label{lem:adding-token}
Let $W\subseteq V(T)$ be connected, let $r\in W$, and let $A,B$ be $k$-PVCs of $T[W]$ with $r\notin A\cup B$.
If $A\sevstep{W}B$, then $A+r\sevstep{W}B+r$.
\end{lemma}

\begin{proof}
We use induction on $|W|$ to replace each part of a TS-sequence during which $r$ becomes occupied and then becomes unoccupied.
The assertion is immediate for $|W|=1$.
Let a TS-sequence from $A$ to $B$ be fixed.
Every interval in which $r$ remains unoccupied extends to a TS-sequence after $r$ is added to all its members, since supersets of $k$-PVCs are $k$-PVCs.
It remains to handle an interval consisting of a slide from a neighbor $i$ to $r$, a sequence keeping $r$ occupied, and a slide from $r$ to a neighbor $j$.
Let $A_0$ be the $k$-PVC immediately after the first slide and $B_0$ the $k$-PVC immediately before the last slide.
The required replacement joins $A_0+i$ to $B_0+j$.
For every neighbor $y$ of $r$, the occupied interval restricts to a TS-sequence from $A_0\cap W^r_y$ to $B_0\cap W^r_y$, by \cref{lem:fixing-occupied-set}(c).

\begin{itemize}
\item \textbf{Case 1: $i=j$.}
Both endpoints of the restricted sequence in $W^r_i$ omit $i$.
Since $|W^r_i|<|W|$, the induction hypothesis joins $(A_0\cap W^r_i)+i$ to $(B_0\cap W^r_i)+i$ inside $T[W^r_i]$.
We execute that sequence with $r$ occupied, followed by the restricted sequences in every other branch.
All these sequences extend by \cref{lem:fixing-occupied-set}(b), and the result is $B_0+i$.

\item \textbf{Case 2: $i\neq j$.}
Starting at $A_0+i$, we keep $r$ occupied and execute the restricted sequence in $W^r_j$ only.
The resulting set is
\[
 Z_1=(A_0+i-W^r_j)+(B_0\cap W^r_j).
\]
The vertex $j$ is unoccupied in $Z_1$, so the proposed slide from $r$ to $j$ gives
\[
 Z_2=(A_0-r+i-W^r_j)+(B_0\cap W^r_j)+j.
\]
To check that $Z_2$ is a $k$-PVC, we compare it with the $k$-PVC $A_0-r+i$ immediately before the original arrival at $r$.
A path on $k$ vertices outside $W^r_j$ is still met by that set.
A path wholly inside $W^r_j$ meets $(B_0\cap W^r_j)+j$, and a path crossing the boundary of $W^r_j$ contains the newly occupied vertex $j$.
Thus, the slide to $j$ is legal.
The subsequent slide from $i$ to $r$ is also legal, because $r$ becomes occupied and every branch restriction is a $k$-PVC.
At this point, the restriction to $W^r_i$ is $A_0\cap W^r_i$, the restriction to $W^r_j$ is $(B_0\cap W^r_j)+j$, and every other branch still has its restriction from $A_0$.
Executing the restricted sequences in $W^r_i$ and in all branches other than $W^r_j$, with $r$ occupied, ends at $B_0+j$.
\end{itemize}
The replacements of all intervals concatenate to the required TS-sequence from $A+r$ to $B+r$.
\end{proof}

The next consequence allows a projected TS-sequence to add tokens at a boundary vertex, provided its token count never exceeds the initial count.

\begin{corollary}\label{cor:root-removal-history}
Let $W\subseteq V(T)$ be connected, let $r\in W$, and let $Z_0,\dots,Z_\ell$ be $k$-PVCs of $T[W]$.
Suppose that consecutive sets differ by a slide, by a removal at $r$, or by the addition of a token at $r$, and that $|Z_i|\leq |Z_0|$ for every $i$.
Then $Z_0\rmstep{W}{r}Z_i$ for every $i$.
\end{corollary}

\begin{proof}
We follow the sequence and undo each addition at $r$ with the last removal before it, using \cref{lem:adding-token}.
We show that $Z_0\rmstep{W}{r}Z_i$ by induction on $i$; the case $i=0$ is trivial.
Let $i<\ell$; by the induction hypothesis, slides and removals at $r$ lead from $Z_0$ to $Z_i$, with every intermediate set a $k$-PVC of $T[W]$.
If $Z_{i+1}$ arises from $Z_i$ by a slide or by a removal at $r$, these operations followed by that step lead to $Z_{i+1}$.
Otherwise, $Z_{i+1}=Z_i+r$ with $r\notin Z_i$, so $|Z_i|<|Z_{i+1}|\leq|Z_0|$, and these operations contain a removal at $r$, as slides keep the number of tokens.
Let $C$ be the set just before the last of these removals.
Then $Z_0\rmstep{W}{r}C$, the vertex $r$ lies in $C$, the set $C-r$ is a $k$-PVC of $T[W]$, and $C-r\sevstep{W}Z_i$, since only slides follow that removal.
By \cref{lem:adding-token}, applied to $C-r$ and $Z_i$, we have $C\sevstep{W}Z_i+r=Z_{i+1}$, so $Z_0\rmstep{W}{r}Z_{i+1}$.
\end{proof}

Removing tokens cannot allow a vertex that was never occupied to become occupied.

\begin{lemma}\label{lem:never-occupied-persist}
Let $W\subseteq V(T)$, let $r\in W$, and let $Z$ be a $k$-PVC of $T[W]$.
If $Z\rmstep{W}{r}Y$, then $Y$ is contained in a member of $[Z]_W$ and
$\NeverOcc_W(Z)\subseteq\NeverOcc_W(Y)$.
\end{lemma}

\begin{proof}
We follow the slides and removals while maintaining a containing $k$-PVC reachable from $Z$.
Initially, that containing set is $Z$ itself.
Suppose that the current set $A$ is contained in a $k$-PVC $B\in[Z]_W$.
A removal from $A$ preserves the inclusion without changing $B$.
For a slide from $x$ to $y$ in $A$, if $y\in B$, then $B$ still contains $A-x+y$.
If $y\notin B$, then the same slide replaces $B$ with $B-x+y$, which contains $A-x+y$ and is a $k$-PVC because it contains a $k$-PVC.
Thus, every set obtained by the operations is contained in a member of $[Z]_W$.
The same argument applies after appending an arbitrary TS-sequence starting at $Y$.
Every vertex occupied in a member of $[Y]_W$ is therefore occupied in a member of $[Z]_W$, which gives the stated inclusion.
\end{proof}

The vertices that are never occupied determine the smallest possible value of $\varepsilon$ at a specified vertex.

\begin{lemma}\label{lem:min-eps-never-occupied}
Let $W\subseteq V(T)$ be connected, let $S$ be a $k$-PVC of $T[W]$, and let $r\in W$.
If $r\notin\NeverOcc_W(S)$, then $\min\{\varepsilon_W(S',r):S'\in[S]_W\}=0$.
Otherwise, this minimum equals the number of vertices of a longest path of $T[\NeverOcc_W(S)]$ starting at $r$.
Moreover, some $S'\in[S]_W$ attains the minimum and, if $r\notin S'$, the component of $T[W-S']$ containing $r$ lies in $\NeverOcc_W(S)$.
\end{lemma}

\begin{proof}
We attain the minimum by occupying all vertices just outside the component of never-occupied vertices that contains $r$.
If $r\notin\NeverOcc_W(S)$, some reachable $k$-PVC contains $r$, giving the minimum zero and the last assertion.
Suppose that $r\in\NeverOcc_W(S)$, and let $C$ be the vertex set of its component in $T[\NeverOcc_W(S)]$.
Every vertex of $C$ remains unoccupied, giving the asserted lower bound.
Each component $Q$ of $T[W-C]$ has a unique edge to $C$; let $z_Q\in Q$ be its endpoint outside $C$.
The uniqueness follows because $T[W]$ is a tree and $C$ is connected.
Moreover, $z_Q\notin\NeverOcc_W(S)$, since otherwise $z_Q$ would belong to $C$.

We occupy the vertices $z_Q$ one at a time and keep each treated component fixed afterwards.
From the current reachable $k$-PVC, there is a TS-sequence to one containing the next $z_Q$.
No slide of this sequence meets $C$, since its vertices are never occupied.
We retain only the slides in components not yet fixed.
Every resulting set is a $k$-PVC: a path wholly inside a fixed component meets its fixed restriction, a path crossing its boundary contains its occupied vertex $z_Q$, and a path avoiding all fixed components has the same occupation as in the original sequence.
Every retained slide also has its original source and target occupations.
Thus, the retained slides legally occupy the next $z_Q$ without changing the earlier components.
After all components have been treated, every neighbor of $C$ outside $C$ is occupied.
The component of unoccupied vertices containing $r$ is then exactly $C\subseteq\NeverOcc_W(S)$, which attains the lower bound and gives the last assertion.
\end{proof}

A rigid token adjacent in $T[W]$ to a vertex without a rigid token must be the only token on a path whose other vertices can never be occupied.

\begin{lemma}\label{lem:rigid-path}
Let $W\subseteq V(T)$ be connected and let $S$ be a $k$-PVC of $T[W]$.
If $r\in\Rigid_W(S)$ has a neighbor in $T[W]$ outside $\Rigid_W(S)$, then some path on $k$ vertices of $T[W]$ contains $r$ and has every other vertex in $\NeverOcc_W(S)$.
In particular, such a path exists whenever $S\neq W$ and $\Rigid_W(S)\neq\emptyset$.
\end{lemma}

\begin{proof}
We prepare the branches at $r$ so that the unoccupied vertices joined to their roots are never occupied, and then use that the token on $r$ cannot slide to an unoccupied neighbor.
Let $z\in N_{T[W]}(r)-\Rigid_W(S)$, and let a reachable $k$-PVC omit $z$.
Since $r$ stays occupied, \cref{lem:fixing-occupied-set} identifies the possible restrictions to its branches with their independent sets of reachable $k$-PVCs.
In particular, the vertices never occupied in each branch are exactly the vertices of $\NeverOcc_W(S)$ in that branch.
Keeping $z$ unoccupied, we apply the last assertion of \cref{lem:min-eps-never-occupied} in every other branch at its root, independently; this gives $A\in[S]_W$ with $z\notin A$ such that, for every neighbor $y\neq z$ of $r$ with $y\notin A$, the component of $T[W^r_y-A]$ containing $y$ lies in $\NeverOcc_W(S)$.
As $r\in\Rigid_W(S)$, the slide $r\to z$ is not legal at $A$, so some path on $k$ vertices of $T[W]$ avoids $A-r+z$.
The path meets $A$, and only at $r$, so it contains $r$.
Its other vertices are joined to $r$ by vertices outside $A-r+z$, so each of them lies, for some neighbor $y\neq z$ of $r$ with $y\notin A$, in the component of $T[W^r_y-A]$ containing $y$, hence in $\NeverOcc_W(S)$.
Finally, if $S\neq W$, then $W-\Rigid_W(S)$ is nonempty.
When $\Rigid_W(S)$ is also nonempty, connectivity supplies an edge from that set to its complement, so the first assertion applies.
\end{proof}

A restriction to an arbitrary subtree need not retain movable tokens.
The maximum-count hypothesis in the next consequence supplies the property we will use.

\begin{corollary}\label{cor:maximum-side}
Let $W\subseteq V(T)$ be connected, let $uv$ be an edge of $T[W]$, and let $Q=W^v_u$.
Suppose that $S$ is a $k$-PVC of $T[W]$ with $\Rigid_W(S)=\emptyset$.
Let $M=\max\{|S'\cap Q|:S'\in[S]_W\}$.
If $M<|Q|$, then every $S'\in[S]_W$ with $|S'\cap Q|=M$ satisfies $\Rigid_Q(S'\cap Q)=\emptyset$.
\end{corollary}

\begin{proof}
We project a TS-sequence onto $Q$ and apply the persistence of never-occupied vertices.
Let $A=S'\cap Q$ for a restriction attaining $M$.
A TS-sequence starting at $S'$ restricts to slides in $Q$ and additions or removals at $u$ when a slide crosses $uv$.
Every restriction is a $k$-PVC of $T[Q]$ and has at most $M$ tokens.
By \cref{cor:root-removal-history}, each restriction can therefore be obtained from $A$ by slides and removals at $u$.
By \cref{lem:never-occupied-persist}, every vertex of $\NeverOcc_Q(A)$ is unoccupied throughout $[S]_W$.
If $\Rigid_Q(A)\neq\emptyset$, then $M<|Q|$ and \cref{lem:rigid-path} give a path on $k$ vertices in $T[Q]$ with all but one vertex in $\NeverOcc_Q(A)$.
Every member of $[S]_W$ must occupy the remaining vertex to meet that path.
That vertex would belong to $\Rigid_W(S)$, a contradiction.
\end{proof}

When all tokens are movable, adding a token at a fixed vertex keeps all tokens movable unless every vertex becomes occupied.
\Cref{cor:movable-counts} uses this fact to compare consecutive token counts.

\begin{corollary}\label{cor:adding-token-movable}
Let $W\subseteq V(T)$ be connected, let $S$ be a $k$-PVC of $T[W]$ with $\Rigid_W(S)=\emptyset$, and let $r\in W$.
Some $A\in[S]_W$ omits $r$, and $[A+r]_W$ is independent of the choice of $A$.
If $|S|+1<|W|$, then $\Rigid_W(A+r)=\emptyset$.
\end{corollary}

\begin{proof}
We combine \cref{lem:adding-token} with the path forced by a rigid token.
A reachable $A$ omitting $r$ exists because $\Rigid_W(S)=\emptyset$, and \cref{lem:adding-token} proves the independence assertion.
Suppose that $|S|+1<|W|$ and $\Rigid_W(A+r)\neq\emptyset$.
By \cref{lem:rigid-path}, some path on $k$ vertices has all but one vertex in $\NeverOcc_W(A+r)$.
Since removing the token on $r$ from $A+r$ gives $A$, \cref{lem:never-occupied-persist} yields $\NeverOcc_W(A+r)\subseteq\NeverOcc_W(A)$.
The remaining vertex of that path must then be occupied throughout $[A]_W=[S]_W$, contradicting $\Rigid_W(S)=\emptyset$.
\end{proof}

The next lemma shows that restricting a $k$-PVC whose tokens are all movable to one side of an edge leaves the endpoint on that side without a rigid token, unless that side is full.
Other tokens on that side need not be movable.

\begin{lemma}\label{lem:attachment-movable}
Let $W\subseteq V(T)$ be connected, and let $S$ be a $k$-PVC of $T[W]$ with $\Rigid_W(S)=\emptyset$.
For an edge $uv$ of $T[W]$, let $Q=W^u_v$.
If $S\cap Q\neq Q$, then $v\notin \Rigid_Q(S\cap Q)$.
\end{lemma}

\begin{proof}
We carry an invariant along each TS-sequence: its restriction to $Q$ is obtainable from $S\cap Q$ by slides and removals at $v$.
Suppose that $v\in\Rigid_Q(S\cap Q)$, and let $S=S_0,\dots,S_\ell$ be a TS-sequence in $T[W]$.
The invariant holds at $S_0$ with no slides or removals.
Suppose that it holds at $S_i$, and let $Z=S_i\cap Q$.
A slide inside $Q$ appends an internal slide, a slide from $v$ to $u$ appends a removal at $v$, and a slide outside $Q$ leaves the restriction unchanged.
For a slide from $u$ to $v$, we have $v\notin Z$ and $|Z|\leq|S\cap Q|$ by the invariant.
Equality would mean that the sequence supplied by the invariant has no removal, so $Z\in[S\cap Q]_Q$, contrary to rigidity at the unoccupied vertex $v$.
Thus, $|Z|<|S\cap Q|$.
Appending the addition at $v$ keeps every count at most $|S\cap Q|$, so \cref{cor:root-removal-history} converts the resulting sequence into slides and removals at $v$, proving the invariant at $S_{i+1}$.
Consequently, every restriction of a member of $[S]_W$ has at most $|S\cap Q|$ tokens, so $S\cap Q$ attains the maximum count on $Q$.
Since $S\cap Q\neq Q$, \cref{cor:maximum-side} makes all its tokens movable in $T[Q]$, contradicting rigidity at $v$.
\end{proof}
\relax\space \section{Characterization of Reachability}\label{sec:characterization}

This section proves that $[S]_W$ is determined by $\Rigid_W(S)$ and the token counts in the components of $T[W]-\Rigid_W(S)$.
We first reconfigure the restriction to a properly rooted subtree to a prescribed $k$-PVC, while ensuring that every token in the remaining subtree is movable.
The resulting reduction proves the Key Lemma by induction on the order, and fixing the rigid tokens then gives the characterization.

\subsection{Unoccupied Vertices in a Properly Rooted Subtree}\label{sec:unoccupied-in-subtree}

Let $W\subseteq V(T)$ be connected, let $uv$ be an edge of $T[W]$ such that $U=W^v_u$ is properly rooted, and let $W'=W^u_v$.
We use the branches $P_i$, roots $w_i$, heights $a_i$, integers $D,m,d,\gamma(U)$, and prescribed $k$-PVCs $C_h$ from \cref{sec:properly-rooted}.
The first statements also apply to a properly rooted tree $(U,u)$ by itself, with no $W'$.
When $u$ is occupied, every $k$-PVC and every TS-sequence on $W'$ extends with the restriction to $U$ fixed, because every path crossing $uv$ contains $u$.
The same assertion holds for removals of a token at a vertex of $W'$, provided every intermediate restriction is a $k$-PVC of $T[W']$.

We begin by interpreting $\gamma(U)$ as an exact bound and recording when the token on $u$ can be removed from $C_h$.

\begin{lemma}\label{lem:properly-rooted-count}
Let $(U,u)$ be properly rooted.
\begin{enumerate}[(a)]
\item The maximum number of unoccupied vertices in a $k$-PVC of $T[U]$ omitting $u$ is $\gamma(U)$.
\item For every $0\leq h\leq\gamma(U)$, the set $C_h$ is a $k$-PVC with $u\in C_h$ and $|U-C_h|=h$.
If $h<\gamma(U)$, then $C_h-u$ is also a $k$-PVC of $T[U]$.
\end{enumerate}
\end{lemma}

\begin{proof}
We count the branches that must contain a token when $u$ is unoccupied.
\cref{fig:f3-prescribed} illustrates the prescribed sets in part (b).
\begin{enumerate}[(a)]
\item Let $A$ be a $k$-PVC of $T[U]$ omitting $u$.
The heights of the branches containing no token have $\sigma$ at most $k-2$; otherwise, their rooted paths together with $u$ would contain an uncovered path on $k$ vertices.
At most $m$ branches are empty, so each of at least $D-m=d$ branches contains a token.
Thus, $|U-A|\leq |U|-d=\gamma(U)$.
Equality is attained by leaving the $m$ shortest branches empty and placing one token at the root of every other branch, with $u$ unoccupied.
Every path avoiding these tokens either lies in a branch, which has no path on $k$ vertices, or passes through $u$ and has at most $1+\sigma(\{a_1,\dots,a_m\})\leq k-1$ vertices.
\item The set $C_h$ contains $u$ and $|U|-1-h$ other vertices, so it is a $k$-PVC with exactly $h$ unoccupied vertices.
If $h<\gamma(U)$, then $|U|-1-h\geq d$.
When at least $D$ vertices other than $u$ are occupied, every branch root is occupied.
Otherwise, the empty branches are at most the $m$ shortest branches, and every other branch root is occupied.
After the token on $u$ is removed, every unoccupied path through $u$ therefore has at most $k-1$ vertices, as in part (a).
No branch contains a path on $k$ vertices, so $C_h-u$ is a $k$-PVC.
\qedhere
\end{enumerate}
\end{proof}

For a $k$-PVC $S$ with $u\in S$, let $b_i=|S\cap P_i|$ and let $E=\{i:b_i=0\}$.
When the restriction to $W'$ is fixed, let $x=\varepsilon_{W'}(S\cap W',v)$, the \emph{exterior value}; when there is no $W'$, let $x=0$.
The \emph{internal slide condition} used below is
\begin{equation}\label{eq:internal-slide-condition}
\text{there is an } A\in\{1,\dots,D\}\text{ with }b_A<|P_A|
\text{ such that }
\sigma\bigl(\{a_i:i\in E-\{A\}\}\cup\{x\}\bigr)\leq k-2.
\end{equation}
The union in \cref{eq:internal-slide-condition} is a multiset union, so the occurrence of $x$ is included even when it equals a branch height.

\begin{lemma}\label{lem:first-internal-slide}
Let $U=W^v_u$ be properly rooted in a connected set $W$, and let $S$ be a $k$-PVC of $T[W]$ with $u\in S$.
After slides inside $U$ that keep $u$ occupied, a first slide of the token on $u$ to a vertex of $U$ is possible with $W'$ fixed if and only if \cref{eq:internal-slide-condition} holds.
The same equivalence holds for $(U,u)$ by itself with $x=0$.
\end{lemma}

\begin{proof}
We compare the branch counts before the first slide from $u$ with the slide test in \cref{lem:slide-test}.
\begin{itemize}
\item[($\Rightarrow$)] Until the first slide from $u$, every branch retains its number of tokens.
If that slide goes to $w_A$, then $b_A<|P_A|$.
Every empty branch other than $P_A$ still has its entire rooted path of height $a_i$ unoccupied, and $W'$ still has value $x$ at $v$.
The necessary slide test implies \cref{eq:internal-slide-condition}, since omitting the contributions of other branches can only decrease $\sigma$.
\item[($\Leftarrow$)] Let $A$ satisfy \cref{eq:internal-slide-condition}.
By \cref{lem:unrestricted-sliding}, we can make $w_A$ unoccupied because $b_A<|P_A|$, and we can occupy the root of every other nonempty branch.
These preparations are independent and legal while $u$ is occupied, by \cref{lem:fixing-occupied-set}.
The nonzero values in the slide test for the slide from $u$ to $w_A$ are then exactly the heights of the empty branches other than $P_A$ and the value $x$.
Thus, \cref{eq:internal-slide-condition} makes the slide legal.
\qedhere
\end{itemize}
\end{proof}

Before normalizing a restriction, we may need to occupy its root.

\begin{lemma}\label{lem:occupying-root}
Let $U=W^v_u$ be properly rooted in a connected set $W$, and let $S$ be a $k$-PVC of $T[W]$ with $u\notin S$.
There is a TS-sequence using only edges of $T[U]$ from $S$ to a $k$-PVC containing $u$.
The restriction to $W'$ and the number of tokens in $U$ remain unchanged.
The same assertion holds for $(U,u)$ by itself.
\end{lemma}

\begin{proof}
We move a closest token in $U$ toward $u$ along their unique path.
Such a token exists because $T[U]$ contains a path on $k$ vertices.
Every vertex strictly between the chosen token and $u$ is initially unoccupied.
At each step, a newly uncovered path would have to contain the vertex just vacated by that token.
If the path leaves the branch containing that vertex toward $u$, it contains the new position of the token.
If it avoids that position, it lies entirely in a branch of $U-u$, which has no path on $k$ vertices.
Thus, every slide is legal, and the last slide occupies $u$.
\end{proof}

We next show that the numerical condition suffices to reach the prescribed $k$-PVC on $U$.
The proof first increases the number of nonempty branches and then balances their token counts against the prescribed set.

\begin{theorem}\label{thm:normalization}
Let $U=W^v_u$ be properly rooted in a connected set $W$, and let $S$ be a $k$-PVC of $T[W]$ with $u\in S$ satisfying \cref{eq:internal-slide-condition}.
For $h=|U-S|$, a TS-sequence inside $U$, with $W'$ fixed, reaches a $k$-PVC whose restriction to $U$ is $C_h$.
The assertion also holds for $(U,u)$ by itself with $x=0$, and for $S\cap U=U=C_0$ with a zero-slide TS-sequence.
\end{theorem}

\begin{proof}
We use two phases of transfers through $u$, preserving \cref{eq:internal-slide-condition} after each transfer.
The first phase spreads tokens to empty branches, and the second phase corrects the branch counts to those of $C_h$.
\cref{fig:f4-transfer} illustrates one such transfer.

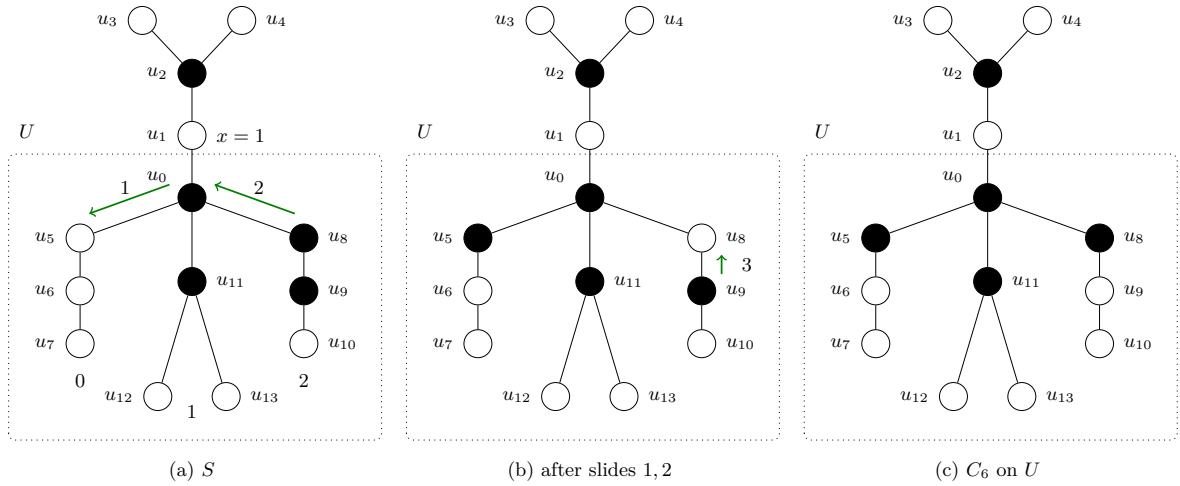
\begin{figure}[ht]
\centering
\begin{adjustbox}{max width=\textwidth}
\begin{tikzpicture}[v/.style={circle,draw=black,fill=white,inner sep=0pt,minimum size=4.5mm},
tok/.style={v,fill=black},rig/.style={v,fill=black,draw=black!30,line width=1.8pt},
nev/.style={v,fill=black!20},line width=.45pt]
\node[tok] (a_u0) at (0.0000,0.0000) {};
\node[v] (a_u1) at (0.0000,1.0000) {};
\node[tok] (a_u2) at (0.0000,2.0000) {};
\node[v] (a_u3) at (-0.8000,2.8500) {};
\node[v] (a_u4) at (0.8000,2.8500) {};
\node[v] (a_u5) at (-1.8000,-0.6500) {};
\node[v] (a_u6) at (-1.8000,-1.5000) {};
\node[v] (a_u7) at (-1.8000,-2.3500) {};
\node[tok] (a_u8) at (1.8000,-0.6500) {};
\node[tok] (a_u9) at (1.8000,-1.5000) {};
\node[v] (a_u10) at (1.8000,-2.3500) {};
\node[tok] (a_u11) at (0.0000,-1.3500) {};
\node[v] (a_u12) at (-0.5500,-3.2000) {};
\node[v] (a_u13) at (0.5500,-3.2000) {};
\draw (a_u0) -- (a_u1);
\draw (a_u1) -- (a_u2);
\draw (a_u2) -- (a_u3);
\draw (a_u2) -- (a_u4);
\draw (a_u0) -- (a_u5);
\draw (a_u5) -- (a_u6);
\draw (a_u6) -- (a_u7);
\draw (a_u0) -- (a_u8);
\draw (a_u8) -- (a_u9);
\draw (a_u9) -- (a_u10);
\draw (a_u0) -- (a_u11);
\draw (a_u11) -- (a_u12);
\draw (a_u11) -- (a_u13);
\node[font=\small,anchor=east,inner sep=1pt] at (-0.3500,0.3300) {$u_{0}$};
\node[font=\small,anchor=east,inner sep=1pt] at (-0.3500,1.0000) {$u_{1}$};
\node[font=\small,anchor=west,inner sep=1pt] at (0.3500,1.0000) {$x=1$};
\node[font=\small,anchor=east,inner sep=1pt] at (-0.3500,2.0000) {$u_{2}$};
\node[font=\small,anchor=east,inner sep=1pt] at (-1.1500,2.8500) {$u_{3}$};
\node[font=\small,anchor=west,inner sep=1pt] at (1.1500,2.8500) {$u_{4}$};
\node[font=\small,anchor=east,inner sep=1pt] at (-2.1500,-0.6500) {$u_{5}$};
\node[font=\small,anchor=east,inner sep=1pt] at (-2.1500,-1.5000) {$u_{6}$};
\node[font=\small,anchor=east,inner sep=1pt] at (-2.1500,-2.3500) {$u_{7}$};
\node[font=\small,anchor=west,inner sep=1pt] at (2.1500,-0.6500) {$u_{8}$};
\node[font=\small,anchor=west,inner sep=1pt] at (2.1500,-1.5000) {$u_{9}$};
\node[font=\small,anchor=west,inner sep=1pt] at (2.1500,-2.3500) {$u_{10}$};
\node[font=\small,anchor=west,inner sep=1pt] at (0.3500,-1.3500) {$u_{11}$};
\node[font=\small,anchor=east,inner sep=1pt] at (-0.9000,-3.2000) {$u_{12}$};
\node[font=\small,anchor=west,inner sep=1pt] at (0.9000,-3.2000) {$u_{13}$};
\draw[dotted,rounded corners=3pt] (-2.9500,-3.9000) rectangle (3.0500,0.7000);
\node[font=\small,anchor=center,inner sep=1pt] at (-2.6500,1.0700) {$U$};
\draw[->,thick,green!50!black] (-0.3626,0.2093) -- (-1.6547,-0.2573);
\node[font=\small,anchor=center,inner sep=1pt] at (-1.0766,0.1641) {$1$};
\draw[->,thick,green!50!black] (1.6547,-0.2573) -- (0.3626,0.2093);
\node[font=\small,anchor=center,inner sep=1pt] at (1.0766,0.1641) {$2$};
\node[font=\small,anchor=center,inner sep=1pt] at (-1.8000,-2.9500) {$0$};
\node[font=\small,anchor=center,inner sep=1pt] at (1.8000,-2.9500) {$2$};
\node[font=\small,anchor=center,inner sep=1pt] at (0.0000,-3.4500) {$1$};
\node[font=\small,anchor=center,inner sep=1pt] at (0.0000,-4.4000) {(a) $S$};
\node[tok] (b_u0) at (6.4000,0.0000) {};
\node[v] (b_u1) at (6.4000,1.0000) {};
\node[tok] (b_u2) at (6.4000,2.0000) {};
\node[v] (b_u3) at (5.6000,2.8500) {};
\node[v] (b_u4) at (7.2000,2.8500) {};
\node[tok] (b_u5) at (4.6000,-0.6500) {};
\node[v] (b_u6) at (4.6000,-1.5000) {};
\node[v] (b_u7) at (4.6000,-2.3500) {};
\node[v] (b_u8) at (8.2000,-0.6500) {};
\node[tok] (b_u9) at (8.2000,-1.5000) {};
\node[v] (b_u10) at (8.2000,-2.3500) {};
\node[tok] (b_u11) at (6.4000,-1.3500) {};
\node[v] (b_u12) at (5.8500,-3.2000) {};
\node[v] (b_u13) at (6.9500,-3.2000) {};
\draw (b_u0) -- (b_u1);
\draw (b_u1) -- (b_u2);
\draw (b_u2) -- (b_u3);
\draw (b_u2) -- (b_u4);
\draw (b_u0) -- (b_u5);
\draw (b_u5) -- (b_u6);
\draw (b_u6) -- (b_u7);
\draw (b_u0) -- (b_u8);
\draw (b_u8) -- (b_u9);
\draw (b_u9) -- (b_u10);
\draw (b_u0) -- (b_u11);
\draw (b_u11) -- (b_u12);
\draw (b_u11) -- (b_u13);
\node[font=\small,anchor=east,inner sep=1pt] at (6.0500,0.3300) {$u_{0}$};
\node[font=\small,anchor=east,inner sep=1pt] at (6.0500,1.0000) {$u_{1}$};
\node[font=\small,anchor=east,inner sep=1pt] at (6.0500,2.0000) {$u_{2}$};
\node[font=\small,anchor=east,inner sep=1pt] at (5.2500,2.8500) {$u_{3}$};
\node[font=\small,anchor=west,inner sep=1pt] at (7.5500,2.8500) {$u_{4}$};
\node[font=\small,anchor=east,inner sep=1pt] at (4.2500,-0.6500) {$u_{5}$};
\node[font=\small,anchor=east,inner sep=1pt] at (4.2500,-1.5000) {$u_{6}$};
\node[font=\small,anchor=east,inner sep=1pt] at (4.2500,-2.3500) {$u_{7}$};
\node[font=\small,anchor=west,inner sep=1pt] at (8.5500,-0.6500) {$u_{8}$};
\node[font=\small,anchor=west,inner sep=1pt] at (8.5500,-1.5000) {$u_{9}$};
\node[font=\small,anchor=west,inner sep=1pt] at (8.5500,-2.3500) {$u_{10}$};
\node[font=\small,anchor=west,inner sep=1pt] at (6.7500,-1.3500) {$u_{11}$};
\node[font=\small,anchor=east,inner sep=1pt] at (5.5000,-3.2000) {$u_{12}$};
\node[font=\small,anchor=west,inner sep=1pt] at (7.3000,-3.2000) {$u_{13}$};
\draw[dotted,rounded corners=3pt] (3.4500,-3.9000) rectangle (9.4500,0.7000);
\node[font=\small,anchor=center,inner sep=1pt] at (3.7500,1.0700) {$U$};
\draw[->,thick,green!50!black] (8.5200,-1.2300) -- (8.5200,-0.9200);
\node[font=\small,anchor=west,inner sep=1pt] at (8.8000,-1.0750) {$3$};
\node[font=\small,anchor=center,inner sep=1pt] at (6.4000,-4.4000) {(b) after slides $1,2$};
\node[tok] (c_u0) at (12.8000,0.0000) {};
\node[v] (c_u1) at (12.8000,1.0000) {};
\node[tok] (c_u2) at (12.8000,2.0000) {};
\node[v] (c_u3) at (12.0000,2.8500) {};
\node[v] (c_u4) at (13.6000,2.8500) {};
\node[tok] (c_u5) at (11.0000,-0.6500) {};
\node[v] (c_u6) at (11.0000,-1.5000) {};
\node[v] (c_u7) at (11.0000,-2.3500) {};
\node[tok] (c_u8) at (14.6000,-0.6500) {};
\node[v] (c_u9) at (14.6000,-1.5000) {};
\node[v] (c_u10) at (14.6000,-2.3500) {};
\node[tok] (c_u11) at (12.8000,-1.3500) {};
\node[v] (c_u12) at (12.2500,-3.2000) {};
\node[v] (c_u13) at (13.3500,-3.2000) {};
\draw (c_u0) -- (c_u1);
\draw (c_u1) -- (c_u2);
\draw (c_u2) -- (c_u3);
\draw (c_u2) -- (c_u4);
\draw (c_u0) -- (c_u5);
\draw (c_u5) -- (c_u6);
\draw (c_u6) -- (c_u7);
\draw (c_u0) -- (c_u8);
\draw (c_u8) -- (c_u9);
\draw (c_u9) -- (c_u10);
\draw (c_u0) -- (c_u11);
\draw (c_u11) -- (c_u12);
\draw (c_u11) -- (c_u13);
\node[font=\small,anchor=east,inner sep=1pt] at (12.4500,0.3300) {$u_{0}$};
\node[font=\small,anchor=east,inner sep=1pt] at (12.4500,1.0000) {$u_{1}$};
\node[font=\small,anchor=east,inner sep=1pt] at (12.4500,2.0000) {$u_{2}$};
\node[font=\small,anchor=east,inner sep=1pt] at (11.6500,2.8500) {$u_{3}$};
\node[font=\small,anchor=west,inner sep=1pt] at (13.9500,2.8500) {$u_{4}$};
\node[font=\small,anchor=east,inner sep=1pt] at (10.6500,-0.6500) {$u_{5}$};
\node[font=\small,anchor=east,inner sep=1pt] at (10.6500,-1.5000) {$u_{6}$};
\node[font=\small,anchor=east,inner sep=1pt] at (10.6500,-2.3500) {$u_{7}$};
\node[font=\small,anchor=west,inner sep=1pt] at (14.9500,-0.6500) {$u_{8}$};
\node[font=\small,anchor=west,inner sep=1pt] at (14.9500,-1.5000) {$u_{9}$};
\node[font=\small,anchor=west,inner sep=1pt] at (14.9500,-2.3500) {$u_{10}$};
\node[font=\small,anchor=west,inner sep=1pt] at (13.1500,-1.3500) {$u_{11}$};
\node[font=\small,anchor=east,inner sep=1pt] at (11.9000,-3.2000) {$u_{12}$};
\node[font=\small,anchor=west,inner sep=1pt] at (13.7000,-3.2000) {$u_{13}$};
\draw[dotted,rounded corners=3pt] (9.8500,-3.9000) rectangle (15.8500,0.7000);
\node[font=\small,anchor=center,inner sep=1pt] at (10.1500,1.0700) {$U$};
\node[font=\small,anchor=center,inner sep=1pt] at (12.8000,-4.4000) {(c) $C_{6}$ on $U$};
\path[use as bounding box] (-3.2,-4.85) rectangle (16.05,3.3);
\end{tikzpicture}
\relax\space \end{adjustbox}
\caption{For $k=4$, a transfer in the proof of \cref{thm:normalization} moves a token from the branch rooted at $u_{8}$ to the branch rooted at $u_{5}$ through $u_{0}$.
The dotted outline encloses $U=T^{u_{1}}_{u_{0}}$, and the restriction to $W'=\{u_{1},u_{2},u_{3},u_{4}\}$ remains fixed.
Black vertices carry tokens, and white vertices are unoccupied.
Panel (a) shows $S=\{u_{0},u_{2},u_{8},u_{9},u_{11}\}$, with $x=\varepsilon_{W'}(S\cap W',u_{1})=1$.
The labels at the branches rooted at $u_{5}$, $u_{8}$, $u_{11}$ give their token counts $0$, $2$, $1$, respectively.
The only empty branch is rooted at $u_{5}$; taking it as $P_A$ gives $\sigma(\{1\})=1\leq 2$ in \cref{eq:internal-slide-condition}.
The numbered arrows are executed in order: $1: u_{0}\to u_{5}$, $2: u_{8}\to u_{0}$, $3: u_{9}\to u_{8}$.
Panel (b) shows the state after the first two slides, and panel (c) the state after the last slide, whose restriction to $U$ is $C_{6}=\{u_{0},u_{5},u_{8},u_{11}\}$.}
\label{fig:f4-transfer}
\end{figure}

By \cref{lem:first-internal-slide}, a first internal slide from $u$ is possible.
After that slide, the restriction to $U$ omits $u$ and still has $h$ unoccupied vertices, so \cref{lem:properly-rooted-count}(a) gives $h\leq\gamma(U)$.
Thus, $C_h$ is defined.
Whenever $u$ is occupied, \cref{lem:unrestricted-sliding,lem:fixing-occupied-set} let us rearrange tokens inside each branch at its current count.
In particular, a nonempty branch can have its root occupied, and a branch that is not full can have its root unoccupied.

For distinct indices $A,B$, suppose that $P_A$ is not full and $P_B$ is nonempty.
A transfer of one token from $P_B$ to $P_A$ is made by first making $w_A$ unoccupied and occupying the roots of all other nonempty branches, then sliding from $u$ to $w_A$ and from $w_B$ to $u$.
The first slide is legal exactly when the empty branches other than $P_A$, together with $x$, have $\sigma$ at most $k-2$.
The second slide restores occupation of $u$ and is legal because every branch contains no path on $k$ vertices.
The transfer increases $b_A$ by one and decreases $b_B$ by one.
If $P_B$ remains nonempty, we occupy its root again by routing inside that branch.
After a transfer, \cref{eq:internal-slide-condition} holds with $B$ as the witness: $P_B$ is not full, and every empty branch other than $P_B$ was empty before and differs from $P_A$.

If there is an empty branch, one of largest height can always be chosen as $P_A$ under \cref{eq:internal-slide-condition}.
Indeed, if the index witnessing the condition is not empty, all empty-branch heights together with $x$ already satisfy the bound.
If that index is empty, removing a largest empty-branch height instead leaves the two largest remaining entries no larger.
The conclusion holds also when some heights are equal.

For every branch $P_i$, let $t_i=|C_h\cap P_i|$.
The branch order is decreasing height, with equal heights ordered by their root labels.
Initially, we occupy the root of every nonempty branch, in increasing root order.
Every root preparation uses the elementary routing procedure to reach the first vertices in label order subject to its required root occupation.
The totals satisfy $\sum_i t_i=\sum_i b_i=|S\cap U|-1$.
The definition of $C_h$ has two possible forms: either every $t_i$ is positive, or every $t_i$ is at most one and the branches with $t_i=1$ form an initial segment of the branch order.

\paragraph{Spreading to empty branches.}
While there is an empty branch and another branch with at least two tokens, the recipient is the first empty branch in the branch order and the donor is the first branch with at least two tokens.
The recipient has largest height among the empty branches, so the preceding argument makes this choice admissible.
The donor remains nonempty after giving one token, and the recipient becomes nonempty.
Thus, each transfer decreases the number of empty branches by one, and at most $D\leq |U|$ transfers occur.
At the end of this phase, either every branch is nonempty or every branch has at most one token.
No claim about the prescribed-count deficit is needed in this phase.

\paragraph{Balancing the prescribed counts.}
As long as some $b_i$ differs from $t_i$, equality of the totals supplies both a deficient branch and a surplus branch.
The recipient $P_A$ is the first branch with $b_A<t_A$ in the branch order, and the donor $P_B$ is the last branch with $b_B>t_B$ in that order.
The recipient is not full, and the donor is nonempty.
We verify the recipient's admissibility in the two possible states left by the spreading phase.
\begin{itemize}
\item \textbf{Case 1: every branch is nonempty.}
The internal slide condition implies $x\leq k-2$, so every nonfull recipient is admissible.
Since $b_A\geq1$ and $t_A>b_A$, some prescribed count is at least two.
The shape of the prescribed counts therefore makes every $t_i$ positive.
The donor retains at least $t_B\geq1$ tokens, so every branch remains nonempty after the transfer.
\item \textbf{Case 2: every branch has at most one token.}
If every $t_i$ were positive, equal totals would force all current and prescribed counts to equal one, contrary to the need for a transfer.
Thus, every $t_i$ is at most one and its positive entries form an initial segment.
The first deficient branch is empty and belongs to this segment.
Every earlier branch of the segment is already occupied, and no empty branch outside the segment precedes it.
Hence, $P_A$ is the first empty branch and has largest height among the empty branches, which makes it admissible.
The recipient changes from zero tokens to one and the donor from one to zero, preserving the property that every branch has at most one token.
\end{itemize}
Each balancing transfer decreases $\sum_i\max(t_i-b_i,0)$ by exactly one.
This nonnegative integer is initially at most $\sum_i t_i\leq |U|$, so at most $|U|$ balancing transfers reach all prescribed counts.
Finally, with $u$ occupied, the branches are routed in increasing root order to their exact restrictions in $C_h$.
All preparations and transfers keep $W'$ fixed and consist of legal slides, as required.
\end{proof}

For the tree $T[U]$ by itself, the next lemma gives all possible token counts with no rigid token and joins $C_h$ to $C_{h+1}$, for $0\leq h<\gamma(U)$, by one removal of the token on $u$ followed by slides.

\begin{lemma}\label{lem:canonical-movable}
Let $(U,u)$ be properly rooted.
\begin{enumerate}[(a)]
\item For every $1\leq h\leq\gamma(U)$, a TS-sequence in $T[U]$ leads from $C_h$ to $C_{h-1}-u$; it changes only one branch and $u$, and its last slide moves the token on $u$ into that branch.
\item For every $1\leq h\leq\gamma(U)$, we have $\Rigid_U(C_h)=\emptyset$.
The possible cardinalities of $k$-PVCs of $T[U]$ whose tokens are all movable are exactly $d,d+1,\dots,|U|-1$.
\item For every $0\leq h<\gamma(U)$, we have $C_h\rmstep{U}{u}C_{h+1}$ by a sequence with exactly one token removal.
\end{enumerate}
\end{lemma}

\begin{proof}
We first move from $C_h$ to $C_{h-1}-u$ inside one branch, and then derive the other two assertions.
\begin{enumerate}[(a)]
\item The set $C_{h-1}$ is $C_h$ together with the next vertex $a$ of the list defining the prescribed sets; let $P_w$ be the branch containing $a$, with root $w$.
Then $w\in C_{h-1}$: either $a=w$, or $a$ is not a branch root, and then all branch roots come before $a$ in the list and lie in $C_h$.
The sets $C_h$ and $C_{h-1}-u$ agree outside $P_w+u$, and $C_{h-1}-u$ has one more token in $P_w$ than $C_h$.
As $T[P_w]$ contains no path on $k$ vertices, \cref{lem:unrestricted-sliding,lem:fixing-occupied-set} turn $C_h\cap P_w$ into $(C_{h-1}\cap P_w)-w$ while $u$ stays occupied.
Then the token on $u$ slides to $w$; the result $C_{h-1}-u$ is a $k$-PVC of $T[U]$ by \cref{lem:properly-rooted-count}(b), as $h-1<\gamma(U)$.
\item By (a), the token on $u$ in $C_h$ is movable.
If some token of $C_h$ were rigid, then, as $C_h\neq U$, \cref{lem:rigid-path} would give a path on $k$ vertices through a rigid vertex whose other vertices are never occupied; the path contains $u$, since every path on $k$ vertices of $T[U]$ does, and $u$ is occupied in $C_h$ and not rigid, a contradiction.
Conversely, if all tokens of a $k$-PVC are movable, then $u$ can be unoccupied in a reachable $k$-PVC.
By \cref{lem:properly-rooted-count}(a), its token count is at least $d$, and it is less than $|U|$ because the full set has no possible slide.
The sets $C_h$ attain every cardinality in this interval.
\item By \cref{lem:properly-rooted-count}(b), removing the token on $u$ from $C_h$ leaves a $k$-PVC, and the reverse of the sequence of (a) for $h+1$ leads from $C_h-u$ to $C_{h+1}$.
Exactly one token has been removed.
\qedhere
\end{enumerate}
\end{proof}

We now determine the largest number of unoccupied vertices that a prescribed properly rooted subtree can contain along a TS-sequence when all tokens are movable.

\begin{theorem}\label{thm:maximum-unoccupied}
Let $W\subseteq V(T)$ be connected, let $U=W^v_u$ be properly rooted, and let $S$ be a $k$-PVC of $T[W]$ with $\Rigid_W(S)=\emptyset$.
Then
\begin{equation}\label{eq:maximum-unoccupied}
\max\{|U-S'|:S'\in[S]_W\}=\min\bigl(\gamma(U),|W-S|\bigr).
\end{equation}
\end{theorem}

\begin{proof}
We bound the count when the token on $u$ first moves, then show that a strict inequality at a maximizing $k$-PVC permits one more token to move from $U$ to $W'$.
Let $q=|W-S|$.
For a reachable $k$-PVC omitting $u$, \cref{lem:properly-rooted-count}(a) bounds its number of unoccupied vertices in $U$ by $\gamma(U)$.
For a reachable $k$-PVC containing $u$, there is a first slide from $u$ because its token is movable.
Until that slide, the number of unoccupied vertices in $U$ does not change.
An internal slide preserves the count, and a slide to $v$ increases it by one.
The resulting restriction to $U$ omits $u$, so the same bound applies to the original count as well.
The total number $q$ is a second upper bound.

Let $h$ be the maximum on the left of \cref{eq:maximum-unoccupied}, and let $A\in[S]_W$ attain it.
By \cref{lem:occupying-root}, we may assume that $u\in A$ without changing $h$.
Suppose that $h<\min(\gamma(U),q)$.

Along a TS-sequence starting at $A$, the first slide from $u$ cannot go to $v$, since that would increase $h$.
Thus, it is internal.
At the $k$-PVC just before that slide, \cref{lem:first-internal-slide,thm:normalization} replace the restriction to $U$ by $C_h$ with $W'$ fixed, which gives $N\in[S]_W$ with $N\cap U=C_h$.
The restriction $N\cap W'$ has $q-h>0$ unoccupied vertices and the largest token count among the restrictions to $W'$ of members of $[S]_W$, so $\Rigid_{W'}(N\cap W')=\emptyset$ by \cref{cor:maximum-side}.
Keeping $C_h$ fixed, we can make $v$ unoccupied by slides in $W'$, using \cref{lem:fixing-occupied-set}.
Since $h<\gamma(U)$, the set $C_h-u$ is a $k$-PVC of $T[U]$ by \cref{lem:properly-rooted-count}(b).
The token on $u$ can now slide to $v$: the restriction to $U$ becomes $C_h-u$, the restriction to $W'$ gains $v$, and every path crossing $uv$ contains the occupied vertex $v$.
The slide increases the number of unoccupied vertices in $U$, contradicting the choice of $h$.
Therefore, equality holds in \cref{eq:maximum-unoccupied}.
\end{proof}

For the induction, the restriction to $U$ must depend only on the tree and the total number of tokens.
The following theorem gives such a restriction and proves that every remaining token is movable, including for every other reachable $k$-PVC with the same restriction to $U$.
For a $k$-PVC $S$ as in the next theorem, let $q=|W-S|$ and let $h^*=\min(\gamma(U),q-1)$.
Since all tokens are movable and $W$ is nonempty, $q\geq 1$.

\begin{theorem}\label{thm:prescribed-subtree}
Let $W\subseteq V(T)$ be connected, let $U=W^v_u$ be properly rooted, let $W'=W^u_v$, and let $S$ be a $k$-PVC of $T[W]$ with $\Rigid_W(S)=\emptyset$.
Some $A\in[S]_W$ satisfies $A\cap U=C_{h^*}$.
Every such $A$ satisfies $\Rigid_{W'}(A\cap W')=\emptyset$, and every TS-sequence in $T[W']$ starting at $A\cap W'$ extends with $C_{h^*}$ fixed.
\end{theorem}

\begin{proof}
We normalize $U$ at a maximum of \cref{thm:maximum-unoccupied} and, when that maximum accounts for all unoccupied vertices of $W$, move one unoccupied vertex back to $W'$.
By \cref{thm:maximum-unoccupied,lem:occupying-root}, some $A\in[S]_W$ contains $u$ and has $\min(\gamma(U),q)$ unoccupied vertices in $U$, the maximum over $[S]_W$.
The first slide from $u$ along a TS-sequence starting at $A$ cannot go to $v$, because that would exceed the maximum, so it is internal, and immediately before it, \cref{lem:first-internal-slide,thm:normalization} take the restriction to $U$ to $C_{\min(\gamma(U),q)}$ with $W'$ fixed.
Let $B$ be the resulting member of $[S]_W$.
\begin{itemize}

\item \textbf{Case 1: $q\leq\gamma(U)$.}
Then $h^*=q-1$, and $B\cap U=C_q$ has all $q$ unoccupied vertices of $W$, so $W'\subseteq B$.
By \cref{lem:canonical-movable}(a), a TS-sequence in $T[U]$ leads from $C_q$ to $C_{q-1}-u$, and it extends with $W'$ fixed by \cref{lem:fixing-occupied-set}, as $v$ is occupied.
Then the token on $v$ slides to $u$: the result has the restriction $C_{q-1}$ to $U$, which contains $u$, and the restriction $W'-v$ to $W'$, which has one unoccupied vertex and is therefore a $k$-PVC of $T[W']$, and every path crossing $uv$ contains $u$.
\item \textbf{Case 2: $q>\gamma(U)$.}
Then $h^*=\gamma(U)$ and $B\cap U=C_{h^*}$.
\end{itemize}

Let $A'\in[S]_W$ satisfy $A'\cap U=C_{h^*}$.
If $q\leq\gamma(U)$, then $A'\cap W'$ has exactly one unoccupied vertex, and by \cref{lem:unrestricted-sliding} all such subsets of $W'$ are $k$-PVCs reachable from one another in $T[W']$, so each vertex can be unoccupied.
If $q>\gamma(U)$, then $A'$ has the maximum $\gamma(U)$ of unoccupied vertices in $U$, so $A'\cap W'$ has the largest token count among the restrictions to $W'$ of members of $[S]_W$ and $q-\gamma(U)\geq 1$ unoccupied vertices, and $\Rigid_{W'}(A'\cap W')=\emptyset$ by \cref{cor:maximum-side}.
In both cases, $u\in C_{h^*}$, so \cref{lem:fixing-occupied-set} extends every TS-sequence on $W'$ while the restriction to $U$ remains fixed.
\end{proof}

\subsection{The Key Lemma}\label{sec:key-lemma}

Demaine et al.~\cite{Demaine2015} delete a safe degree-$1$ vertex, that is, a leaf whose neighbor has at most one neighbor of degree more than one, together with its neighbor and the resulting isolated vertices.
Our proof deletes a properly rooted subtree instead, and the next example shows that, already for $k=3$, the smaller subtrees deleted by Demaine et al.\ do not suffice for this argument.

\begin{example}\label{ex:path7}
Let $k=3$, and let $T$ be the path $a_0a_1\dots a_6$.
A $3$-PVC $S$ of $T$ with two tokens leaves five unoccupied vertices, and every component of $T-S$ has at most two vertices; hence, the $3$-PVCs of $T$ with two tokens are $\{a_1,a_4\}$, $\{a_2,a_4\}$ and $\{a_2,a_5\}$.
They are reachable from one another, since the token on $a_1$ can slide to $a_2$ and then the token on $a_4$ can slide to $a_5$, and all their tokens are movable.
The safe degree-$1$ vertices of $T$ are $a_0$ and $a_6$, and the deletion of Demaine et al.\ removes $\{a_0,a_1\}$ or $\{a_5,a_6\}$, respectively.
Fixing the tokens of a $3$-PVC on $\{a_0,a_1\}$ with $a_1$ occupied leaves at most one token for the path $a_2a_3\dots a_6$.
The empty set is not a $3$-PVC of this path, and the only $3$-PVC of it with one token is $\{a_4\}$, whose token cannot move.
In the other two $3$-PVCs, $a_0$ and $a_1$ are unoccupied, so the path $a_0a_1a_2$ keeps the token on $a_2$ in place, and a slide on the remaining path that vacates $a_2$ does not give a $3$-PVC of $T$.
By symmetry, the same holds for $\{a_5,a_6\}$.
Hence, in every TS-sequence in $T$ that keeps the tokens on a deleted set fixed, some vertex of the remaining path stays occupied: for $\{a_0,a_1\}$, this vertex is $a_4$ if $a_1$ is occupied and $a_2$ otherwise, and symmetrically for $\{a_5,a_6\}$.
Therefore, neither deletion can be used as in the proof of the Key Lemma, which needs a remaining part in which all tokens can be moved by TS-sequences that extend to $T$ while the deleted part stays fixed.
The properly rooted subtree with the vertex set $\{a_0,a_1,a_2\}$ and the root $a_2$ behaves as required: the $3$-PVC $\{a_2,a_4\}$ has two unoccupied vertices in it, and on the remaining path $a_3a_4a_5a_6$, the $3$-PVCs $\{a_4\}$ and $\{a_5\}$ are reachable from one another, so all their tokens are movable.
\end{example}

The preceding theorem replaces part of each $k$-PVC by the same prescribed set and leaves a smaller tree on which all tokens are movable.
That reduction proves the following analogue of the Key Lemma for independent sets of trees in Demaine et al.~\cite{Demaine2015}.

\begin{lemma}[Key Lemma]\label{lem:key}
Let $W\subseteq V(T)$ be connected, and let $S,S'$ be $k$-PVCs of $T[W]$ with $|S|=|S'|$ and $\Rigid_W(S)=\Rigid_W(S')=\emptyset$.
Then $S\sevstep{W}S'$.
\end{lemma}

\begin{proof}
We use induction on $|W|$, normalizing both $k$-PVCs on the same properly rooted subtree before applying the induction hypothesis to its complement.
Let $t=|S|=|S'|$ and $q=|W|-t$.
If $t=0$, both sets are empty.
If $t=1$, the feasible positions of the token are the intersection of the vertex sets of all paths on $k$ vertices, or all of $W$ when there is no such path.
For every two positions in that intersection, the unique path joining them in $T[W]$ lies in every one of those paths.
Thus, every vertex on the joining path is a feasible position, and the token slides along it from $S$ to $S'$.
If $q<k$ or $T[W]$ contains no path on $k$ vertices, \cref{lem:unrestricted-sliding} applies.

In the remaining case, $t\geq 2$, $q\geq k$, and $T[W]$ contains a path on $k$ vertices.
In particular, $|W|\geq k+2$, and \cref{lem:properly-rooted-existence}(b) gives a properly rooted subtree $U=W^v_u$ with nonempty complement $W'=W^u_v$.
A fixed choice, for example the least tuple $(|U|,u,v)$ among such subtrees, depends only on the ordered tree.
By \cref{thm:prescribed-subtree}, both $S$ and $S'$ reach $k$-PVCs $A$ and $A'$ whose restrictions to $U$ are $C_{\min(\gamma(U),q-1)}$.
The restrictions $A\cap W'$ and $A'\cap W'$ have the same number of tokens, all movable in $T[W']$.
Since $0<|W'|<|W|$, the induction hypothesis joins those restrictions by a TS-sequence in $T[W']$.
The sequence extends with the prescribed restriction to $U$ fixed, by \cref{thm:prescribed-subtree}.
Concatenating the sequence from $S$ to $A$, the extended sequence from $A$ to $A'$, and the reverse of the sequence from $S'$ to $A'$ proves $S\sevstep{W}S'$.
\end{proof}

\subsection{The Characterization}\label{sec:characterization-proof}

Fixing the rigid tokens separates a $k$-PVC into restrictions whose tokens are all movable.
The Key Lemma therefore yields a characterization for arbitrary $k$-PVCs of a tree.

\begin{theorem}\label{thm:characterization}
Let $W\subseteq V(T)$ be connected, and let $S,S'$ be $k$-PVCs of $T[W]$.
Then $S\sevstep{W}S'$ if and only if $\Rigid_W(S)=\Rigid_W(S')$ and $|S\cap P|=|S'\cap P|$ for the vertex set $P$ of every component of $T[W]-\Rigid_W(S)$.
\end{theorem}

\begin{figure}[ht]
\centering
\begin{adjustbox}{max width=\textwidth}
\begin{tikzpicture}[v/.style={circle,draw=black,fill=white,inner sep=0pt,minimum size=4.5mm},
tok/.style={v,fill=black},rig/.style={v,fill=black,draw=black!30,line width=1.8pt},
nev/.style={v,fill=black!20},line width=.45pt]
\node[rig] (a_u0) at (0.0000,0.0000) {};
\node[tok] (a_u1) at (0.0000,1.0000) {};
\node[tok] (a_u2) at (0.0000,2.0000) {};
\node[v] (a_u3) at (-0.8000,2.8500) {};
\node[v] (a_u4) at (0.8000,2.8500) {};
\node[v] (a_u5) at (-1.8000,-0.6500) {};
\node[v] (a_u6) at (-1.8000,-1.5000) {};
\node[v] (a_u7) at (-1.8000,-2.3500) {};
\node[v] (a_u8) at (1.8000,-0.6500) {};
\node[v] (a_u9) at (1.8000,-1.5000) {};
\node[v] (a_u10) at (1.8000,-2.3500) {};
\node[tok] (a_u11) at (0.0000,-1.3500) {};
\node[v] (a_u12) at (-0.5500,-3.2000) {};
\node[v] (a_u13) at (0.5500,-3.2000) {};
\draw (a_u0) -- (a_u1);
\draw (a_u1) -- (a_u2);
\draw (a_u2) -- (a_u3);
\draw (a_u2) -- (a_u4);
\draw (a_u0) -- (a_u5);
\draw (a_u5) -- (a_u6);
\draw (a_u6) -- (a_u7);
\draw (a_u0) -- (a_u8);
\draw (a_u8) -- (a_u9);
\draw (a_u9) -- (a_u10);
\draw (a_u0) -- (a_u11);
\draw (a_u11) -- (a_u12);
\draw (a_u11) -- (a_u13);
\node[font=\small,anchor=east,inner sep=1pt] at (-0.3500,0.0000) {$u_{0}$};
\node[font=\small,anchor=east,inner sep=1pt] at (-0.3500,1.0000) {$u_{1}$};
\node[font=\small,anchor=east,inner sep=1pt] at (-0.3500,2.0000) {$u_{2}$};
\node[font=\small,anchor=east,inner sep=1pt] at (-1.1500,2.8500) {$u_{3}$};
\node[font=\small,anchor=west,inner sep=1pt] at (1.1500,2.8500) {$u_{4}$};
\node[font=\small,anchor=east,inner sep=1pt] at (-2.1500,-0.6500) {$u_{5}$};
\node[font=\small,anchor=east,inner sep=1pt] at (-2.1500,-1.5000) {$u_{6}$};
\node[font=\small,anchor=east,inner sep=1pt] at (-2.1500,-2.3500) {$u_{7}$};
\node[font=\small,anchor=west,inner sep=1pt] at (2.1500,-0.6500) {$u_{8}$};
\node[font=\small,anchor=west,inner sep=1pt] at (2.1500,-1.5000) {$u_{9}$};
\node[font=\small,anchor=west,inner sep=1pt] at (2.1500,-2.3500) {$u_{10}$};
\node[font=\small,anchor=west,inner sep=1pt] at (0.3500,-1.3500) {$u_{11}$};
\node[font=\small,anchor=north,inner sep=1pt] at (-0.5500,-3.6700) {$u_{12}$};
\node[font=\small,anchor=north,inner sep=1pt] at (0.5500,-3.6700) {$u_{13}$};
\draw[dotted,rounded corners=3pt] (-1.8000,0.6000) rectangle (1.8000,3.3000);
\node[font=\small,anchor=center,inner sep=1pt] at (0.0000,3.6000) {$2$};
\draw[dotted,rounded corners=3pt] (-2.9500,-2.8500) rectangle (-1.2500,-0.2500);
\node[font=\small,anchor=center,inner sep=1pt] at (-2.1000,-3.1500) {$0$};
\draw[dotted,rounded corners=3pt] (1.2500,-2.8500) rectangle (3.0500,-0.2500);
\node[font=\small,anchor=center,inner sep=1pt] at (2.1500,-3.1500) {$0$};
\draw[dotted,rounded corners=3pt] (-1.1000,-4.2000) rectangle (1.1000,-0.9500);
\node[font=\small,anchor=center,inner sep=1pt] at (0.0000,-4.5000) {$1$};
\node[font=\small,anchor=center,inner sep=1pt] at (0.0000,-5.0500) {(a) $S$};
\node[rig] (b_u0) at (6.4000,0.0000) {};
\node[v] (b_u1) at (6.4000,1.0000) {};
\node[v] (b_u2) at (6.4000,2.0000) {};
\node[tok] (b_u3) at (5.6000,2.8500) {};
\node[tok] (b_u4) at (7.2000,2.8500) {};
\node[v] (b_u5) at (4.6000,-0.6500) {};
\node[v] (b_u6) at (4.6000,-1.5000) {};
\node[v] (b_u7) at (4.6000,-2.3500) {};
\node[v] (b_u8) at (8.2000,-0.6500) {};
\node[v] (b_u9) at (8.2000,-1.5000) {};
\node[v] (b_u10) at (8.2000,-2.3500) {};
\node[v] (b_u11) at (6.4000,-1.3500) {};
\node[tok] (b_u12) at (5.8500,-3.2000) {};
\node[v] (b_u13) at (6.9500,-3.2000) {};
\draw (b_u0) -- (b_u1);
\draw (b_u1) -- (b_u2);
\draw (b_u2) -- (b_u3);
\draw (b_u2) -- (b_u4);
\draw (b_u0) -- (b_u5);
\draw (b_u5) -- (b_u6);
\draw (b_u6) -- (b_u7);
\draw (b_u0) -- (b_u8);
\draw (b_u8) -- (b_u9);
\draw (b_u9) -- (b_u10);
\draw (b_u0) -- (b_u11);
\draw (b_u11) -- (b_u12);
\draw (b_u11) -- (b_u13);
\node[font=\small,anchor=east,inner sep=1pt] at (6.0500,0.0000) {$u_{0}$};
\node[font=\small,anchor=east,inner sep=1pt] at (6.0500,1.0000) {$u_{1}$};
\node[font=\small,anchor=east,inner sep=1pt] at (6.0500,2.0000) {$u_{2}$};
\node[font=\small,anchor=east,inner sep=1pt] at (5.2500,2.8500) {$u_{3}$};
\node[font=\small,anchor=west,inner sep=1pt] at (7.5500,2.8500) {$u_{4}$};
\node[font=\small,anchor=east,inner sep=1pt] at (4.2500,-0.6500) {$u_{5}$};
\node[font=\small,anchor=east,inner sep=1pt] at (4.2500,-1.5000) {$u_{6}$};
\node[font=\small,anchor=east,inner sep=1pt] at (4.2500,-2.3500) {$u_{7}$};
\node[font=\small,anchor=west,inner sep=1pt] at (8.5500,-0.6500) {$u_{8}$};
\node[font=\small,anchor=west,inner sep=1pt] at (8.5500,-1.5000) {$u_{9}$};
\node[font=\small,anchor=west,inner sep=1pt] at (8.5500,-2.3500) {$u_{10}$};
\node[font=\small,anchor=west,inner sep=1pt] at (6.7500,-1.3500) {$u_{11}$};
\node[font=\small,anchor=north,inner sep=1pt] at (5.8500,-3.6700) {$u_{12}$};
\node[font=\small,anchor=north,inner sep=1pt] at (6.9500,-3.6700) {$u_{13}$};
\draw[dotted,rounded corners=3pt] (4.6000,0.6000) rectangle (8.2000,3.3000);
\node[font=\small,anchor=center,inner sep=1pt] at (6.4000,3.6000) {$2$};
\draw[dotted,rounded corners=3pt] (3.4500,-2.8500) rectangle (5.1500,-0.2500);
\node[font=\small,anchor=center,inner sep=1pt] at (4.3000,-3.1500) {$0$};
\draw[dotted,rounded corners=3pt] (7.6500,-2.8500) rectangle (9.4500,-0.2500);
\node[font=\small,anchor=center,inner sep=1pt] at (8.5500,-3.1500) {$0$};
\draw[dotted,rounded corners=3pt] (5.3000,-4.2000) rectangle (7.5000,-0.9500);
\node[font=\small,anchor=center,inner sep=1pt] at (6.4000,-4.5000) {$1$};
\node[font=\small,anchor=center,inner sep=1pt] at (6.4000,-5.0500) {(b) $S'$};
\node[rig] (c_u0) at (12.8000,0.0000) {};
\node[tok] (c_u1) at (12.8000,1.0000) {};
\node[tok] (c_u2) at (12.8000,2.0000) {};
\node[tok] (c_u3) at (12.0000,2.8500) {};
\node[v] (c_u4) at (13.6000,2.8500) {};
\node[v] (c_u5) at (11.0000,-0.6500) {};
\node[v] (c_u6) at (11.0000,-1.5000) {};
\node[v] (c_u7) at (11.0000,-2.3500) {};
\node[v] (c_u8) at (14.6000,-0.6500) {};
\node[v] (c_u9) at (14.6000,-1.5000) {};
\node[v] (c_u10) at (14.6000,-2.3500) {};
\node[v] (c_u11) at (12.8000,-1.3500) {};
\node[v] (c_u12) at (12.2500,-3.2000) {};
\node[v] (c_u13) at (13.3500,-3.2000) {};
\draw (c_u0) -- (c_u1);
\draw (c_u1) -- (c_u2);
\draw (c_u2) -- (c_u3);
\draw (c_u2) -- (c_u4);
\draw (c_u0) -- (c_u5);
\draw (c_u5) -- (c_u6);
\draw (c_u6) -- (c_u7);
\draw (c_u0) -- (c_u8);
\draw (c_u8) -- (c_u9);
\draw (c_u9) -- (c_u10);
\draw (c_u0) -- (c_u11);
\draw (c_u11) -- (c_u12);
\draw (c_u11) -- (c_u13);
\node[font=\small,anchor=east,inner sep=1pt] at (12.4500,0.0000) {$u_{0}$};
\node[font=\small,anchor=east,inner sep=1pt] at (12.4500,1.0000) {$u_{1}$};
\node[font=\small,anchor=east,inner sep=1pt] at (12.4500,2.0000) {$u_{2}$};
\node[font=\small,anchor=east,inner sep=1pt] at (11.6500,2.8500) {$u_{3}$};
\node[font=\small,anchor=west,inner sep=1pt] at (13.9500,2.8500) {$u_{4}$};
\node[font=\small,anchor=east,inner sep=1pt] at (10.6500,-0.6500) {$u_{5}$};
\node[font=\small,anchor=east,inner sep=1pt] at (10.6500,-1.5000) {$u_{6}$};
\node[font=\small,anchor=east,inner sep=1pt] at (10.6500,-2.3500) {$u_{7}$};
\node[font=\small,anchor=west,inner sep=1pt] at (14.9500,-0.6500) {$u_{8}$};
\node[font=\small,anchor=west,inner sep=1pt] at (14.9500,-1.5000) {$u_{9}$};
\node[font=\small,anchor=west,inner sep=1pt] at (14.9500,-2.3500) {$u_{10}$};
\node[font=\small,anchor=west,inner sep=1pt] at (13.1500,-1.3500) {$u_{11}$};
\node[font=\small,anchor=north,inner sep=1pt] at (12.2500,-3.6700) {$u_{12}$};
\node[font=\small,anchor=north,inner sep=1pt] at (13.3500,-3.6700) {$u_{13}$};
\draw[dotted,rounded corners=3pt] (11.0000,0.6000) rectangle (14.6000,3.3000);
\node[font=\small,anchor=center,inner sep=1pt] at (12.8000,3.6000) {$3$};
\draw[dotted,rounded corners=3pt] (9.8500,-2.8500) rectangle (11.5500,-0.2500);
\node[font=\small,anchor=center,inner sep=1pt] at (10.7000,-3.1500) {$0$};
\draw[dotted,rounded corners=3pt] (14.0500,-2.8500) rectangle (15.8500,-0.2500);
\node[font=\small,anchor=center,inner sep=1pt] at (14.9500,-3.1500) {$0$};
\draw[dotted,rounded corners=3pt] (11.7000,-4.2000) rectangle (13.9000,-0.9500);
\node[font=\small,anchor=center,inner sep=1pt] at (12.8000,-4.5000) {$0$};
\node[font=\small,anchor=center,inner sep=1pt] at (12.8000,-5.0500) {(c) $S''$};
\path[use as bounding box] (-3.25,-5.45) rectangle (16.05,3.85);
\end{tikzpicture}
\relax\space \end{adjustbox}
\caption{The $4$-PVCs of the tree in \cref{fig:f1-basic} illustrate \cref{thm:characterization}.
Black vertices carry tokens, thick gray outlines mark rigid tokens, and white vertices are unoccupied.
Dotted outlines enclose the components of $T-\Rigid(S)$, and the labels give their token counts.
Panels (a), (b), and (c) show $S=\{u_{0},u_{1},u_{2},u_{11}\}$, $S'=\{u_{0},u_{3},u_{4},u_{12}\}$, and $S''=\{u_{0},u_{1},u_{2},u_{3}\}$, respectively, all with rigid set $\{u_{0}\}$.
In component order $\{u_{1},u_{2},u_{3},u_{4}\}; \{u_{5},u_{6},u_{7}\}; \{u_{8},u_{9},u_{10}\}; \{u_{11},u_{12},u_{13}\}$, the counts for $S$ and $S'$ are $(2,0,0,1)$, whereas those for $S''$ are $(3,0,0,0)$.
Thus, $S'$ is reachable from $S$, and $S''$ is not.}
\label{fig:f2-components}
\end{figure}
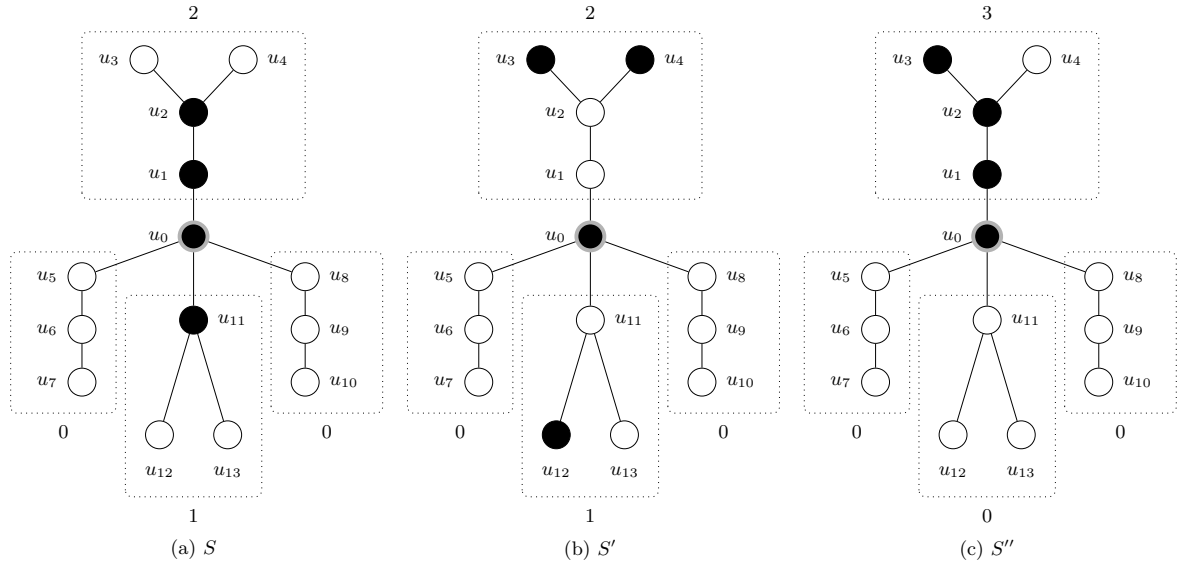

\cref{fig:f2-components} illustrates how the component counts distinguish reachable and unreachable $k$-PVCs with the same rigid set.

\begin{proof}
We fix the vertices with rigid tokens and use \cref{lem:fixing-occupied-set} in both directions.
Let $R=\Rigid_W(S)$.
\begin{itemize}
\item[($\Rightarrow$)] If $S\sevstep{W}S'$, their sets of reachable $k$-PVCs agree, so $\Rigid_W(S')=R$.
Every TS-sequence starting at $S$ keeps $R$ occupied.
By \cref{lem:fixing-occupied-set}(c), it restricts to a TS-sequence on each component of $T[W]-R$, and such a sequence preserves the token count there.
\item[($\Leftarrow$)] Suppose that the rigid sets and the stated counts agree.
For every vertex $x\in W-R$, some member of $[S]_W$ omits $x$.
Projecting a TS-sequence to that member onto the component $P$ containing $x$ shows that $x\notin\Rigid_P(S\cap P)$.
Thus, $\Rigid_P(S\cap P)=\emptyset$ for every such $P$, and the same argument gives $\Rigid_P(S'\cap P)=\emptyset$.
The token counts on $P$ agree, so \cref{lem:key} joins $S\cap P$ to $S'\cap P$ inside $T[P]$.
By \cref{lem:fixing-occupied-set}(b), we execute these sequences one component at a time while keeping $R$ occupied.
The resulting $k$-PVC agrees with $S'$ on every component and on $R$, and is therefore $S'$.
\qedhere
\end{itemize}
\end{proof}

For $k=2$, \cref{thm:characterization} and the characterization of Demaine et al.~\cite{Demaine2015} describe the same relation, but they delete different vertices.
The next remark shows how the two deleted sets correspond.

\begin{remark}\label{rem:demaine-k2}
Let $k=2$, let $S$ be a $2$-PVC of $T$, let $I=V(T)-S$, and let $J=\NeverOcc(S)$; we denote by $N(J)$ the set of vertices outside $J$ with a neighbor in $J$, and set $N[J]=J\cup N(J)$.
A slide of $S$ from $u$ to $v$ is a slide of the independent set $I$ from $v$ to $u$, so $J$ is the set of vertices with rigid tokens of $I$ in the sense of Demaine et al.~\cite{Demaine2015}; they delete $J$ together with its neighbors and compare the numbers of tokens of $I$ in the components of $T-N[J]$.
We claim that $\Rigid(S)$ consists of the vertices of $N(J)$ and of the components of $T-N[J]$ that contain no vertex of $I$.
Hence, the components of $T-\Rigid(S)$ are the sets $\{x\}$ with $x\in J$, which contain no token of $S$, and the components $Q$ of $T-N[J]$ with $Q\cap I\neq\emptyset$, on which $S$ has $|Q|-|I\cap Q|$ tokens.
We prove the claim by fixing $N(J)$ and examining the remaining components.
Every vertex of $J$ is unoccupied throughout $[S]$.
A vertex adjacent to a vertex of $J$ is occupied in every member of $[S]$, because the edge joining the two vertices must meet the $2$-PVC; thus, $N(J)\subseteq\Rigid(S)$.
By \cref{lem:fixing-occupied-set} with $X=N(J)$, the set $[S]$ consists of the unions of $N(J)$ with independently chosen members of $[S\cap Q]_Q$ for the vertex sets $Q$ of the components of $T-N(J)$, and these sets are the sets $\{x\}$ with $x\in J$ and the vertex sets of the components of $T-N[J]$.
The singleton components $\{x\}$ with $x\in J$ contain no token of $S$.
Let $Q$ now be the vertex set of a component of $T-N[J]$.
\begin{itemize}
\item \textbf{Case 1: $Q\cap I=\emptyset$.}
Then $S\cap Q=Q$, so no slide is possible in $T[Q]$.
Hence, $Q\subseteq\Rigid(S)$.
\item \textbf{Case 2: $Q\cap I\neq\emptyset$.}
Suppose that $\Rigid_Q(S\cap Q)\neq\emptyset$.
By \cref{lem:rigid-path}, some edge of $T[Q]$ has an end in $\NeverOcc_Q(S\cap Q)$.
The product description of $[S]$ implies that this end belongs to $J$, contradicting $Q\cap J=\emptyset$.
Thus, $\Rigid_Q(S\cap Q)=\emptyset$, and no vertex of $Q$ belongs to $\Rigid(S)$.
\end{itemize}
\end{remark}

We finish with a consequence used to analyze token removals in \cref{sec:algorithm}.
For a nonempty connected set $P$, let $F(P)$ be the set of cardinalities of $k$-PVCs $S$ of $T[P]$ with $\Rigid_P(S)=\emptyset$.
When all such $k$-PVCs of cardinality $t$ are reachable from one another, we denote their common set of reachable $k$-PVCs by $\Omega(P,t)$.

\begin{corollary}\label{cor:movable-counts}
For every nonempty connected set $P$, the set $F(P)$ is a nonempty interval of integers ending at $|P|-1$, and $\Omega(P,t)$ exists for every $t\in F(P)$.
For every $r\in P$ and $t\in F(P)$, adding a token at $r$ successively gives $\Omega(P,t+1),\dots,\Omega(P,|P|-1)$ and then the full $k$-PVC $P$.
Equivalently, for $t\leq t'\leq |P|$, every $k$-PVC in $\Omega(P,t)$ can be obtained from every $k$-PVC in $\Omega(P,t')$ by slides and exactly $t'-t$ removals at $r$, where the set at $t'=|P|$ consists only of $P$.
\end{corollary}

\begin{proof}
We combine uniqueness from the Key Lemma with the effect of adding a token in \cref{cor:adding-token-movable}.
The sets with exactly one unoccupied vertex are all $k$-PVCs and are reachable from one another by \cref{lem:unrestricted-sliding}.
Each vertex is unoccupied in one of them, so $|P|-1\in F(P)$, including when $|P|=1$.
The full set $P$ has no legal slide and has a rigid token on every vertex, so $|P|\notin F(P)$.
By \cref{lem:key}, all $k$-PVCs of each count $t\in F(P)$ with movable tokens have the same set $\Omega(P,t)$ of reachable $k$-PVCs.
For every fixed $r$, \cref{cor:adding-token-movable} gives a unique set of reachable $k$-PVCs after adding a token, and its tokens are all movable unless it is the full set.
Hence, $F(P)$ is upward closed up to $|P|-1$, and these successive additions give the asserted sets $\Omega(P,t+1),\dots,\Omega(P,|P|-1)$.
Reversing each addition edge gives a token removal at $r$; TS-sequences within each set of reachable $k$-PVCs connect the required representatives before and after the removal.
The count difference fixes the number of removals at exactly $t'-t$.
\end{proof}
\relax\space \section{Algorithm for Trees}\label{sec:algorithm}

By \cref{thm:characterization}, deciding reachability reduces to computing the vertices with rigid tokens.
We first define two integers for every rooted subtree and show how they determine the minimum of $\varepsilon$ when all tokens are movable.
The proof counts the tokens that can be removed at the root, first from the set in which every vertex is occupied and then from an arbitrary $k$-PVC whose tokens are all movable.
We then give \cref{alg:compute-rigid}, which removes vertices from a set of candidates for rigid tokens, and prove its correctness.
Finally, we analyze its running time and prove \cref{thm:decision}.

\subsection{A Recurrence for Rooted Subtrees}\label{subsec:recurrence}

Throughout this subsection, $P\subseteq V(T)$ is nonempty and connected, and $r\in P$.
For each neighbor $y$ of $r$ in $T[P]$, the rooted subtree $(P^r_y,y)$ supplies two integers $c_y=c(P^r_y,y)$ and $\tau_y=\tau(P^r_y,y)$.
The following recurrence defines $c(P,r)$ and $\tau(P,r)$ by induction on $|P|$.
Let $D$ be the number of positive members of the multiset $\{\tau_y:y\in N_{T[P]}(r)\}$, and let $a_1,\ldots,a_D$ list those members in nondecreasing order.
Let $m$ be the largest integer in $\{0,\ldots,D\}$ such that $\sigma(\{a_1,\ldots,a_m\})\leq k-2$.
For each $j$, choosing the $j$ smallest members minimizes the sum of the two largest chosen members.
Consequently, $m$ is also the maximum cardinality of a submultiset of the positive members whose two largest members sum to at most $k-2$.
With the maximum of an empty multiset interpreted as zero, the recurrence is
\begin{equation}\label{eq:ctau-recurrence}
\begin{aligned}
\delta&=D-m,&
\eta&=\begin{cases}
1+\max_y\tau_y,&D=m,\\
0,&D>m,
\end{cases}\\
c(P,r)&=\sum_y c_y+\delta,&
\tau(P,r)&=\eta.
\end{aligned}
\end{equation}

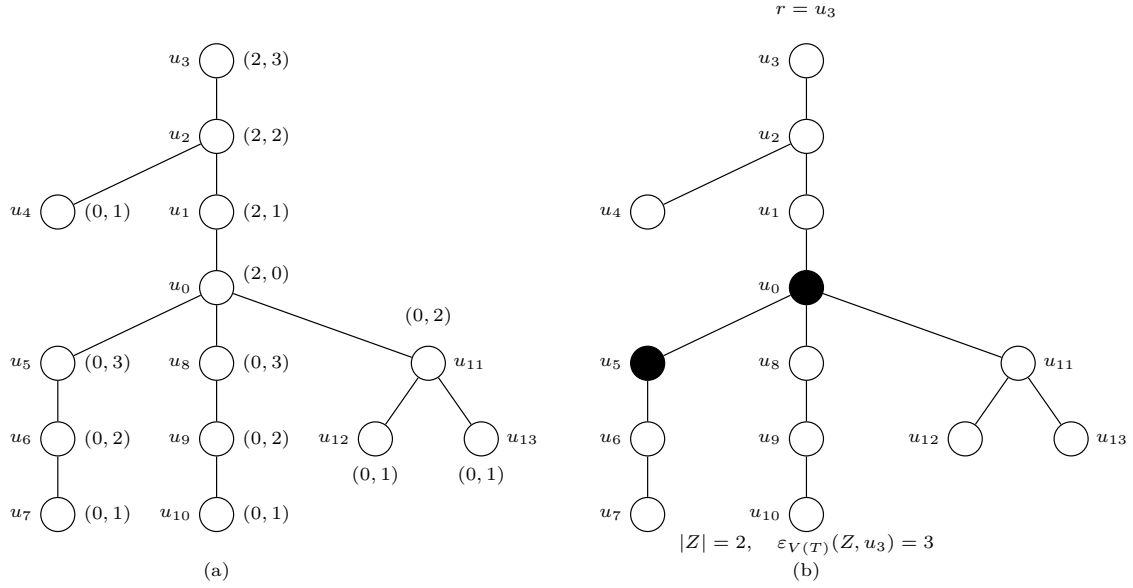
\begin{figure}[ht]
\centering
\begin{adjustbox}{max width=\textwidth}
\begin{tikzpicture}[v/.style={circle,draw=black,fill=white,inner sep=0pt,minimum size=4.5mm},
tok/.style={v,fill=black},rig/.style={v,fill=black,draw=black!30,line width=1.8pt},
nev/.style={v,fill=black!20},line width=.45pt]
\path[use as bounding box] (-3.20,-7.15) rectangle (12.35,1.05);
\node[v] (a_u0) at (0.0000,-3.0000) {};
\node[v] (a_u1) at (0.0000,-2.0000) {};
\node[v] (a_u2) at (0.0000,-1.0000) {};
\node[v] (a_u3) at (0.0000,0.0000) {};
\node[v] (a_u4) at (-2.1000,-2.0000) {};
\node[v] (a_u5) at (-2.1000,-4.0000) {};
\node[v] (a_u6) at (-2.1000,-5.0000) {};
\node[v] (a_u7) at (-2.1000,-6.0000) {};
\node[v] (a_u8) at (0.0000,-4.0000) {};
\node[v] (a_u9) at (0.0000,-5.0000) {};
\node[v] (a_u10) at (0.0000,-6.0000) {};
\node[v] (a_u11) at (2.8000,-4.0000) {};
\node[v] (a_u12) at (2.1000,-5.0000) {};
\node[v] (a_u13) at (3.5000,-5.0000) {};
\draw (a_u0) -- (a_u1);
\draw (a_u1) -- (a_u2);
\draw (a_u2) -- (a_u3);
\draw (a_u2) -- (a_u4);
\draw (a_u0) -- (a_u5);
\draw (a_u5) -- (a_u6);
\draw (a_u6) -- (a_u7);
\draw (a_u0) -- (a_u8);
\draw (a_u8) -- (a_u9);
\draw (a_u9) -- (a_u10);
\draw (a_u0) -- (a_u11);
\draw (a_u11) -- (a_u12);
\draw (a_u11) -- (a_u13);
\node[font=\scriptsize,anchor=east,inner sep=1pt] at (-0.2900,-3.0000) {$u_{0}$};
\node[font=\scriptsize,anchor=east,inner sep=1pt] at (-0.2900,-2.0000) {$u_{1}$};
\node[font=\scriptsize,anchor=east,inner sep=1pt] at (-0.2900,-1.0000) {$u_{2}$};
\node[font=\scriptsize,anchor=east,inner sep=1pt] at (-0.2900,0.0000) {$u_{3}$};
\node[font=\scriptsize,anchor=east,inner sep=1pt] at (-2.3900,-2.0000) {$u_{4}$};
\node[font=\scriptsize,anchor=east,inner sep=1pt] at (-2.3900,-4.0000) {$u_{5}$};
\node[font=\scriptsize,anchor=east,inner sep=1pt] at (-2.3900,-5.0000) {$u_{6}$};
\node[font=\scriptsize,anchor=east,inner sep=1pt] at (-2.3900,-6.0000) {$u_{7}$};
\node[font=\scriptsize,anchor=east,inner sep=1pt] at (-0.2900,-4.0000) {$u_{8}$};
\node[font=\scriptsize,anchor=east,inner sep=1pt] at (-0.2900,-5.0000) {$u_{9}$};
\node[font=\scriptsize,anchor=east,inner sep=1pt] at (-0.2900,-6.0000) {$u_{10}$};
\node[font=\scriptsize,anchor=west,inner sep=1pt] at (3.0900,-4.0000) {$u_{11}$};
\node[font=\scriptsize,anchor=east,inner sep=1pt] at (1.8100,-5.0000) {$u_{12}$};
\node[font=\scriptsize,anchor=west,inner sep=1pt] at (3.7900,-5.0000) {$u_{13}$};
\node[font=\scriptsize,anchor=west,inner sep=1pt] at (0.3000,-2.8200) {$(2,0)$};
\node[font=\scriptsize,anchor=west,inner sep=1pt] at (0.3000,-2.0000) {$(2,1)$};
\node[font=\scriptsize,anchor=west,inner sep=1pt] at (0.3000,-1.0000) {$(2,2)$};
\node[font=\scriptsize,anchor=west,inner sep=1pt] at (0.3000,0.0000) {$(2,3)$};
\node[font=\scriptsize,anchor=west,inner sep=1pt] at (-1.8000,-2.0000) {$(0,1)$};
\node[font=\scriptsize,anchor=west,inner sep=1pt] at (-1.8000,-4.0000) {$(0,3)$};
\node[font=\scriptsize,anchor=west,inner sep=1pt] at (-1.8000,-5.0000) {$(0,2)$};
\node[font=\scriptsize,anchor=west,inner sep=1pt] at (-1.8000,-6.0000) {$(0,1)$};
\node[font=\scriptsize,anchor=west,inner sep=1pt] at (0.3000,-4.0000) {$(0,3)$};
\node[font=\scriptsize,anchor=west,inner sep=1pt] at (0.3000,-5.0000) {$(0,2)$};
\node[font=\scriptsize,anchor=west,inner sep=1pt] at (0.3000,-6.0000) {$(0,1)$};
\node[font=\scriptsize,anchor=south,inner sep=1pt] at (2.8000,-3.5500) {$(0,2)$};
\node[font=\scriptsize,anchor=north,inner sep=1pt] at (2.1000,-5.3200) {$(0,1)$};
\node[font=\scriptsize,anchor=north,inner sep=1pt] at (3.5000,-5.3200) {$(0,1)$};
\node[tok] (b_u0) at (7.8000,-3.0000) {};
\node[v] (b_u1) at (7.8000,-2.0000) {};
\node[v] (b_u2) at (7.8000,-1.0000) {};
\node[v] (b_u3) at (7.8000,0.0000) {};
\node[v] (b_u4) at (5.7000,-2.0000) {};
\node[tok] (b_u5) at (5.7000,-4.0000) {};
\node[v] (b_u6) at (5.7000,-5.0000) {};
\node[v] (b_u7) at (5.7000,-6.0000) {};
\node[v] (b_u8) at (7.8000,-4.0000) {};
\node[v] (b_u9) at (7.8000,-5.0000) {};
\node[v] (b_u10) at (7.8000,-6.0000) {};
\node[v] (b_u11) at (10.6000,-4.0000) {};
\node[v] (b_u12) at (9.9000,-5.0000) {};
\node[v] (b_u13) at (11.3000,-5.0000) {};
\draw (b_u0) -- (b_u1);
\draw (b_u1) -- (b_u2);
\draw (b_u2) -- (b_u3);
\draw (b_u2) -- (b_u4);
\draw (b_u0) -- (b_u5);
\draw (b_u5) -- (b_u6);
\draw (b_u6) -- (b_u7);
\draw (b_u0) -- (b_u8);
\draw (b_u8) -- (b_u9);
\draw (b_u9) -- (b_u10);
\draw (b_u0) -- (b_u11);
\draw (b_u11) -- (b_u12);
\draw (b_u11) -- (b_u13);
\node[font=\scriptsize,anchor=east,inner sep=1pt] at (7.5100,-3.0000) {$u_{0}$};
\node[font=\scriptsize,anchor=east,inner sep=1pt] at (7.5100,-2.0000) {$u_{1}$};
\node[font=\scriptsize,anchor=east,inner sep=1pt] at (7.5100,-1.0000) {$u_{2}$};
\node[font=\scriptsize,anchor=east,inner sep=1pt] at (7.5100,0.0000) {$u_{3}$};
\node[font=\scriptsize,anchor=east,inner sep=1pt] at (5.4100,-2.0000) {$u_{4}$};
\node[font=\scriptsize,anchor=east,inner sep=1pt] at (5.4100,-4.0000) {$u_{5}$};
\node[font=\scriptsize,anchor=east,inner sep=1pt] at (5.4100,-5.0000) {$u_{6}$};
\node[font=\scriptsize,anchor=east,inner sep=1pt] at (5.4100,-6.0000) {$u_{7}$};
\node[font=\scriptsize,anchor=east,inner sep=1pt] at (7.5100,-4.0000) {$u_{8}$};
\node[font=\scriptsize,anchor=east,inner sep=1pt] at (7.5100,-5.0000) {$u_{9}$};
\node[font=\scriptsize,anchor=east,inner sep=1pt] at (7.5100,-6.0000) {$u_{10}$};
\node[font=\scriptsize,anchor=west,inner sep=1pt] at (10.8900,-4.0000) {$u_{11}$};
\node[font=\scriptsize,anchor=east,inner sep=1pt] at (9.6100,-5.0000) {$u_{12}$};
\node[font=\scriptsize,anchor=west,inner sep=1pt] at (11.5900,-5.0000) {$u_{13}$};
\node[font=\scriptsize,anchor=center,inner sep=1pt] at (0.0000,-6.7500) {(a)};
\node[font=\scriptsize,anchor=center,inner sep=1pt] at (7.8000,-6.7500) {(b)};
\node[font=\scriptsize,anchor=center,inner sep=1pt] at (7.8000,0.6500) {$r=u_{3}$};
\node[font=\scriptsize,anchor=center,inner sep=1pt] at (7.8000,-6.4000) {$|Z|=2,\quad\varepsilon_{V(T)}(Z,u_{3})=3$};
\end{tikzpicture}
\relax\space \end{adjustbox}
\caption{The values of the recurrence in \cref{eq:ctau-recurrence} for $k=4$ on $T$ rooted at $r=u_{3}$.
In panel (a), every vertex is labeled with the pair $(c,\tau)$ of its rooted subtree, and no tokens are shown.
At $u_{0}$, the positive child values are $2,3,3$, so $D=3$, $m=1$, $\delta=2$, and $\eta=0$.
Panel (b) shows the set $Z=\{u_{0},u_{5}\}$, reached from $V(T)$ by slides and removals at $r$; black vertices carry tokens, and white vertices are unoccupied.
The set attains both bounds of \cref{thm:root-removals}: $|Z|=2$ and $\varepsilon_{V(T)}(Z,r)=3$.}
\label{fig:f7-recurrence}
\end{figure}

\cref{fig:f7-recurrence} illustrates the recurrence and a set attained by slides and removals at the root.

In particular, the pair for a single vertex is $(c(\{r\},r),\tau(\{r\},r))=(0,1)$.
The recurrence always gives $0\leq\tau(P,r)\leq k-1$: when $D=m$, the largest positive member is at most $k-2$, and when $D>m$, the value is zero.
It also gives $0\leq c(P,r)\leq |P|-1$.
Indeed, if $r$ has $b$ children, the induction hypothesis and $\delta\leq D\leq b$ give
\[
c(P,r)\leq\sum_y(|P^r_y|-1)+b=|P|-1.
\]

The values $c_y$ occur only through their sum.
Thus, changing one $c_y$ by an integer $x$ while keeping every $\tau_y$ fixed changes $c(P,r)$ by $x$ and leaves $\tau(P,r)$ unchanged.
A child with $\tau_y=0$ contributes only $c_y$ to the recurrence.
Both observations continue to hold when the change is passed along the path to another root.

The recurrence takes a simple form for the two smallest values of $k$.

\begin{remark}\label{rem:recurrence-small-k}
For $k=2$, every positive value $\tau_y$ equals $1$, which exceeds $k-2=0$, so $m=0$.
Thus, $c(P,r)$ is $\sum_y c_y$ plus the number of children $y$ with $\tau_y=1$, and $\tau(P,r)=1$ if no child has $\tau_y=1$ and $\tau(P,r)=0$ otherwise.
For a graph $G$, let $\alpha(G)$ denote its \emph{independence number}, the maximum cardinality of an independent set of $G$, with value zero for the empty graph.
For $k=2$, we have $c(P,r)=\alpha(T[P]-r)$ and $\tau(P,r)=\alpha(T[P])-\alpha(T[P]-r)$.
We prove these identities by induction on $|P|$.
For $P=\{r\}$, they follow from $(c(P,r),\tau(P,r))=(0,1)$.
Suppose that the identities hold for every child $y$.
Then $\alpha(T[P^r_y])=c_y+\tau_y$ and $\alpha(T[P^r_y]-y)=c_y$.
Consequently, $\alpha(T[P]-r)=\sum_y(c_y+\tau_y)=c(P,r)$.
The maximum cardinality of an independent set containing $r$ is $1+\sum_y c_y$, so $\alpha(T[P])=\max\{\alpha(T[P]-r),1+\sum_yc_y\}$.
This maximum exceeds $\alpha(T[P]-r)$ by one exactly when no child has $\tau_y=1$.
In particular, $\tau(P,r)=1$ exactly when $r$ belongs to every maximum independent set of $T[P]$.
For $k=3$, the positive values are $1$ and $2$, and $m=1$ if some child has $\tau_y=1$ and $m=0$ otherwise.
Thus, $c(P,r)=\sum_y c_y+D-m$, and $\tau(P,r)$ is $1$ if no child has a positive value, $2$ if exactly one child has a positive value and this value is $1$, and $0$ otherwise.
\end{remark}

We will use the following direct interpretation on subtrees that contain no path on $k$ vertices.

\begin{lemma}\label{lem:ctau-path-free}
For every integer $k\geq2$, every connected set $P\subseteq V(T)$ such that $T[P]$ contains no path on $k$ vertices, and every $r\in P$, the recurrence gives $c(P,r)=0$ and $\tau(P,r)$ equal to the number of vertices of a longest path in $T[P]$ starting at $r$.
\end{lemma}

\begin{proof}
We compare the recurrence with the rooted heights, by induction on $|P|$.
For a single vertex, the assertion follows from the pair $(0,1)$.
For each child $y$ of $r$, the induction hypothesis gives $c_y=0$ and $\tau_y$ equal to the rooted height at $y$.
The two largest child heights sum to at most $k-2$, since otherwise their paths together with $r$ would contain a path on $k$ vertices.
Hence, $D=m$, $\delta=0$, and $\eta=1+\max_y\tau_y$.
The last expression is exactly the rooted height at $r$.
\end{proof}

\subsection{Removing Tokens at the Root}\label{subsec:root-removals}

We now interpret $c(P,r)$ as the minimum number of tokens remaining after slides and removals at $r$ from the $k$-PVC $P$.
Among the resulting $k$-PVCs with that many tokens, the minimum of $\varepsilon$ at $r$ is $\tau(P,r)$.
The relation $\rmstep{P}{r}$ includes any number of slides and removals at $r$, including zero removals.

\begin{theorem}\label{thm:root-removals}
For every integer $k\geq2$, every nonempty connected set $P\subseteq V(T)$, and every $r\in P$, the following statements hold.
\begin{enumerate}[(a)]
\item Every $k$-PVC $Z$ with $P\rmstep{P}{r}Z$ satisfies $|Z|\geq c(P,r)$.
\item Every such $Z$ with $|Z|=c(P,r)$ satisfies $\varepsilon_P(Z,r)\geq\tau(P,r)$.
\item Some $k$-PVC $Z$ with $P\rmstep{P}{r}Z$ satisfies both $|Z|=c(P,r)$ and $\varepsilon_P(Z,r)=\tau(P,r)$.
\end{enumerate}
\end{theorem}

\begin{proof}
We prove (a) and (b) together by induction on $|P|$, bounding the number of unoccupied vertices in the child subtrees; (c) is proved after \cref{thm:min-eps-formula}, whose proof uses only (a) and (b).
\cref{fig:f7-recurrence} illustrates a set attaining both bounds in part (c).
For a single vertex, removing its token gives the empty set, with $c=0$ and $\varepsilon=\tau=1$.
For a child $y$ of $r$, let $P_y=P^r_y$, $c_y=c(P_y,y)$, $\tau_y=\tau(P_y,y)$, and $q_y=|P_y|-c_y$.
The induction hypothesis says that at most $q_y$ tokens can be removed at $y$ from $P_y$, and that $\varepsilon$ at $y$ is at least $\tau_y$ after $q_y$ removals.
In particular, $q_y\geq1$, since a set in which all vertices are occupied permits one removal.
Let $D$ and $m$ be the integers used in \cref{eq:ctau-recurrence} at $r$.

\begin{enumerate}[(a)]
\item We follow an arbitrary sequence of slides and removals at $r$ from $P$ to $Z$.
Its restriction to $P_y$ consists of internal slides and of additions or removals at $y$ when a token crosses $ry$.
The number of tokens in that restriction never exceeds $|P_y|$.
By \cref{cor:root-removal-history}, every restriction reached in this way can also be reached from $P_y$ by slides and removals at $y$ alone.
Consequently, it has at most $q_y$ unoccupied vertices.
Whenever it has exactly $q_y$ unoccupied vertices, its value of $\varepsilon$ at $y$ is at least $\tau_y$.

\begin{itemize}
\item \textbf{Case 1: $r$ is unoccupied at the state under consideration.}
Among the children with $\tau_y>0$, those whose subtrees have exactly $q_y$ unoccupied vertices must have their values $\tau_y$ in a multiset with $\sigma\leq k-2$.
Otherwise, their unoccupied paths together with $r$ would contain an unoccupied path on $k$ vertices.
There are at most $m$ such children.
Each of the other $D-m$ children with positive $\tau_y$ has at most $q_y-1$ unoccupied vertices.
Hence, the total number of unoccupied vertices is at most
\begin{equation}\label{eq:root-removal-bound}
1+\sum_yq_y-D+m=|P|-c(P,r).
\end{equation}

\item \textbf{Case 2: $r$ is occupied at the state under consideration.}
If no slide has yet brought a token to $r$, then $r$ has never been unoccupied, and no token has left any $P_y$ through $r$.
Every child subtree is then full, so the bound holds.
Otherwise, immediately before the last slide bringing a token to $r$, at most $m$ children with positive $\tau_y$ had exactly $q_y$ unoccupied vertices.
That slide changes the number of unoccupied vertices in only its source subtree.
No later operation involving $r$ occurs before the occupied state under consideration, since each departure or removal would require another arrival to occupy $r$ again.
Thus, at most $\min(D,m+1)$ children with positive $\tau_y$ have exactly $q_y$ unoccupied vertices.
The total number of unoccupied vertices is at most
\[
\sum_yq_y-D+\min(D,m+1)
\leq1+\sum_yq_y-D+m.
\]
\end{itemize}
Both possible occupations of $r$ satisfy \cref{eq:root-removal-bound}, which proves $|Z|\geq c(P,r)$.

\item Suppose that the terminal configuration $Z$ satisfies $|Z|=c(P,r)$.
If $D>m$, then $\tau(P,r)=0$, and the assertion follows from the nonnegativity of $\varepsilon$.
If $D=m$, the bound for an occupied $r$ is $\sum_yq_y$, which is one smaller than \cref{eq:root-removal-bound}.
Therefore, $r$ is unoccupied, and equality requires exactly $q_y$ unoccupied vertices in every $P_y$.
The induction hypothesis now gives
\[
\varepsilon_P(Z,r)
=1+\max_y\varepsilon_{P_y}(Z\cap P_y,y)
\geq1+\max_y\tau_y
=\tau(P,r).
\]

\end{enumerate}
\end{proof}

\subsection{The Minimum of \texorpdfstring{$\varepsilon$}{epsilon}}\label{subsec:min-eps}

To extend \cref{thm:root-removals} to every $k$-PVC whose tokens are all movable, we again remove a properly rooted subtree.
The next lemma computes the effect of that removal on $c$ and $\tau$.

\begin{lemma}\label{lem:ctau-additivity}
For every integer $k\geq2$, every nonempty connected set $P\subseteq V(T)$, and every edge $uv$ of $T[P]$ such that $U=P^v_u$ is properly rooted, we have
\[
c(U,u)=|U|-\gamma(U),\qquad \tau(U,u)=0.
\]
Moreover, for every $r\in P-U$,
\begin{equation}\label{eq:ctau-additivity}
c(P,r)=|U|-\gamma(U)+c(P-U,r),
\qquad
\tau(P,r)=\tau(P-U,r).
\end{equation}
\end{lemma}

\begin{proof}
We first evaluate the recurrence at $u$ and then follow the change along the path to $r$.
Every child subtree of $u$ in $U$ contains no path on $k$ vertices.
By \cref{lem:ctau-path-free}, its pair is $(0,a_i)$, where $a_i$ is its rooted height.
Since $T[U]$ contains a path on $k$ vertices, not all these heights have $\sigma\leq k-2$.
Thus, the recurrence gives $c(U,u)=D-m=d=|U|-\gamma(U)$ and $\tau(U,u)=0$.
Deleting $U$ removes a child with value $\tau=0$ from the recurrence at $v$.
It subtracts $d$ from $c$ there and does not change $\tau$.
At every subsequent vertex of the $vr$-path, one incoming value of $c$ decreases by $d$, while every incoming value of $\tau$ is unchanged.
The additivity observation after \cref{eq:ctau-recurrence} gives the same change at each vertex of the path, including $r$.
\end{proof}

We will also join two $k$-PVCs whose tokens are all movable, using occupied vertices at the joining edge.

\begin{lemma}\label{lem:occupied-root-gluing}
For every integer $k\geq2$, every connected set $P\subseteq V(T)$, and every edge $uv$ of $T[P]$, let $U=P^v_u$ and $W'=P^u_v$.
If $A$ and $B$ are $k$-PVCs of $T[U]$ and $T[W']$, respectively, with $u\in A$, $v\in B$, and $\Rigid_U(A)=\Rigid_{W'}(B)=\emptyset$, then $A\cup B$ is a $k$-PVC of $T[P]$ and $\Rigid_P(A\cup B)=\emptyset$.
\end{lemma}

\begin{proof}
We keep one end of the joining edge occupied while performing TS-sequences on the other side.
Every path that meets both sides contains $u$ and $v$, so $A\cup B$ is a $k$-PVC.
With $B$ fixed, every TS-sequence in $[A]_U$ is valid in $P$, because $v$ remains occupied.
Hence, every vertex of $U$ can be made unoccupied.
Starting again at $A\cup B$ and keeping $A$ fixed shows the same assertion for every vertex of $W'$, since $u$ remains occupied.
Therefore, no vertex belongs to $\Rigid_P(A\cup B)$.
\end{proof}

The next theorem shows that the lower bounds of \cref{thm:root-removals} are attained from every $k$-PVC whose tokens are all movable.
The number of removable tokens is consequently $|S|-c(P,r)$.

\begin{theorem}\label{thm:min-eps-formula}
For every integer $k\geq2$, every nonempty connected set $P\subseteq V(T)$, every $r\in P$, and every $k$-PVC $S$ of $T[P]$ with $\Rigid_P(S)=\emptyset$, the following statements hold.
\begin{enumerate}[(a)]
\item Every $k$-PVC $Z$ with $S\rmstep{P}{r}Z$ satisfies $|Z|\geq c(P,r)$.
\item Every such $Z$ with $|Z|=c(P,r)$ satisfies $\varepsilon_P(Z,r)\geq\tau(P,r)$.
\item Some $k$-PVC $Z$ with $S\rmstep{P}{r}Z$ satisfies both $|Z|=c(P,r)$ and $\varepsilon_P(Z,r)=\tau(P,r)$.
\end{enumerate}
\end{theorem}

\begin{proof}
We obtain the lower bounds by adding tokens at $r$, and prove attainment by induction on $|P|$ after removing a properly rooted subtree that avoids $r$.
Let $t=|S|$ and $q=|P|-t$.
\begin{enumerate}[(a)]
\item By \cref{cor:movable-counts}, $P\rmstep{P}{r}S$.
Concatenating those operations with $S\rmstep{P}{r}Z$ gives $P\rmstep{P}{r}Z$.
Part (a) of \cref{thm:root-removals} yields $|Z|\geq c(P,r)$.
\item The same concatenation and part (b) of \cref{thm:root-removals} give $\varepsilon_P(Z,r)\geq\tau(P,r)$ whenever $|Z|=c(P,r)$.
\item We distinguish the following cases for the induction proving attainment.
\begin{itemize}
\item \textbf{Case 1: $T[P]$ contains no path on $k$ vertices.}
As long as a token remains, \cref{lem:unrestricted-sliding} permits a TS-sequence that occupies $r$.
Removing that token leaves a $k$-PVC, since every subset of $P$ is a $k$-PVC.
After $t$ removals, the empty set remains.
By \cref{lem:ctau-path-free}, its token count is $c(P,r)=0$, and its value of $\varepsilon$ at $r$ is the rooted height $\tau(P,r)$.

\item \textbf{Case 2: $T[P]$ contains a path on $k$ vertices, and $T[P-r]$ does not.}
The definitions of $d$, $\gamma$ and $C_h$ apply to the whole rooted tree $(P,r)$, with no exterior.
The computation in the first paragraph of the proof of \cref{lem:ctau-additivity} gives $(c(P,r),\tau(P,r))=(d,0)$.
By \cref{lem:canonical-movable}, the number $t$ lies between $d$ and $|P|-1$, and all tokens of $C_q$ are movable.
By \cref{lem:key}, there is a TS-sequence from $S$ to $C_q$.
Repeatedly applying the last assertion of \cref{lem:canonical-movable} removes tokens at $r$ until $C_{\gamma(P)}$ is reached.
The latter $k$-PVC has $d$ tokens and occupies $r$, so it attains both bounds.

\item \textbf{Case 3: $T[P-r]$ contains a path on $k$ vertices.}
By \cref{lem:properly-rooted-existence}(a), there is an edge $uv$ such that $U=P^v_u$ is properly rooted and $r\notin U$.
Let $W'=P^u_v$, $\gamma=\gamma(U)$, and $d=|U|-\gamma$.
We first prove attainment for all $k$-PVCs at the current order whose number of unoccupied vertices is greater than $\gamma$.

\begin{itemize}
\item \textbf{Case 3.1: $q>\gamma$.}
By \cref{thm:prescribed-subtree}, a TS-sequence reaches a $k$-PVC that agrees with $C_\gamma$ on $U$ and whose restriction to $W'$ has all tokens movable.
That restriction has $t-d$ tokens and is not full, since it has $q-\gamma\geq1$ unoccupied vertices.
The induction hypothesis on the strictly smaller tree $T[W']$ removes $(t-d)-c(W',r)$ tokens at $r$ and reaches a $k$-PVC $B$ with
\[
|B|=c(W',r),\qquad \varepsilon_{W'}(B,r)=\tau(W',r).
\]
Every slide and removal in these operations is valid in $P$ with $C_\gamma$ fixed, because $u$ stays occupied.
An unoccupied path starting at $r$ cannot enter $U$ for the same reason.
Thus, the resulting $k$-PVC $Z=C_\gamma\cup B$ satisfies
\[
|Z|=d+c(W',r)=c(P,r),\qquad
\varepsilon_P(Z,r)=\tau(W',r)=\tau(P,r),
\]
where the last equalities follow from \cref{eq:ctau-additivity}.

\item \textbf{Case 3.2: $q\leq\gamma$ and $|W'|\geq2$.}
Let $A=C_\gamma$ on $U$, and let $B$ be a $k$-PVC of $T[W']$ with exactly one unoccupied vertex other than $v$.
All tokens of $A$ are movable by \cref{lem:canonical-movable}.
All tokens of $B$ are movable by \cref{lem:unrestricted-sliding}, since every position of its single unoccupied vertex is permitted.
The two incident vertices $u$ and $v$ are occupied.
By \cref{lem:occupied-root-gluing}, all tokens of $A\cup B$ are movable.
Its token count $t_0=d+|W'|-1$ gives $\gamma+1$ unoccupied vertices in $P$.
The preceding $q>\gamma$ argument, already proved at the current order using only the induction hypothesis on $W'$, gives operations from $A\cup B$ attaining the bounds.
Since $q\leq\gamma$, we have $t\geq d+|W'|=t_0+1$.
By \cref{cor:movable-counts}, adding tokens at $r$ from $A\cup B$ reaches, at count $t$, the unique set of pairwise reachable $k$-PVCs whose tokens are all movable.
That set is $[S]_P$ by \cref{lem:key}.
Reversing the additions, with their intervening TS-sequences, gives $S\rmstep{P}{r}A\cup B$ using exactly $t-t_0$ removals.
Following the operations just established from $A\cup B$ proves attainment from $S$.

\item \textbf{Case 3.3: $q\leq\gamma$ and $|W'|=1$.}
The complementary set is $W'=\{r\}$, and $v=r$.
The $k$-PVC $A=C_\gamma+r$ has $t_0=d+1$ tokens.
Keeping $r$ occupied allows every vertex of $U$ to become unoccupied, since all tokens of $C_\gamma$ are movable in $U$.
In particular, a TS-sequence in $U$ can make $u$ unoccupied.
The token on $r$ can then slide to $u$: its new position meets every path crossing the edge $ru$, and the newly unoccupied vertex $r$ has no other neighbor.
Hence, all tokens of $A$ are movable in $P$.
At $A$ itself, removing the token on $r$ leaves $C_\gamma$, a $k$-PVC of $T[P]$, and its value of $\varepsilon$ at $r$ is one because $u$ is occupied.
The singleton pair and \cref{eq:ctau-additivity} give
\[
c(P,r)=d+c(\{r\},r)=d,
\qquad
\tau(P,r)=\tau(\{r\},r)=1.
\]
The assumption $q\leq\gamma$ is exactly $t\geq d+1=t_0$.
As in the preceding case, \cref{cor:movable-counts,lem:key} give $S\rmstep{P}{r}A$ with $t-t_0$ removals.
The final removal at $r$ therefore attains both bounds.
\end{itemize}
\end{itemize}
The induction uses only smaller trees, except for the explicitly completed $q>\gamma$ case at the current order.
\end{enumerate}
\end{proof}

\begin{proof}[Proof of \cref{thm:root-removals}(c)]
We remove the token on $r$ from $P$ and apply \cref{thm:min-eps-formula}(c).
The set $P-r$ has one unoccupied vertex, so it is a $k$-PVC of $T[P]$, and by \cref{lem:unrestricted-sliding}, all such sets are reachable from one another, so $\Rigid_P(P-r)=\emptyset$.
\Cref{thm:min-eps-formula}(c) for $S=P-r$ gives a $k$-PVC $Z$ with $P-r\rmstep{P}{r}Z$, $|Z|=c(P,r)$ and $\varepsilon_P(Z,r)=\tau(P,r)$, and the removal at $r$ followed by these operations gives $P\rmstep{P}{r}Z$.
\end{proof}

For a nonempty connected set $P$, a nonnegative integer $t$, and $r\in P$, let
\begin{equation}\label{eq:lambda}
\lambda(P,t,r)=
\begin{cases}
\tau(P,r),&t=c(P,r),\\
0,&t\neq c(P,r).
\end{cases}
\end{equation}
The following consequence is the form used by \cref{alg:compute-rigid}.

\begin{corollary}\label{cor:min-eps-formula}
For every integer $k\geq2$, every nonempty connected set $P\subseteq V(T)$, and every $k$-PVC $S$ of $T[P]$ with $\Rigid_P(S)=\emptyset$, we have, for every $r\in P$,
\begin{equation}\label{eq:min-eps-formula}
\min\{\varepsilon_P(S',r):S'\in[S]_P\}=\lambda(P,|S|,r).
\end{equation}
Moreover,
\begin{equation}\label{eq:never-occupied-formula}
\NeverOcc_P(S)=\{r\in P:|S|=c(P,r)\text{ and }\tau(P,r)>0\}.
\end{equation}
\end{corollary}

\begin{proof}
We separate the cases according to whether a token can still be removed at $r$.
By \cref{thm:min-eps-formula}, $|S|\geq c(P,r)$.
\begin{itemize}
\item \textbf{Case 1: $|S|>c(P,r)$.}
The operations attaining $c(P,r)$ include a first removal at $r$.
The preparation before that removal is a TS-sequence from $S$ to a $k$-PVC occupying $r$.
The minimum in \cref{eq:min-eps-formula} is therefore zero.
\item \textbf{Case 2: $|S|=c(P,r)$.}
No removal is possible, so the $k$-PVCs reached by $\rmstep{P}{r}$ are precisely the members of $[S]_P$.
Parts (b) and (c) of \cref{thm:min-eps-formula} give the minimum $\tau(P,r)$.
\end{itemize}
Finally, a vertex is never occupied exactly when its minimum value of $\varepsilon$ is positive, which proves \cref{eq:never-occupied-formula}.
\end{proof}

\subsection{Computing the Rigid Tokens}\label{subsec:rigid-algorithm}

Like the linear-time algorithm of Demaine et al.~\cite{Demaine2015} for independent sets, \cref{alg:compute-rigid} starts with all tokens as candidates for rigid tokens and repeatedly removes a candidate whose token can move while all other candidates stay occupied.
Demaine et al.~\cite{Demaine2015} delete such a token from the independent set without changing the rigidity of the other tokens; deleting a token from a $k$-PVC may destroy the $k$-PVC property, so \cref{alg:compute-rigid} keeps $S$ fixed and removes the vertex only from the set of candidates.

Let $W\subseteq V(T)$ be nonempty and connected, and let $S$ be a $k$-PVC of $T[W]$.
For $B\subseteq S$ and $w\in W-B$, let $C_B(w)$ be the vertex set of the component of $T[W]-B$ containing $w$, and let
\[
\lambda_B(w)=\lambda(C_B(w),|S\cap C_B(w)|,w).
\]

\begin{definition}\label{def:slide-condition}
For $B\subseteq S$, $v\in B$, and a neighbor $z\in W-B$ of $v$, the pair $(v,z)$ satisfies the \emph{slide condition} for $B$ if
\begin{equation}\label{eq:algorithm-slide-condition}
\sigma\bigl(\{\lambda_B(w):w\in N_{T[W]}(v)-B-z\}\bigr)\leq k-2.
\end{equation}
The braces in \cref{eq:algorithm-slide-condition} denote a multiset, with one member for each indicated neighbor.
\end{definition}

The condition matches \cref{lem:slide-test} after making $z$ unoccupied and minimizing $\varepsilon$ independently at the other neighbors outside $B$.
The proof below establishes that these preparations are available whenever the condition is used.

\begin{algorithm}[t]
\caption{Computing the rigid tokens}\label{alg:compute-rigid}
\KwIn{An integer $k\geq2$, a nonempty connected set $W\subseteq V(T)$, and a $k$-PVC $S$ of $T[W]$}
\KwOut{$\Rigid_W(S)$ and $\NeverOcc_W(S)$}
$B\gets S$\;
\While{true}{
  \ForEach{vertex set $P$ of a component of $T[W]-B$}{
    $t_P\gets |S\cap P|$\;
    Compute $(c(P,r),\tau(P,r))$ for all $r\in P$ by \cref{eq:ctau-recurrence}\;
    $H_P\gets\{r\in P:t_P=c(P,r)\text{ and }\tau(P,r)>0\}$\;
  }
  Find a pair $(v,z)$ satisfying \cref{eq:algorithm-slide-condition}, if one exists\;
  \If{no such pair exists}{
    \Return{$\bigl(B,\bigcup_P H_P\bigr)$}\;
  }
  $B\gets B-v$\;
}
\end{algorithm}

\cref{alg:compute-rigid} uses the formula for the minimum of $\varepsilon$ only on restrictions for which all tokens are movable.
That property follows from the invariant of the proof below, which each removal preserves.

\begin{theorem}\label{thm:rigid-algorithm}
For every integer $k\geq2$, every nonempty connected set $W\subseteq V(T)$, and every $k$-PVC $S$ of $T[W]$, \cref{alg:compute-rigid} terminates after at most $|S|+1$ rounds and returns $(\Rigid_W(S),\NeverOcc_W(S))$.
\end{theorem}

\begin{proof}
We prove that every removal preserves the invariant, and then use the first slide from a remaining candidate to prove completeness at termination.
At the start of each round, the invariant is that every vertex of $W-B$ is unoccupied in some $k$-PVC reachable from $S$ by slides that keep every vertex of $B$ occupied.
Initially, $B=S$, and every vertex of $W-S$ is unoccupied in $S$ itself.
The invariant gives
\begin{equation}\label{eq:algorithm-invariant}
\Rigid_P(S\cap P)=\emptyset
\quad\text{for the vertex set $P$ of every component of $T[W]-B$,}
\end{equation}
since, by \cref{lem:fixing-occupied-set}(c), the sequence for a vertex $x\in P$ restricts to a TS-sequence in $T[P]$ from $S\cap P$ to a set omitting $x$.

Suppose that the invariant holds for $B$.
While $B$ stays occupied, TS-sequences inside different components of $T[W]-B$ can be performed independently by \cref{lem:fixing-occupied-set}.
By \cref{eq:algorithm-invariant} and \cref{cor:min-eps-formula}, \cref{alg:compute-rigid} computes the actual minimum of $\varepsilon$ for each restriction $S\cap P$.

Suppose that $(v,z)$ satisfies the slide condition.
Distinct neighbors of $v$ outside $B$ lie in distinct components of $T[W]-B$, since a path joining two of them in that graph would create a cycle through $v$.
The invariant permits a TS-sequence making $z$ unoccupied in its component.
For every other neighbor $w$ of $v$ outside $B$, \cref{cor:min-eps-formula} permits a TS-sequence reaching value $\lambda_B(w)$ at $w$ in its component.
All these preparations can be performed independently while keeping $B$ occupied.
After the preparations, the token on $v$ can slide to $z$ by \cref{lem:slide-test} and \cref{eq:algorithm-slide-condition}.
Equivalently, every newly unoccupied path must pass through $v$, avoid the now occupied vertex $z$ and the vertices of $B-v$, and have at most $1+\sigma(\{\lambda_B(w):w\in N_{T[W]}(v)-B-z\})\leq k-1$ vertices.

The preparations and the slide form a TS-sequence from $S$ that keeps every vertex of $B-v$ occupied and makes $v$ unoccupied, and the sequences of the invariant for the vertices of $W-B$ keep $B$, hence $B-v$, occupied.
So the invariant holds for $B-v$.

Every round except the last removes one vertex from $B\subseteq S$, so at most $|S|+1$ rounds occur.
Suppose that the final set $B$ contains a vertex that is not rigid.
A TS-sequence from $S$ making such a vertex unoccupied has a first slide that makes some vertex of $B$ unoccupied.
Let that slide move a token from $v\in B$ to $z$, and let $S_1$ be the preceding $k$-PVC.
Until that slide, all vertices of $B$ stay occupied, so every restriction to a component $P$ of $T[W]-B$ stays in $[S\cap P]_P$.
In particular, $z\notin B$, and, for every other neighbor $w\in W-B$ of $v$,
\[
\lambda_B(w)\leq
\varepsilon_{C_B(w)}(S_1\cap C_B(w),w).
\]
Because all vertices of $B$ are occupied, an unoccupied path starting at $w$ cannot leave $C_B(w)$.
The actual values on the right therefore satisfy the necessary slide test at $v$.
Decreasing entries of a multiset cannot increase $\sigma$, so $(v,z)$ satisfies \cref{eq:algorithm-slide-condition}.
That conclusion contradicts termination.
Hence, $B\subseteq\Rigid_W(S)$.
The invariant already shows that every vertex outside $B$ can be made unoccupied globally, so $B=\Rigid_W(S)$.

With the true rigid set fixed, \cref{lem:fixing-occupied-set} identifies $[S]_W$ with the independent choices of its restrictions to the components of $T[W]-B$.
By \cref{eq:never-occupied-formula}, each $H_P$ is exactly the set of vertices never occupied in its component.
No vertex of $B$ is never occupied.
Therefore, their union is $\NeverOcc_W(S)$, as required.
\end{proof}

\begin{figure}[ht]
\centering
\begin{adjustbox}{max width=\textwidth}
\begin{tikzpicture}[v/.style={circle,draw=black,fill=white,inner sep=0pt,minimum size=4.5mm},
tok/.style={v,fill=black},rig/.style={v,fill=black,draw=black!30,line width=1.8pt},
nev/.style={v,fill=black!20},line width=.45pt]
\node[rig] (a_u0) at (0.0000,0.0000) {};
\node[rig] (a_u1) at (0.0000,0.8200) {};
\node[rig] (a_u2) at (0.0000,1.6400) {};
\node[v] (a_u3) at (-0.6560,2.3370) {};
\node[v] (a_u4) at (0.6560,2.3370) {};
\node[v] (a_u5) at (-1.4760,-0.5330) {};
\node[v] (a_u6) at (-1.4760,-1.2300) {};
\node[v] (a_u7) at (-1.4760,-1.9270) {};
\node[v] (a_u8) at (1.4760,-0.5330) {};
\node[v] (a_u9) at (1.4760,-1.2300) {};
\node[v] (a_u10) at (1.4760,-1.9270) {};
\node[rig] (a_u11) at (0.0000,-1.1070) {};
\node[v] (a_u12) at (-0.4510,-2.6240) {};
\node[v] (a_u13) at (0.4510,-2.6240) {};
\draw (a_u0) -- (a_u1);
\draw (a_u1) -- (a_u2);
\draw (a_u2) -- (a_u3);
\draw (a_u2) -- (a_u4);
\draw (a_u0) -- (a_u5);
\draw (a_u5) -- (a_u6);
\draw (a_u6) -- (a_u7);
\draw (a_u0) -- (a_u8);
\draw (a_u8) -- (a_u9);
\draw (a_u9) -- (a_u10);
\draw (a_u0) -- (a_u11);
\draw (a_u11) -- (a_u12);
\draw (a_u11) -- (a_u13);
\node[font=\scriptsize,anchor=east,inner sep=1pt] at (-0.2900,0.0000) {$u_{0}$};
\node[font=\scriptsize,anchor=east,inner sep=1pt] at (-0.2900,0.8200) {$u_{1}$};
\node[font=\scriptsize,anchor=east,inner sep=1pt] at (-0.2900,1.6400) {$u_{2}$};
\node[font=\scriptsize,anchor=east,inner sep=1pt] at (-0.9460,2.3370) {$u_{3}$};
\node[font=\scriptsize,anchor=west,inner sep=1pt] at (0.9460,2.3370) {$u_{4}$};
\node[font=\scriptsize,anchor=east,inner sep=1pt] at (0.3660,2.3370) {$1$};
\node[font=\scriptsize,anchor=east,inner sep=1pt] at (-1.7660,-0.5330) {$u_{5}$};
\node[font=\scriptsize,anchor=west,inner sep=1pt] at (-1.1860,-0.5330) {$3$};
\node[font=\scriptsize,anchor=east,inner sep=1pt] at (-1.7660,-1.2300) {$u_{6}$};
\node[font=\scriptsize,anchor=east,inner sep=1pt] at (-1.7660,-1.9270) {$u_{7}$};
\node[font=\scriptsize,anchor=west,inner sep=1pt] at (1.7660,-0.5330) {$u_{8}$};
\node[font=\scriptsize,anchor=east,inner sep=1pt] at (1.1860,-0.5330) {$3$};
\node[font=\scriptsize,anchor=west,inner sep=1pt] at (1.7660,-1.2300) {$u_{9}$};
\node[font=\scriptsize,anchor=west,inner sep=1pt] at (1.7660,-1.9270) {$u_{10}$};
\node[font=\scriptsize,anchor=west,inner sep=1pt] at (0.2900,-1.1070) {$u_{11}$};
\node[font=\scriptsize,anchor=east,inner sep=1pt] at (-0.7410,-2.6240) {$u_{12}$};
\node[font=\scriptsize,anchor=west,inner sep=1pt] at (0.7410,-2.6240) {$u_{13}$};
\node[font=\scriptsize,anchor=south,inner sep=1pt] at (-0.6560,2.6970) {$1$};
\node[font=\scriptsize,anchor=east,inner sep=1pt] at (-0.7710,-2.3240) {$1$};
\node[font=\scriptsize,anchor=west,inner sep=1pt] at (0.7710,-2.3240) {$1$};
\draw[->,thick,green!50!black] (0.0480,2.0559) -- (-0.2379,2.3597);
\node[font=\small,anchor=center,inner sep=1pt] at (0.0000,-3.1160) {(a) $B_{0}$};
\node[rig] (b_u0) at (5.5760,0.0000) {};
\node[rig] (b_u1) at (5.5760,0.8200) {};
\node[tok] (b_u2) at (5.5760,1.6400) {};
\node[v] (b_u3) at (4.9200,2.3370) {};
\node[v] (b_u4) at (6.2320,2.3370) {};
\node[v] (b_u5) at (4.1000,-0.5330) {};
\node[v] (b_u6) at (4.1000,-1.2300) {};
\node[v] (b_u7) at (4.1000,-1.9270) {};
\node[v] (b_u8) at (7.0520,-0.5330) {};
\node[v] (b_u9) at (7.0520,-1.2300) {};
\node[v] (b_u10) at (7.0520,-1.9270) {};
\node[rig] (b_u11) at (5.5760,-1.1070) {};
\node[v] (b_u12) at (5.1250,-2.6240) {};
\node[v] (b_u13) at (6.0270,-2.6240) {};
\draw (b_u0) -- (b_u1);
\draw (b_u1) -- (b_u2);
\draw (b_u2) -- (b_u3);
\draw (b_u2) -- (b_u4);
\draw (b_u0) -- (b_u5);
\draw (b_u5) -- (b_u6);
\draw (b_u6) -- (b_u7);
\draw (b_u0) -- (b_u8);
\draw (b_u8) -- (b_u9);
\draw (b_u9) -- (b_u10);
\draw (b_u0) -- (b_u11);
\draw (b_u11) -- (b_u12);
\draw (b_u11) -- (b_u13);
\node[font=\scriptsize,anchor=east,inner sep=1pt] at (5.2860,0.0000) {$u_{0}$};
\node[font=\scriptsize,anchor=east,inner sep=1pt] at (5.2860,0.8200) {$u_{1}$};
\node[font=\scriptsize,anchor=east,inner sep=1pt] at (5.2860,1.6400) {$u_{2}$};
\node[font=\scriptsize,anchor=west,inner sep=1pt] at (5.8660,1.6400) {$0$};
\node[font=\scriptsize,anchor=east,inner sep=1pt] at (4.6300,2.3370) {$u_{3}$};
\node[font=\scriptsize,anchor=west,inner sep=1pt] at (6.5220,2.3370) {$u_{4}$};
\node[font=\scriptsize,anchor=east,inner sep=1pt] at (3.8100,-0.5330) {$u_{5}$};
\node[font=\scriptsize,anchor=west,inner sep=1pt] at (4.3900,-0.5330) {$3$};
\node[font=\scriptsize,anchor=east,inner sep=1pt] at (3.8100,-1.2300) {$u_{6}$};
\node[font=\scriptsize,anchor=east,inner sep=1pt] at (3.8100,-1.9270) {$u_{7}$};
\node[font=\scriptsize,anchor=west,inner sep=1pt] at (7.3420,-0.5330) {$u_{8}$};
\node[font=\scriptsize,anchor=east,inner sep=1pt] at (6.7620,-0.5330) {$3$};
\node[font=\scriptsize,anchor=west,inner sep=1pt] at (7.3420,-1.2300) {$u_{9}$};
\node[font=\scriptsize,anchor=west,inner sep=1pt] at (7.3420,-1.9270) {$u_{10}$};
\node[font=\scriptsize,anchor=west,inner sep=1pt] at (5.8660,-1.1070) {$u_{11}$};
\node[font=\scriptsize,anchor=east,inner sep=1pt] at (4.8350,-2.6240) {$u_{12}$};
\node[font=\scriptsize,anchor=west,inner sep=1pt] at (6.3170,-2.6240) {$u_{13}$};
\node[font=\scriptsize,anchor=east,inner sep=1pt] at (4.8050,-2.3240) {$1$};
\node[font=\scriptsize,anchor=west,inner sep=1pt] at (6.3470,-2.3240) {$1$};
\draw[->,thick,green!50!black] (5.8960,1.0900) -- (5.8960,1.3700);
\node[font=\small,anchor=center,inner sep=1pt] at (5.5760,-3.1160) {(b) $B_{1}$};
\node[rig] (c_u0) at (0.0000,-6.3960) {};
\node[tok] (c_u1) at (0.0000,-5.5760) {};
\node[tok] (c_u2) at (0.0000,-4.7560) {};
\node[v] (c_u3) at (-0.6560,-4.0590) {};
\node[v] (c_u4) at (0.6560,-4.0590) {};
\node[v] (c_u5) at (-1.4760,-6.9290) {};
\node[v] (c_u6) at (-1.4760,-7.6260) {};
\node[v] (c_u7) at (-1.4760,-8.3230) {};
\node[v] (c_u8) at (1.4760,-6.9290) {};
\node[v] (c_u9) at (1.4760,-7.6260) {};
\node[v] (c_u10) at (1.4760,-8.3230) {};
\node[rig] (c_u11) at (0.0000,-7.5030) {};
\node[v] (c_u12) at (-0.4510,-9.0200) {};
\node[v] (c_u13) at (0.4510,-9.0200) {};
\draw (c_u0) -- (c_u1);
\draw (c_u1) -- (c_u2);
\draw (c_u2) -- (c_u3);
\draw (c_u2) -- (c_u4);
\draw (c_u0) -- (c_u5);
\draw (c_u5) -- (c_u6);
\draw (c_u6) -- (c_u7);
\draw (c_u0) -- (c_u8);
\draw (c_u8) -- (c_u9);
\draw (c_u9) -- (c_u10);
\draw (c_u0) -- (c_u11);
\draw (c_u11) -- (c_u12);
\draw (c_u11) -- (c_u13);
\node[font=\scriptsize,anchor=east,inner sep=1pt] at (-0.2900,-6.3960) {$u_{0}$};
\node[font=\scriptsize,anchor=east,inner sep=1pt] at (-0.2900,-5.5760) {$u_{1}$};
\node[font=\scriptsize,anchor=west,inner sep=1pt] at (0.2900,-5.5760) {$0$};
\node[font=\scriptsize,anchor=east,inner sep=1pt] at (-0.2900,-4.7560) {$u_{2}$};
\node[font=\scriptsize,anchor=east,inner sep=1pt] at (-0.9460,-4.0590) {$u_{3}$};
\node[font=\scriptsize,anchor=west,inner sep=1pt] at (0.9460,-4.0590) {$u_{4}$};
\node[font=\scriptsize,anchor=east,inner sep=1pt] at (-1.7660,-6.9290) {$u_{5}$};
\node[font=\scriptsize,anchor=west,inner sep=1pt] at (-1.1860,-6.9290) {$3$};
\node[font=\scriptsize,anchor=east,inner sep=1pt] at (-1.7660,-7.6260) {$u_{6}$};
\node[font=\scriptsize,anchor=east,inner sep=1pt] at (-1.7660,-8.3230) {$u_{7}$};
\node[font=\scriptsize,anchor=west,inner sep=1pt] at (1.7660,-6.9290) {$u_{8}$};
\node[font=\scriptsize,anchor=east,inner sep=1pt] at (1.1860,-6.9290) {$3$};
\node[font=\scriptsize,anchor=west,inner sep=1pt] at (1.7660,-7.6260) {$u_{9}$};
\node[font=\scriptsize,anchor=west,inner sep=1pt] at (1.7660,-8.3230) {$u_{10}$};
\node[font=\scriptsize,anchor=west,inner sep=1pt] at (0.2900,-7.5030) {$u_{11}$};
\node[font=\scriptsize,anchor=east,inner sep=1pt] at (-0.7410,-9.0200) {$u_{12}$};
\node[font=\scriptsize,anchor=west,inner sep=1pt] at (0.7410,-9.0200) {$u_{13}$};
\node[font=\scriptsize,anchor=east,inner sep=1pt] at (-0.7710,-8.7200) {$1$};
\node[font=\scriptsize,anchor=west,inner sep=1pt] at (0.7710,-8.7200) {$1$};
\draw[->,thick,green!50!black] (-0.3837,-7.6706) -- (-0.6808,-8.6700);
\node[font=\small,anchor=center,inner sep=1pt] at (0.0000,-9.5120) {(c) $B_{2}$};
\node[rig] (d_u0) at (5.5760,-6.3960) {};
\node[tok] (d_u1) at (5.5760,-5.5760) {};
\node[tok] (d_u2) at (5.5760,-4.7560) {};
\node[v] (d_u3) at (4.9200,-4.0590) {};
\node[v] (d_u4) at (6.2320,-4.0590) {};
\node[nev] (d_u5) at (4.1000,-6.9290) {};
\node[nev] (d_u6) at (4.1000,-7.6260) {};
\node[nev] (d_u7) at (4.1000,-8.3230) {};
\node[nev] (d_u8) at (7.0520,-6.9290) {};
\node[nev] (d_u9) at (7.0520,-7.6260) {};
\node[nev] (d_u10) at (7.0520,-8.3230) {};
\node[tok] (d_u11) at (5.5760,-7.5030) {};
\node[v] (d_u12) at (5.1250,-9.0200) {};
\node[v] (d_u13) at (6.0270,-9.0200) {};
\draw (d_u0) -- (d_u1);
\draw (d_u1) -- (d_u2);
\draw (d_u2) -- (d_u3);
\draw (d_u2) -- (d_u4);
\draw (d_u0) -- (d_u5);
\draw (d_u5) -- (d_u6);
\draw (d_u6) -- (d_u7);
\draw (d_u0) -- (d_u8);
\draw (d_u8) -- (d_u9);
\draw (d_u9) -- (d_u10);
\draw (d_u0) -- (d_u11);
\draw (d_u11) -- (d_u12);
\draw (d_u11) -- (d_u13);
\node[font=\scriptsize,anchor=east,inner sep=1pt] at (5.2860,-6.3960) {$u_{0}$};
\node[font=\scriptsize,anchor=east,inner sep=1pt] at (5.2860,-5.5760) {$u_{1}$};
\node[font=\scriptsize,anchor=west,inner sep=1pt] at (5.8660,-5.5760) {$0$};
\node[font=\scriptsize,anchor=east,inner sep=1pt] at (5.2860,-4.7560) {$u_{2}$};
\node[font=\scriptsize,anchor=east,inner sep=1pt] at (4.6300,-4.0590) {$u_{3}$};
\node[font=\scriptsize,anchor=west,inner sep=1pt] at (6.5220,-4.0590) {$u_{4}$};
\node[font=\scriptsize,anchor=east,inner sep=1pt] at (3.8100,-6.9290) {$u_{5}$};
\node[font=\scriptsize,anchor=west,inner sep=1pt] at (4.3900,-6.9290) {$3$};
\node[font=\scriptsize,anchor=east,inner sep=1pt] at (3.8100,-7.6260) {$u_{6}$};
\node[font=\scriptsize,anchor=east,inner sep=1pt] at (3.8100,-8.3230) {$u_{7}$};
\node[font=\scriptsize,anchor=west,inner sep=1pt] at (7.3420,-6.9290) {$u_{8}$};
\node[font=\scriptsize,anchor=east,inner sep=1pt] at (6.7620,-6.9290) {$3$};
\node[font=\scriptsize,anchor=west,inner sep=1pt] at (7.3420,-7.6260) {$u_{9}$};
\node[font=\scriptsize,anchor=west,inner sep=1pt] at (7.3420,-8.3230) {$u_{10}$};
\node[font=\scriptsize,anchor=west,inner sep=1pt] at (5.8660,-7.5030) {$u_{11}$};
\node[font=\scriptsize,anchor=east,inner sep=1pt] at (5.2860,-7.5030) {$0$};
\node[font=\scriptsize,anchor=east,inner sep=1pt] at (4.8350,-9.0200) {$u_{12}$};
\node[font=\scriptsize,anchor=west,inner sep=1pt] at (6.3170,-9.0200) {$u_{13}$};
\node[font=\small,anchor=center,inner sep=1pt] at (5.5760,-9.5120) {(d) $B_{3}$};
\path[use as bounding box] (-2.2960,-9.9220) rectangle (7.8720,3.1160);
\end{tikzpicture}
\relax\space \end{adjustbox}
\caption{The run of \cref{alg:compute-rigid} on $S=\{u_{0},u_{1},u_{2},u_{11}\}$ with $k=4$.
Panels (a)--(d) show the candidate set $B_i$ after $i$ removals, for $i\in\{0,\ldots,3\}$.
Black vertices carry the tokens of the fixed input $S$; thick gray outlines mark the candidates in $B_i$, while the other black vertices have been removed from the candidate set.
White vertices are unoccupied, and the light gray vertices in the final panel are never occupied.
Values beside neighbors outside $B_i$ are $\lambda_{B_i}$; each arrow gives a pair satisfying the slide condition and indicates removal of its source from the candidate set, without changing $S$.
The successive pairs are $(u_{2},u_{3})$, $(u_{1},u_{2})$, $(u_{11},u_{12})$; the output is $(\Rigid(S),\NeverOcc(S))=(\{u_{0}\},\{u_{5},u_{6},u_{7},u_{8},u_{9},u_{10}\})$.}
\label{fig:f8-candidates}
\end{figure}
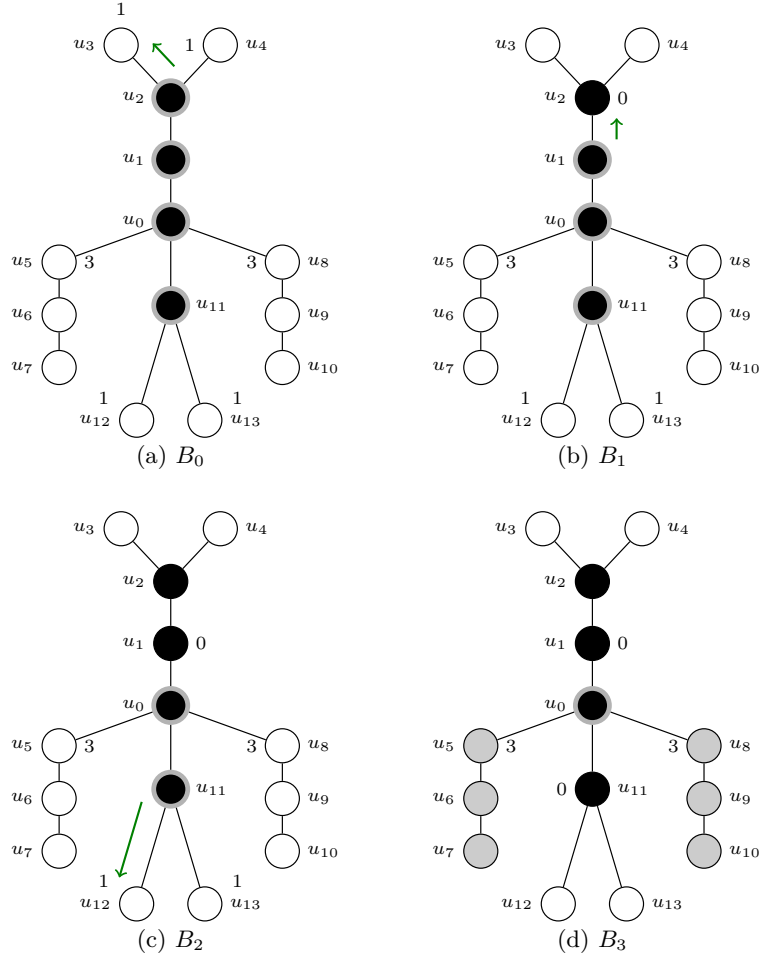

\cref{fig:f8-candidates} illustrates the successive candidate sets in \cref{alg:compute-rigid}.

\begin{remark}\label{rem:any-order}
The choice of a pair satisfying the slide condition in \cref{alg:compute-rigid} is arbitrary.
The invariant is preserved for every such pair, and the proof at termination uses only the absence of another pair.
Thus, every removal order returns the same sets $\Rigid_W(S)$ and $\NeverOcc_W(S)$.
\end{remark}

\subsection{Running Time and Proof of the Decision Theorem}\label{subsec:running-time}

We describe how to compute every pair $(c(P,r),\tau(P,r))$ in one component in $O(k|P|)$ time, and how to check all slide conditions in linear time per round.
These bounds give the claimed running time after at most $n+1$ rounds.

For each vertex set $P$ of a component, each round of \cref{alg:compute-rigid} roots $T[P]$ temporarily at an arbitrary vertex and makes two passes.
The first pass, from the leaves toward the temporary root, computes the pair for every subtree below an edge.
The second pass, from the temporary root toward the leaves, computes the pair on the other side of every edge.
At a vertex $x$, a sum stores the incoming values of $c$, and a histogram with indices $0,\ldots,k-1$ stores the multiplicities of the incoming values of $\tau$.
Initializing the histograms takes $O(k|P|)$ time, and accumulating their entries and the sums takes $O(|P|)$ time.

The histogram determines $D$, $m$, and $\eta$ in $O(k)$ time.
To find $m$, the positive indices are processed in increasing order while retaining the largest selected value and the number of selected entries.
For an index $s$ of positive multiplicity, its first occurrence is permitted exactly when its sum with the previous largest selected value, or zero if none was selected, is at most $k-2$.
After that first occurrence, all further occurrences of $s$ are permitted exactly when $2s\leq k-2$.
Thus, each index takes constant time: either all of its occurrences are included, or the longest permitted prefix ends after zero or one occurrence at that index.
After the first failed inclusion, no longer prefix is permitted.
The scan therefore computes exactly the value $m$ of \cref{eq:ctau-recurrence}, without listing every occurrence separately for each edge.

In the second pass, all pairs entering a vertex $x$ are known when $x$ is processed.
The pair sent toward a neighbor $y$ is obtained by subtracting the incoming value of $c$ from $y$ and one occurrence of its value of $\tau$, and then evaluating the recurrence on the remaining histogram.
The subtracted entry is restored before processing another neighbor.
Each evaluation takes $O(k)$ time.
There are two ordered ends of each edge and one full-root evaluation for each vertex, so the two passes take $O(k|P|)$ time in total.
They also compute $(c(P,r),\tau(P,r))$ for every possible root $r\in P$.
Over all components of $T[W]-B$, the work is $O(kn)$, where $n=|W|$.
Finding the components, counting the original tokens in them, and computing the values $\lambda_B$ and the sets $H_P$ take $O(n)$ additional time.

For a candidate $v\in B$, a scan of its neighbors outside $B$ keeps the three largest values $\lambda_B(w)$, with multiplicity and their corresponding vertices.
For every proposed target $z$, removing its occurrence from those three, if present, leaves the two largest values among the other neighbors; missing values count as zero.
If $z$ is not among those three vertices, removing it does not change the two largest values.
Thus, each pair $(v,z)$ is tested in constant time after the scan at $v$.
The sum of the vertex degrees in the tree is $2(n-1)$, so finding all candidates for removal takes $O(n)$ time per round.
For $2\leq k\leq n$, one round of \cref{alg:compute-rigid} consequently takes $O(kn)$ time and uses polynomial space, where $n=|W|$.

\begin{proof}[Proof of \cref{thm:decision}]
The decision algorithm computes the two rigid sets and applies the characterization of reachability.
For $k>n$, every subset is a $k$-PVC, so \cref{lem:unrestricted-sliding} gives reachability exactly when $|S|=|S'|$.
Suppose that $2\leq k\leq n$.
The tree condition and the two $k$-PVC conditions can be checked in $O(n)$ time.
For the latter checks, a pass through each component of $T-S$ and of $T-S'$ records the two largest downward path lengths at every vertex and rejects if their sum together with that vertex is at least $k$.
The input satisfies the $k$-PVC conditions exactly when neither unoccupied forest contains a path on $k$ vertices.

The decision algorithm runs \cref{alg:compute-rigid} for $S$ and for $S'$ with $W=V(T)$.
By \cref{thm:rigid-algorithm}, the returned first sets are $\Rigid_W(S)$ and $\Rigid_W(S')$.
If they differ, it rejects.
Otherwise, with their common value denoted by $R$, it accepts exactly when $|S\cap P|=|S'\cap P|$ for the vertex set $P$ of every component of $T-R$.
By \cref{thm:characterization}, this condition is equivalent to reachability.
Comparing the rigid sets and the component counts takes $O(n)$ time.
There are at most $n+1$ rounds per input $k$-PVC, each taking $O(kn)$ time by the preceding analysis.
The total running time is therefore $O(kn^2)$.
\end{proof}
\relax\space \section{Finding a TS-Sequence}\label{sec:construction}

We first describe short branch operations and a normalization procedure, then give two procedures for choosing a $k$-PVC with a prescribed property at one vertex.
We then connect the two supplied endpoints by recursion on smaller subtrees.
Both endpoints first receive the same prescribed restriction on one side of an edge, and a single recursive call connects their remaining restrictions.
Finally, we bound the length of the output and the running time in terms of the order of the tree.
Throughout this section, all choices among vertices follow the fixed order from the preliminaries.

\subsection{Elementary Procedures and Prescribed Targets}\label{subsec:construction-preparation}

Under the hypotheses of \cref{lem:unrestricted-sliding}, the following explicit routing procedure joins two sets of the same cardinality on a tree with $p$ vertices.
This leaf-pruning procedure appears in Kornhauser~\cite{Kornhauser1984}.
The procedure fixes a leaf at a time, always choosing the least remaining leaf.
If its occupation differs from the target, a breadth-first search finds a nearest vertex with the desired occupation.
Every intermediate vertex of the joining path has the opposite occupation, so successive slides along the path give the leaf its desired occupation.
The leaf is then fixed, and the procedure continues on the remaining tree.
The token counts in the remaining tree agree with the target counts, so the search always succeeds when the leaf differs from its target.
At a stage with $j$ vertices, at most $j-1$ slides are needed.
Thus, the procedure uses at most $p(p-1)/2$ slides.
It applies when the tree contains no path on $k$ vertices or when fewer than $k$ vertices are unoccupied.

The same search occupies the root of a nonempty branch of a properly rooted subtree, or makes the root unoccupied when that branch is not full.
In the first case, a nearest token moves toward the root.
In the second case, successive tokens move toward a nearest unoccupied vertex, until the root is unoccupied.
These moves are legal while the root of the properly rooted subtree is occupied, since the branch contains no path on $k$ vertices.
When the root $u$ of the properly rooted subtree itself is unoccupied, a nearest token inside that subtree can move to $u$ by \cref{lem:occupying-root}.
Indeed, at each slide, a path through the newly unoccupied vertex either meets the token at its new position or stays inside one branch, where no path on $k$ vertices exists.
This procedure leaves the entire exterior fixed.

When two branch configurations differ only by exchanging one occupied vertex with one unoccupied vertex, we route the tokens along the path joining those vertices.
For a connected vertex set $B$ and vertices $s,t\in B$, the procedure $\mathsc{Exchange}(B,S,s,t)$ returns the empty list if $s=t$.
Otherwise, it scans the unique path from $t$ to $s$.
Starting at $t$, it moves each token encountered along that path back to the position vacated by the preceding token, with the first token moving to $t$.
The scan continues until it has moved the token initially at $s$.
The next lemma gives the conditions under which these slides are legal in the ambient tree.

\begin{lemma}\label{lem:path-exchange}
Let $k\geq2$, let $T[P]$ be a tree, let $U=P^v_u$ be properly rooted, and let $B$ be a branch of $T[U]-u$.
Suppose that $S$ is a $k$-PVC of $T[P]$ with $u\in S$ and that $s,t\in B$.
If $s\in S$ and $t\notin S$, then $\text{\normalfont\textsc{Exchange}}(B,S,s,t)$ is a TS-sequence from $S$ to $S-s+t$ with exactly $\operatorname{dist}_{T[B]}(s,t)$ slides.
Every slide has both ends in $B$, so every prefix fixes the entire restriction outside $B$.
If $s=t$, the returned list is empty, regardless of the occupation of that vertex.
\end{lemma}

\begin{proof}
We partition the path into intervals ending at its initially occupied vertices.
The case $s=t$ follows from the first test of the procedure.
In the other case, let $x_0,\dots,x_d$ be the path, with $x_0=t$ and $x_d=s$.
Let $j_1,\dots,j_r$ be its initially occupied indices, listed in increasing order, so that $j_r=d$, and let $j_0=0$.
For $1\leq i\leq r$, the token at $x_{j_i}$ moves successively to $x_{j_i-1},\dots,x_{j_{i-1}}$.
The destination $x_{j_{i-1}}$ is vacant: this holds initially for $x_0$, and the preceding interval has just vacated it when $i>1$.
The intermediate vertices of the interval were initially unoccupied and have not yet been visited.
Thus, every slide goes from an occupied vertex to an adjacent unoccupied vertex.
The vertex $u$ stays occupied, and $B$ contains no path on $k$ vertices, so \cref{lem:fixing-occupied-set} makes every slide legal in $T[P]$.
Each initially occupied internal path vertex is vacated and then occupied by the following interval.
Consequently, the only final occupation changes are at $s$ and $t$, and the number of slides is $\sum_{i=1}^r(j_i-j_{i-1})=d$.
All these slides stay inside $B$, proving the assertion for every prefix.
\end{proof}

We now specify the normalization used by the pair construction.
Its endpoint is the same prescribed set $C_h$ as in \cref{thm:normalization}.
For a properly rooted side $U=P^v_u$, let $P_i$ be its branches, let $w_i$ be their roots, and let $a_i$ be their heights.
The branch order is decreasing height, with equal heights ordered by their root labels, as in the definition of $C_h$.
For a configuration $S$ containing $u$, let $a=|U|$ and $h=|U-S|$.
Let $b_i=|S\cap P_i|$ and $t_i=|C_h\cap P_i|$.
The procedure $\mathsc{Normalize}(P,U,u,S)$ returns immediately when $U$ is full.
Otherwise, it first occupies every nonempty branch root by the nearest-token procedure, processing the roots in increasing label order.
While the branch counts differ from the prescribed counts, it chooses the first empty branch as recipient if there is one, and otherwise the first deficient branch, for which $b_i<t_i$.
Its donor is the last surplus branch, for which $b_i>t_i$.
It makes the recipient root unoccupied by the nearest-unoccupied-vertex procedure, slides from $u$ to that root and from the donor root to $u$, and occupies the donor root again if the donor remains nonempty.
Finally, it routes each branch to its exact restriction in $C_h$, processing the roots in increasing label order.
All searches visit neighbors in increasing vertex order.

The occupied roots make each transfer use only two branch preparations.
The following lemma proves both the required ambient correctness and the quadratic length bound.

\begin{lemma}\label{lem:fast-normalization}
Let $k\geq2$, let $T[P]$ be a tree, let $U=P^v_u$ be properly rooted, and let $S$ be a $k$-PVC of $T[P]$ with $u\in S$.
Suppose that $S\cap U=U$ or that \cref{eq:internal-slide-condition} holds with the exterior value $x=\varepsilon_{P^u_v}(S\cap P^u_v,v)$.
The procedure $\text{\normalfont\textsc{Normalize}}(P,U,u,S)$ returns a TS-sequence of at most $2a^2$ slides from $S$ to $(S-U)\cup C_h$.
Every prefix fixes the restriction outside $U$.
\end{lemma}

\begin{proof}
We maintain the internal slide condition and occupied roots in nonempty branches at the boundaries of each transfer, and decrease the total deficit from the prescribed branch counts at each transfer.
The full case returns the empty list and has $C_h=C_0=U$.
Otherwise, \cref{lem:first-internal-slide} permits an internal departure from $u$.
The restriction after that departure omits $u$ and has $h$ unoccupied vertices, so \cref{lem:properly-rooted-count}(a) gives $h\leq\gamma(U)$.
Thus, $C_h$ is defined and has the same token count on $U$ as $S$.

The initial root preparations are legal because $u$ stays occupied and every branch contains no path on $k$ vertices.
They preserve all branch counts and the exterior, hence also the internal slide condition.
At the start of the transfer loop, every nonempty branch has an occupied root and $\sum_i b_i=\sum_i t_i$.
The definition of $C_h$ gives two possibilities: every $t_i$ is positive, or the $t_i$ are zero or one and their positive entries form an initial segment of the branch order.

If the counts differ, equality of the totals supplies a surplus branch and a deficient branch.
When there is an empty branch, the first such branch is deficient.
Indeed, this follows at once if every $t_i$ is positive.
In the other case, if the first empty branch were beyond the positive initial segment, every prescribed positive branch would contain at least one token, and equality of the totals would force all counts to agree.
If there is no empty branch, the first deficient branch exists by equality of the totals.
In either case, the recipient is not full, and the last surplus branch is a distinct nonempty donor.

We next verify that the chosen recipient $P_A$ permits a departure from $u$.
If there is an empty branch, $P_A$ has largest height among the empty branches.
Removing its height from the tested multiset leaves the two largest entries no larger than when any other empty height is removed.
If the witness in \cref{eq:internal-slide-condition} is nonempty, the entire multiset of empty-branch heights already satisfies that condition.
Thus, the condition holds with $A$ in either case.
If no branch is empty, the condition reduces to $x\leq k-2$, and every nonfull recipient is admissible.

Making $w_A$ unoccupied is legal while $u$ stays occupied.
Every other nonempty branch root remains occupied, so the admissibility just proved makes the slide from $u$ to $w_A$ legal.
The slide from the donor root $w_B$ to $u$ is then legal because it occupies $u$ and the branch $P_B$ contains no path on $k$ vertices.
If $P_B$ still has a token, its nearest token is moved to $w_B$.
This restoration is legal with $u$ occupied and reestablishes the occupied-root invariant.
The recipient count increases by one, the donor count decreases by one, and all other counts stay fixed.
After the transfer, $P_B$ is not full, and every empty branch other than $P_B$ was empty before the transfer and differs from $P_A$.
Therefore, \cref{eq:internal-slide-condition} holds with $B$ as its next witness.

The nonnegative integer $\Delta=\sum_i\max(t_i-b_i,0)$ decreases by exactly one at each transfer.
Its initial value is at most $\sum_i t_i=|C_h|-1\leq a-1$, so $a-1$ iterations suffice.
If no donor or recipient exists, equality of the totals already gives $b_i=t_i$ for every branch.
Thus, the loop with $a-1$ iterations as its limit ends with all prescribed counts attained.
The final branch routings are legal with $u$ occupied and reach the exact sets $C_h\cap P_i$.
Every emitted slide has both ends in $U$, so the complete list and all its prefixes preserve the exterior.

For the length bound, let $r_i=|P_i|$.
The branches partition $U-u$, so $\sum_i r_i=a-1$, and distinct branches have total order at most $a-1$.
The initial root preparations use at most $\sum_i r_i=a-1$ slides.
A transfer from $P_B$ to $P_A$ uses at most $(r_A-1)+2+(r_B-1)=r_A+r_B\leq a-1$ slides, including donor restoration when needed.
Finally, the branch routings use at most $\sum_i r_i^2\leq(\sum_i r_i)^2=(a-1)^2$ slides.
The complete length is therefore at most
\[
(a-1)+(a-1)^2+(a-1)^2\leq2a^2.
\]
\end{proof}

\Cref{lem:attachment-movable} supplies the movability needed when a preparation is restricted to one side of an edge.

We next choose a target before constructing a TS-sequence to it.
For a connected set $P$, a vertex $r\in P$, and an integer $t$, suppose that some $k$-PVC of $T[P]$ has $t$ tokens and all its tokens are movable.
There are two possible requests: the target must leave $r$ unoccupied, or it must minimize $\varepsilon_P(Z,r)$ among $k$-PVCs $Z$ with $t$ movable tokens.
By \cref{lem:key}, all these $k$-PVCs are reachable from one another, so the latter request is also a minimum over each of their sets $[Z]_P$.

\begin{lemma}\label{lem:explicit-targets}
If $k,t$ are integers with $k\geq2$, $P\subseteq V(T)$ is connected, $r\in P$, and some $k$-PVC of $T[P]$ has $t$ tokens, all movable,
then a $k$-PVC $Z$ of $T[P]$ with $t$ tokens, all movable, can be constructed in time polynomial in $|P|+k$ subject to either prescribed requirement: $r\notin Z$, or $\varepsilon_P(Z,r)$ is minimum among all such $k$-PVCs.
\end{lemma}

\begin{proof}
We give a recursive construction that keeps $r$ in the remaining subtree, and prove its properties by induction on $|P|$.
Since all tokens of some $t$-token $k$-PVC are movable, $t<|P|$.
Let $q=|P|-t$.

\begin{itemize}
\item \textbf{Case 1: $T[P]$ contains no path on $k$ vertices.}
Every $t$-element subset is a $k$-PVC, and \cref{lem:unrestricted-sliding} joins every two such subsets.
Since $t<|P|$, each vertex can be left unoccupied, so every token is movable.
For the first request, the target consists of the first $t$ vertices of $P-r$.
For the second request with $t>0$, it consists of $r$ and the first $t-1$ other vertices, attaining $\varepsilon=0$.
When $t=0$, the only target is the empty set, which attains the only possible value of $\varepsilon$.

\item \textbf{Case 2: $T[P]$ contains a path on $k$ vertices.}
The tree $T[P]$ is rooted at $r$.
A deepest vertex whose descendant subtree contains a path on $k$ vertices determines a properly rooted subtree $U$ with root $u$.
Among vertices of maximum depth, the least triple $(|U|,u,v)$ resolves ties, where $v$ is the parent of $u$ when $U\neq P$.
Every branch of $T[U]-u$ contains no path on $k$ vertices.

\begin{itemize}
\item \textbf{Case 2.1: $U=P$.}
Then $u=r$.
By \cref{lem:canonical-movable}, $d\leq t\leq |P|-1$, where $d=|P|-\gamma(P)$.
For the minimum request, the target is $C_q$.
Its root is occupied, so it attains zero, and all its tokens are movable by \cref{lem:canonical-movable}.
For the request that $r$ be unoccupied, the target first occupies the roots of the $d$ longest branches, using the fixed order to resolve ties.
It then occupies the least unused vertices other than $r$ until its cardinality is $t$.
The set of the roots of the $d$ longest branches is a $k$-PVC: its entirely unoccupied branches are the $m$ shortest branches, whose heights have $\sigma$ at most $k-2$, so every unoccupied path through $r$ has at most $k-1$ vertices, as in the proof of \cref{lem:properly-rooted-count}(a).
The target contains this set and is therefore a $k$-PVC as well.
The procedure for occupying $r$ gives a TS-sequence from this target to a $k$-PVC with $r$ occupied.
The reverse of the last slide is an internal slide from $r$.
Thus, the resulting $k$-PVC satisfies \cref{eq:internal-slide-condition} with exterior value zero and reaches $C_q$ by \cref{thm:normalization}.
All tokens of the original target are therefore movable by \cref{lem:canonical-movable}.

\item \textbf{Case 2.2: $U\neq P$.}
Let $W=P\setminus U$.
The set $W$ is connected and contains $r$.
Except for the final special case below, the target places $C_{h^*}$ on $U$, where $h^*=\min(\gamma(U),q-1)$, and recursively constructs the requested target on $W$ with $t-|C_{h^*}|$ tokens.
To justify this recursive call, \cref{thm:prescribed-subtree} supplies a reachable $k$-PVC with $C_{h^*}$ on $U$ and all tokens movable in its restriction to $W$.
The induction hypothesis therefore applies at the stated count.
By \cref{lem:key}, the resulting target on $W$ is reachable from that restriction.
These moves lift with $u$ occupied by \cref{lem:fixing-occupied-set}.
Hence, the combined target has all tokens movable in $T[P]$.
The argument certifies reachability of the target; the procedure computes only its vertex set and does not seek the TS-sequence used in this justification.
The first request is satisfied because $r$ remains in $W$ and is unoccupied in the recursive target.

For the minimum request, three cases remain.
\begin{itemize}
\item \textbf{Case 2.2.1: $q>\gamma(U)$.}
Then $h^*=\gamma(U)$ and $|C_{h^*}|=d$, where $d=|U|-\gamma(U)$.
By \cref{lem:ctau-additivity}, $c(P,r)=d+c(W,r)$ and $\tau(P,r)=\tau(W,r)$.
The formula in \cref{cor:min-eps-formula} gives the same minimum at count $t$ on $P$ and at count $t-d$ on $W$.
An unoccupied path from $r$ cannot cross the occupied vertex $u$, so the recursive target attains that minimum in $T[P]$ as well.

\item \textbf{Case 2.2.2: $q\leq\gamma(U)$ and $|W|\geq 2$.}
The recursive target on $W$ has exactly one unoccupied vertex.
Such a $k$-PVC can occupy $r$, so its minimum at $r$ is zero.
The induction hypothesis attains zero, which is also the minimum in $T[P]$.

\item \textbf{Case 2.2.3: $q\leq\gamma(U)$ and $W=\{r\}$.}
The target is instead $C_q+r$.
Its cardinality is $t$, and it occupies $r$, attaining zero.
Every vertex of $U$ can be made unoccupied while $r$ remains occupied, by \cref{lem:canonical-movable,lem:fixing-occupied-set}.
After $u$ is made unoccupied in this manner, the slide from $r$ to $u$ is legal: $u$ becomes occupied, and $r$ is a leaf of $T[P]$.
Thus, the token on $r$ is movable as well.
\end{itemize}
\end{itemize}
\end{itemize}

Every recursive call removes a nonempty set $U$ and recurses only once on the smaller set $W$.
There are at most $|P|$ calls.
The subtrees, branch heights, ordered branch lists, and sets $C_h$ are computed by tree traversals and finite lists of vertices.
Each call therefore takes polynomial time, proving the running-time assertion.
\end{proof}

\subsection{The Recursive Construction}\label{subsec:construction-recursion}

The procedure $\mathsc{ConnectPair}(P,S,S')$ constructs a TS-sequence between its two supplied endpoints.
It first fixes the common rigid set.
When no token is rigid, it puts the same prescribed set on a properly rooted subtree in each endpoint and makes one recursive call between the remaining restrictions.
For a list $A$ of slides, its inverse $A^{-1}$ reverses the list and exchanges the source and destination of every slide.
The concatenation of lists is denoted by $A\cdot B$.
Every returned list is retained in full, including repeated slides and consecutive inverse slides.

We first specify how a recursive call prepares the exterior of a properly rooted subtree.
Let $Q\subsetneq P$ be connected and let $r\in Q$.
To make $r$ unoccupied, the preparation returns the empty list if $r$ is already unoccupied.
Otherwise, its precondition is $r\notin\Rigid_Q(S\cap Q)$.
It computes $R=\Rigid_Q(S\cap Q)$ and the vertex set $A$ of the component of $T[Q]-R$ containing $r$.
All tokens of $S\cap A$ are movable by \cref{lem:fixing-occupied-set}.
The preparation constructs the target of \cref{lem:explicit-targets} on $A$, with the same token count and with $r$ unoccupied, and calls $\mathsc{ConnectPair}$ once to reach that target.
The resulting list lifts with $R$ and the other components fixed.

To minimize $\varepsilon_Q(S\cap Q,r)$, the preparation returns the empty list if $r$ is occupied.
Otherwise, it uses the minimum target of \cref{lem:explicit-targets} in the component $A$ defined above.
Every unoccupied path from $r$ avoids $R$, so the minimum in $A$ is also the minimum in $Q$.
The exact product assertion of \cref{lem:fixing-occupied-set} justifies fixing the other components.
Each preparation therefore makes at most one recursive call, on $A\subseteq Q\subsetneq P$.
When the current restriction already equals the chosen target, that call is omitted.

The next algorithm states the pair recursion.
The components in its rigid case are processed in increasing order of their greatest vertices.
In its one-token case, the unique token moves along the path to the unique token of $S'$.
Both positions meet every path on $k$ vertices, and their joining path also lies in every such path, so every intermediate position is feasible.

\begin{algorithm}[t]
\caption{$\mathsc{ConnectPair}(P,S,S')$}\label{alg:connect-pair}
\KwIn{two reachable $k$-PVCs $S,S'$ of the tree $T[P]$}
\KwOut{a list of slides from $S$ to $S'$}
\If{$S=S'$}{\Return{the empty list}\;}
$R\gets\Rigid_P(S)$\;
\If{$R\neq\emptyset$}{
For each component $Q$ of $T[P]-R$, call $\mathsc{ConnectPair}(Q,S\cap Q,S'\cap Q)$, keeping all other vertices fixed\;
\Return{the concatenation of these lists}\;
}
$q\gets |P|-|S|$\;
\If{$q<k$ or $T[P]$ contains no path on $k$ vertices}{
\Return{the elementary routing list from $S$ to $S'$}\;
}
\If{$|S|=1$}{\Return{the path of the token from its position in $S$ to its position in $S'$}\;}
Choose the properly rooted side $U=P^v_u$ by the least triple $(|U|,u,v)$; let $W=P^u_v$\;
$(A,\alpha)\gets\mathsc{Settle}(P,U,W,u,v,S)$\;
$(B,\beta)\gets\mathsc{Settle}(P,U,W,u,v,S')$\;
$\rho\gets\mathsc{ConnectPair}(W,A\cap W,B\cap W)$, omitting the call if the two restrictions agree\;
\Return{$\alpha\cdot\rho\cdot\beta^{-1}$}\;
\end{algorithm}

In the nonbase case, $|S|\geq2$, $q\geq k$, and $T[P]$ contains a path on $k$ vertices.
The existence of the proper side $U$ follows from \cref{lem:properly-rooted-existence}.
Its choice depends only on $T[P]$, $k$, and the vertex order.
Let $h^*=\min(\gamma(U),q-1)$.
The procedure $\mathsc{Settle}$ returns a reachable endpoint whose restriction to $U$ is exactly $C_{h^*}$.
Its early return tests equality with this set, rather than only the number of unoccupied vertices.
In \cref{alg:construct-sequence}, $S$ is updated after every slide, and $\alpha$ records those slides in their execution order.

\begin{algorithm}[t]
\caption{$\mathsc{Settle}(P,U,W,u,v,S)$}\label{alg:construct-sequence}
\KwIn{a $k$-PVC $S$ of $T[P]$ with no rigid token and a properly rooted side $U=P^v_u$, with $W=P^u_v$}
\KwOut{an endpoint $A$ with $A\cap U=C_{h^*}$ and its list $\alpha$ from $S$}
$q\gets|P|-|S|$; $h^*\gets\min(\gamma(U),q-1)$; $\alpha\gets\langle\rangle$\;
\If{$S\cap U=C_{h^*}$}{\Return{$(S,\alpha)$}\;}
Occupy $u$ by the nearest-token procedure inside $U$\;
\For{$i=1,2$}{
\If{$S\cap U=U$}{break\;}
Minimize $x=\varepsilon_W(S\cap W,v)$ by preparation in $W$\;
\If{\cref{eq:internal-slide-condition} holds for $x$}{
Normalize $U$ by \cref{lem:fast-normalization}, keeping $W$ fixed; break\;
}
Occupy the root of every nonempty branch of $T[U]-u$, in increasing root order\;
Make $v$ unoccupied by preparation in $W$\;
Slide $u$ to $v$, then occupy $u$ by the nearest-token procedure inside $U$\;
}
$h\gets|U-S|$\;
\While{$h<h^*$}{
Make $v$ unoccupied by preparation in $W$; slide $u$ to $v$\;
Let $z$ be the unique vertex of $C_h-C_{h+1}$, let $B$ contain $z$, and let $w$ be its root\;
Slide $w$ to $u$; execute $\mathsc{Exchange}(B,S,z,w)$; $h\gets h+1$\;
}
\If{$h>h^*$}{
Let $z$ be the unique vertex of $C_{h-1}-C_h$, let $B$ contain $z$, and let $w$ be its root\;
Execute $\mathsc{Exchange}(B,S,w,z)$\;
Slide $u$ to $w$, then slide $v$ to $u$\;
}
\Return{$(S,\alpha)$}\;
\end{algorithm}

The normalizations in \cref{alg:construct-sequence} use the single transfer loop of \cref{lem:fast-normalization}.
The two calls to $\mathsc{Exchange}$ return the empty list when $z=w$.
All nearest-token and nearest-unoccupied-vertex preparations use breadth-first search, visiting neighbors in increasing vertex order.

The following lemma justifies all changes made by $\mathsc{Settle}$, including the two-step bound in its first stage and the final return of one token.
Its hypothesis on smaller calls will be supplied by induction in the pair construction.

\begin{lemma}\label{lem:settle-correctness}
Let $k\geq2$, let $T[P]$ be a tree, let $U=P^v_u$ be properly rooted, and let $W=P^u_v$.
Suppose that $S$ is a $k$-PVC of $T[P]$ with no rigid token and that every recursive pair call on a proper subset of $P$ returns a TS-sequence between its supplied reachable endpoints.
Then $\text{\normalfont\textsc{Settle}}(P,U,W,u,v,S)$ terminates and returns a TS-sequence from $S$ to its returned endpoint $A$, with $A\cap U=C_{\min(\gamma(U),|P-S|-1)}$.
Every token of $A\cap W$ is movable in $T[W]$.
\end{lemma}

\begin{proof}
We first reach a prescribed set on $U$, then adjust its number of unoccupied vertices to $h^*$.
Every operation below is legal, so every intermediate configuration remains reachable from $S$ and has no rigid token in $T[P]$.
The early return has the required restriction by its exact equality test.
Otherwise, \cref{lem:occupying-root} occupies $u$ without changing the number of unoccupied vertices in $U$ or the exterior.

\paragraph{The first normalization.}
If $U$ is full, its restriction is already $C_0$.
Otherwise, the minimum preparation attains $x$ by \cref{lem:explicit-targets,lem:key} and the hypothesis on smaller pair calls.
If \cref{eq:internal-slide-condition} holds, \cref{lem:fast-normalization} reaches $C_h$ for the current count $h$ while preserving the entire exterior.
Suppose that this condition fails at the attained minimum.
Before a first slide from $u$, branch counts are unchanged and the exterior stays in its current reachability class.
Every possible exterior value is at least $x$, and increasing that value cannot make the failed condition hold.
Thus, \cref{lem:first-internal-slide} rules out an internal first slide from $u$.
Since its token is movable in $T[P]$, a first slide from $u$ must be possible and must go to $v$.
The branches that are empty before this slide remain empty after it, so their heights have $\sigma$ at most $k-2$.
The exterior cannot be full, since then it could not change while $u$ remained occupied and the first slide to $v$ would be impossible.
By \cref{lem:attachment-movable}, the preparation can therefore make $v$ unoccupied inside $W$.

The algorithm occupies every nonempty branch root and makes $v$ unoccupied.
The slide from $u$ to $v$ is legal: paths crossing $uv$ meet $v$, and an unoccupied path through $u$ inside $U$ has at most $1+\sigma\leq k-1$ vertices.
The nearest-token procedure occupies $u$ again inside $U$.
This failure step increases $h$ by one and leaves $v$ occupied.
Consequently, the second minimum preparation is empty and its exterior value is zero.

To justify normalization at that value, suppose that $u$ is occupied and $U$ is not full in any reachable state.
A first slide from $u$ is possible because its token is movable.
If that slide is internal, \cref{lem:first-internal-slide} gives the internal slide condition, which remains true after the exterior value is decreased to zero.
If the slide goes to $v$, the entirely empty branch heights have $\sigma\leq k-2$.
At least one branch is not full, and choosing it as the receiving branch gives the same condition with exterior value zero.
Thus, the second step normalizes successfully.
There is at most one failure and at most one normalization, within the two stated steps.
The algorithm now recomputes $h$ from the resulting configuration, whose restriction to $U$ is $C_h$.

\paragraph{Increasing the count to $h^*$.}
Suppose that $h<h^*$.
Then $h<\gamma(U)$, and $W$ is not full because it has $q-h>0$ unoccupied vertices.
The attachment lemma and a smaller pair call make $v$ unoccupied.
By \cref{lem:properly-rooted-count}(b), $C_h-u$ is a $k$-PVC of $T[U]$, so sliding $u$ to $v$ is legal.
The sets $C_h$ and $C_{h+1}$ differ in exactly one vertex $z$.
Let $B$ contain $z$ and let $w$ be its root.
The definition of the prescribed sets puts $w$ in $C_h$, because branch roots precede other branch vertices.
The slide from $w$ to $u$ is legal: it occupies $u$, and $B$ contains no path on $k$ vertices.
The two slides have replaced the restriction $C_h$ by $C_h-w$ and added $v$ to the prepared exterior.
If $z=w$, this restriction is already $C_{h+1}$, and the exchange is empty.
Otherwise, $z$ is occupied and $w$ is unoccupied, so \cref{lem:path-exchange} moves a token from $z$ to $w$ and leaves every other occupation unchanged.
The restriction on $U$ is then exactly $C_h-z=C_{h+1}$, while the prepared exterior retains its added token at $v$.
Each iteration increases $h$ by one, so the loop reaches $h^*$ if it starts below that value.

\paragraph{Returning one token when the initial count is larger.}
By \cref{thm:maximum-unoccupied}, the count after the first stage is at most $\min(\gamma(U),q)$.
If that count exceeds $h^*$, it is therefore exactly $q$, with $q\leq\gamma(U)$ and $W$ full.
The increasing loop makes no move.
Let $z$ be the unique vertex of $C_{q-1}-C_q$, let $B$ contain $z$, and let $w$ be the root of $B$.
The root $w$ lies in $C_{q-1}$.
If $z=w$, the exchange is empty, and $C_q=C_{q-1}-w$ already holds.
Otherwise, $w\in C_q$ and $z\notin C_q$, so \cref{lem:path-exchange} exchanges their occupations inside $B$ with $u$ occupied.
In both cases, the resulting restriction is exactly $C_{q-1}-w$.
After the slide from $u$ to $w$, the restriction on $U$ is exactly $C_{q-1}-u$.
That set is a $k$-PVC by \cref{lem:properly-rooted-count}(b), since $q-1<\gamma(U)$, and the full exterior meets every path crossing $uv$.
The slide from $v$ to $u$ then reaches exactly
\[
C_{q-1}\cup(W-v).
\]
It is legal because $u$ becomes occupied and the only unoccupied vertex in $W$ is $v$.
The restriction to $U$ is $C_{h^*}$ without any further normalization.

All preparations use smaller pair calls, the first stage has at most two steps, and the increasing loop has at most $|U|$ steps.
Thus, the procedure terminates with the asserted actual list and endpoint.
Finally, \cref{thm:prescribed-subtree} applies to every reachable endpoint with the exact prescribed restriction and gives movability of every token in $A\cap W$.
\end{proof}

We can now prove that the pair recursion reaches the supplied target.
No canonical endpoint for the whole tree is needed.

\begin{theorem}\label{thm:pair-correctness}
For every integer $k\geq2$, every tree $T[P]$, and every reachable pair $S,S'$ of $k$-PVCs of $T[P]$, the procedure $\text{\normalfont\textsc{ConnectPair}}(P,S,S')$ terminates and returns a TS-sequence from $S$ to $S'$.
On a tree, a wrapper that first checks the cover conditions and the decision test of \cref{thm:decision} rejects precisely the invalid or unreachable pairs and otherwise returns this TS-sequence.
\end{theorem}

\begin{proof}
We use strong induction on $|P|$ for the actual list returned between the two endpoints.
Equal inputs return the empty list.
For unequal inputs, reachability gives the same rigid set $R$ and the same token count in each component of $T[P]-R$.
If $R\neq\emptyset$, each component is smaller, and the exact product statement of \cref{lem:fixing-occupied-set} makes its two restrictions reachable.
The induction hypothesis supplies their pair lists.
These lists lift with all other vertices fixed and concatenate to a TS-sequence ending at $S'$.

Suppose that $R=\emptyset$.
The unrestricted-routing cases are justified by \cref{lem:unrestricted-sliding}, and the one-token case was justified before \cref{alg:connect-pair}.
In the remaining case, both calls to $\mathsc{Settle}$ use the same $U$, the same $q$, and the same $h^*$.
Every recursive preparation is on a subset of $W$, which is smaller than $P$, so the induction hypothesis supplies the assumption in \cref{lem:settle-correctness}.
The returned lists $\alpha$ and $\beta$ lead from $S$ to $A$ and from $S'$ to $B$, with $A\cap U=B\cap U=C_{h^*}$.
The restrictions $A\cap W$ and $B\cap W$ have equal cardinalities and no rigid tokens.
By \cref{lem:key}, they are reachable in $T[W]$.
The induction hypothesis supplies the list $\rho$ between these exact restrictions.
Since $u$ remains occupied, that list lifts to a TS-sequence from $A$ to $B$ with $C_{h^*}$ fixed.
Therefore, the returned list $\alpha\cdot\rho\cdot\beta^{-1}$ is a TS-sequence from $S$ to $S'$.
Every pair call has smaller order, and the local routines terminate by \cref{lem:settle-correctness}, proving termination as well.
The assertion about rejection follows from the cover checks and \cref{thm:decision}.
\end{proof}

\subsection{The Length of the Output}\label{subsec:construction-length}

We first count the recursive preparations made while settling one endpoint.
The cutoff at $q-1$ and the empty second minimum preparation are both used in this count.
Let $a=|U|$, let $g=\gamma(U)$, and let $h_0$ be the initial number of unoccupied vertices in $U$.
Let $\delta$ be one if the first stage takes its failure step and zero otherwise.
The first stage ends with $h_0+\delta$ unoccupied vertices, where $\delta\leq1$ and $\delta=0$ if $h_0=0$.
If $h_0=0$, that stage makes no recursive preparation; otherwise, it makes at most $1+\delta$.
The subsequent loop makes at most one preparation for each increase of $h$.
For a real number $x$, let $x_+=\max(x,0)$.

The next bound counts the complete lists from smaller pair calls, including repetitions.
It also bounds the slides emitted locally by a settling stage.
For a properly rooted side of order $a$, let
\begin{equation}\label{eq:pair-local-cost}
\ell(a)=4a^2.
\end{equation}

\begin{lemma}\label{lem:settle-cost}
In a nonbase call of \cref{alg:connect-pair}, each call to $\text{\normalfont\textsc{Settle}}$ makes at most $g$ smaller pair calls on subsets of $W$ and emits at most $\ell(a)$ slides in addition to their returned lists.
\end{lemma}

\begin{proof}
We count the recursive preparations and then use the branch partition to bound the local operations.
An early return satisfies both bounds.
Otherwise, \cref{thm:maximum-unoccupied} gives $h_0+\delta\leq\min(g,q)$, and $q\geq k\geq2$ in a nonbase call.
The total number of preparations is at most
\begin{equation}\label{eq:pair-preparations}
\begin{cases}0,&h_0=0,\\1+\delta,&h_0>0\end{cases}
+\bigl(\min(g,q-1)-h_0-\delta\bigr)_+
\leq g.
\end{equation}
If $h_0=0$, the inequality follows from $\delta=0$.
If $h_0>0$ and $h_0+\delta\leq h^*$, its left side is $1+h^*-h_0\leq g$.
In the remaining case, $h_0+\delta=q\leq g$, and the left side is $1+\delta\leq2\leq q\leq g$.
The final return uses no recursive preparation.

Proper rooting gives $a\geq k\geq2$ and $g\leq a-1$.
The initial occupation of $u$ uses at most $a-1$ slides.
If the first stage fails once, the occupation of all nonempty branch roots uses at most the sum of their orders, which is $a-1$.
That failure also uses one slide from $u$ to $v$ and at most $a-1$ slides to occupy $u$ again, for a total of at most $2a-1$ local slides.
The single successful normalization uses at most $2a^2$ slides by \cref{lem:fast-normalization}.
Each increasing step uses two attachment slides and a path exchange in one branch $B$.
By \cref{lem:path-exchange}, this costs at most $2+(|B|-1)\leq a$, because $|B|\leq a-1$.
There are at most $h^*\leq g\leq a-1$ such steps, and the final return, if taken, has the same bound of $a$ slides.
Consequently, the total local cost is at most
\[
(a-1)+(2a-1)+2a^2+a(a-1)+a
=3a^2+3a-2
\leq4a^2,
\]
since $(a-1)(a-2)\geq0$.
\end{proof}

For the length induction, let $r=\sqrt{3}$ and define the real potential $F$ on the nonnegative integers by
\begin{equation}\label{eq:pair-potential}
F(p)=r^p-1.
\end{equation}
For an ambient order $n\geq1$, the corresponding natural-number budget is
\begin{equation}\label{eq:pair-budget}
B_n(p)=\left\lfloor8n^2F(p)\right\rfloor.
\end{equation}
These quantities are used only in the length proof; the construction evaluates neither square roots nor floors.

The potential accounts for both the components of a rigid decomposition and the smaller calls on the complement of a properly rooted side.
The following sum inequality applies to every finite list $p_1,\dots,p_m$ of nonnegative integers, including the empty list.

\begin{lemma}\label{lem:pair-potential}
The function $F$ is nonnegative and nondecreasing, and $F(0)=0$.
For nonnegative integers $c,d$ and an integer $a\geq2$, it satisfies
\begin{align}
F(c)+F(d)&\leq F(c+d),\label{eq:pair-superadd}\\
1+(2a-1)F(d)&\leq F(a+d),\label{eq:pair-branch}\\
F(d)&\leq3^{\lceil d/2\rceil}.\label{eq:pair-envelope}
\end{align}
The budgets are nondecreasing, satisfy $B_n(0)=0$, and obey
\begin{align}
\sum_i B_n(p_i)&\leq B_n\!\left(\sum_i p_i\right),\label{eq:pair-budget-sum}\\
8n^2+(2a-1)B_n(d)&\leq B_n(a+d),\label{eq:pair-budget-branch}\\
p^2&\leq B_n(p)\qquad(0\leq p\leq n),\label{eq:pair-budget-base}\\
B_n(p)&\leq8n^2\,3^{\lceil p/2\rceil}.\label{eq:pair-budget-envelope}
\end{align}
The sum in \cref{eq:pair-budget-sum} may be empty.
\end{lemma}

\begin{proof}
We first prove the real inequalities and then transfer them to integer budgets.
Since $r>1$, nonnegativity, monotonicity and $F(0)=0$ follow from \cref{eq:pair-potential}.
Moreover,
\[
F(c+d)-F(c)-F(d)=(r^c-1)(r^d-1)\geq0,
\]
which proves \cref{eq:pair-superadd}.
For every integer $a\geq2$, we have $2a-1\leq r^a$.
The cases $a=2$ and $a=3$ follow from $r^2=3$ and $r^3=3\sqrt{3}\geq5$, respectively.
The implication from $a$ to $a+2$ follows from $2(a+2)-1\leq3(2a-1)$ and $r^{a+2}=3r^a$.
Therefore,
\[
F(a+d)-(2a-1)F(d)
=\bigl(r^a-(2a-1)\bigr)r^d+2a-2
\geq1,
\]
proving \cref{eq:pair-branch}.
Finally, $d\leq2\lceil d/2\rceil$ gives $F(d)\leq r^d\leq3^{\lceil d/2\rceil}$.

Because $8n^2F(p)\geq0$, an integer $m\geq0$ satisfies $m\leq B_n(p)$ exactly when $m\leq8n^2F(p)$.
This observation transfers monotonicity and the value at zero to the budgets.
For \cref{eq:pair-budget-sum}, each summand is at most $8n^2F(p_i)$, and repeated use of \cref{eq:pair-superadd} bounds their sum by $8n^2F(\sum_i p_i)$.
The left side is an integer, so the floor gives the asserted inequality, including the empty sum.
Similarly, \cref{eq:pair-branch} gives
\[
8n^2+(2a-1)B_n(d)
\leq8n^2\bigl(1+(2a-1)F(d)\bigr)
\leq8n^2F(a+d),
\]
and the integer left side is at most $B_n(a+d)$.
For \cref{eq:pair-budget-base}, the case $p=0$ is equality.
When $1\leq p\leq n$, we have $F(p)\geq\sqrt{3}-1>1/8$, so $p^2\leq n^2\leq8n^2F(p)$, and the floor again gives the result.
The final bound follows from \cref{eq:pair-envelope} and $B_n(p)\leq8n^2F(p)$.
\end{proof}

The quadratic local cost now gives an explicit length bound for the actual returned list.

\begin{lemma}\label{lem:output-length}
For every integer $k\geq2$, every tree $T$ of order $n\geq1$, and every reachable pair $S,S'$ of $k$-PVCs of $T$, the list returned by $\text{\normalfont\textsc{ConnectPair}}(V(T),S,S')$ has length at most
\[
8n^2\,3^{\lceil n/2\rceil}.
\]
\end{lemma}

\begin{proof}
We use strong induction on $p\leq n$ to bound the list for every reachable pair on a connected set of order $p$ by $B_n(p)$.
The correctness of these actual lists was established in \cref{thm:pair-correctness}.
Equal inputs return no slides.
In an elementary or one-token case, the output has at most $p^2$ slides, which is at most $B_n(p)$ by \cref{eq:pair-budget-base}.
For a nonempty rigid set, the component calls have disjoint vertex sets with total order at most $p-1$.
Their endpoints are valid and reachable by \cref{lem:fixing-occupied-set}.
The induction hypothesis, \cref{eq:pair-budget-sum} and monotonicity bound their concatenation by $B_n(p)$, including when there are no components.

In the remaining case, let $a=|U|$ and $b=|W|$, so $a+b=p$ and $a\geq k\geq2$.
Every recursive preparation is on a subset of $W$ with valid reachable endpoints, and its length is at most $B_n(b)$ by the induction hypothesis and monotonicity.
By \cref{lem:settle-cost}, the two settling stages use at most $2\gamma(U)$ such lists, and the middle pair call contributes at most one more.
Since $\gamma(U)\leq a-1$, there are at most $2a-1$ retained smaller lists.
The two settling stages emit at most $8a^2\leq8n^2$ local slides.
Reversing the second list does not change its length.
Thus, \cref{eq:pair-budget-branch} bounds the complete list by
\[
8n^2+(2a-1)B_n(b)\leq B_n(a+b)=B_n(p).
\]
Taking $p=n$ and applying \cref{eq:pair-budget-envelope} proves the stated bound.
\end{proof}

\subsection{Running Time}\label{subsec:construction-time}

We analyze the algorithm with exact tree computations and explicit lists of slides.
Let $n=|V(T)|$ and let $L$ be the length of its actual output, before any shortening.
Vertex sets are Boolean arrays.
We count unit-cost array accesses, arithmetic operations on integer labels and counts, and reads and writes of slide records.
These are arithmetic-operation bounds, not bit-complexity bounds.
The cost of reading the encoded input is added separately; in particular, a binary encoding of $k$ contributes $O(\log(k+1))$ additional time.
Child lists are assembled by one pass over each list occurrence, using tail pointers or arrays with a final concatenation.
Repeatedly copying an accumulated prefix is unnecessary.
The bound below includes all retained child lists and their reversals.

When $k>n$, the algorithm first uses elementary routing, before any computation whose cost depends on $k$.
For unequal equal-size inputs, both sets are nonfull and every token is movable by \cref{lem:unrestricted-sliding}, so this shortcut returns exactly the list that the pair recursion would return.
Equal inputs return the empty list.
For the remainder of the analysis, suppose that $k\leq n$.
Only calls to $\mathsc{ConnectPair}$ are nodes of the recursive call tree; target synthesis and local routines are counted within their calling node.
Every child acts on a strictly smaller vertex set, so the depth is at most $n$.
A rigid split has at most $n$ children, and \cref{lem:settle-cost} bounds the number of children of a nonbase call by $2\gamma(U)+1\leq2n-1$.
Every nonempty returned list occurs in full in the final list, possibly reversed.
At any fixed depth, these occurrences are disjoint, so there are at most $L$ calls with nonempty output at that depth.
There are therefore at most $nL$ nonempty calls.
By \cref{thm:pair-correctness}, an empty pair list has equal endpoints, and the equality guard returns before making any recursive call.
All empty calls are leaves.
Attaching the at most $O(n)$ empty children of each nonempty call gives
\begin{equation}\label{eq:construction-calls}
\text{the number of recursive calls is }O\bigl(n^2(L+1)\bigr).
\end{equation}

The work at one node, apart from reading its returned child lists, is $O(n^6)$ operations.
The following bounds give explicit ways to obtain every required quantity.
\begin{itemize}
\item Components, paths, diameters, and rooted heights are computed by tree traversals, within $O(n^2)$ operations using Boolean arrays.
The unoccupied forest contains a path on $k$ vertices exactly when one of its components has diameter at least $k-1$.
Testing every ordered attaching edge and its branches therefore finds the least properly rooted side within $O(n^4)$ operations.
Branch heights, the sorted branch order, $\gamma(U)$, and each $C_h$ are computed within this bound.
\item A target computation in \cref{lem:explicit-targets} has at most $n$ recursive levels, each with at most $O(n^4)$ operations.
All $O(n)$ target computations at the node therefore use $O(n^6)$ operations.
\item The rigid sets are evaluated by \cref{alg:compute-rigid} and the rooted recurrence of \cref{eq:ctau-recurrence}.
There are $O(n)$ evaluations at a node, each using $O(kn^2)$ operations, for a total of $O(n^4)$ when $k\leq n$.
\item An elementary routing uses at most $n$ searches and costs $O(n^3)$ operations.
A nearest-token or nearest-unoccupied-vertex preparation uses one search and at most $n-1$ slides, within $O(n^2)$ operations.
The normalizer has at most $n$ initial root preparations, at most $n-1$ transfers with two such preparations each, and at most $n$ final elementary branch routings.
Scanning the ordered branch counts selects each recipient and donor within $O(n^2)$ operations.
Thus, a normalization costs $O(n^4)$ operations.
Each path exchange finds one path, scans its occupations, and emits at most $n-1$ slides, within $O(n^2)$ operations.
There is at most one normalization in each settling stage, and all other exchanges and root preparations also fit within $O(n^4)$ operations per node.
\end{itemize}

These computations test paths by diameters and rooted heights and evaluate the branch and rigid formulas by sorted lists and the rooted recurrence.
In particular, the largest compatible set of branch heights is obtained from a sorted prefix, as in \cref{eq:ctau-recurrence}, instead of enumerating subsets of branches.
Exact values of $\varepsilon$ are rooted heights in the unoccupied forest.
These evaluations preserve the tests and fixed choices of the construction without enumerating configurations.
The running-time bound applies to this realization of the algorithm.
The Lean definitions use finite path and subset enumerations as reference definitions for some of these quantities; their literal evaluation is not claimed to satisfy this time bound.

The sum of the lengths of all returned lists is at most $nL$, since each final slide occurrence belongs to at most one call at each depth.
One pass over each child-list occurrence, including reversals and endpoint updates, therefore uses $O(nL)$ additional operations.
Together with \cref{eq:construction-calls}, the local bound gives
\begin{equation}\label{eq:construction-time}
O\bigl(n^8(L+1)\bigr).
\end{equation}
The intermediate estimate concerns the complete retained list; the proof below substitutes the length bound to obtain a running time in terms of $n$ alone.
It does not assert a matching lower bound on the length of that list or on shortest TS-sequences.

\begin{proof}[Proof of \cref{thm:construction}]
We combine pair correctness with the length and running-time estimates.
By \cref{thm:pair-correctness}, the algorithm returns a TS-sequence between the supplied endpoints.
Its length is at most $8n^2\,3^{\lceil n/2\rceil}$ by \cref{lem:output-length}.
When $k\leq n$, \cref{eq:construction-time} gives $O(n^8(L+1))$ operations.
Since $3^{\lceil n/2\rceil}\leq\sqrt{3}\,3^{n/2}$, substituting the length bound gives $O(n^{10}3^{n/2})$ operations.
When $k>n$, every subset is a $k$-PVC, and direct elementary routing uses at most $n(n-1)/2$ slides in $O(n^3)$ time, satisfying the stated length and time bounds.
For pairs not known to be valid and reachable, the initial cover checks and the decision algorithm reject precisely the other pairs; their time is also within the stated bound.
\end{proof}

\relax\space \section{Concluding Remarks}\label{sec:conclusion}

In this paper, we have shown that Token Sliding for $k$-PVCs of trees of order $n$ can be decided in $O(kn^2)$ time for every integer $k\geq2$, and we have given a direct recursive construction with at most $8n^2\,3^{\lceil n/2\rceil}$ slides and running time $O(n^{10}3^{n/2})$.
The decision algorithm computes the vertices with rigid tokens and uses the token counts in the components remaining after their deletion.

It remains open whether TS-sequences of polynomial length between reachable $k$-PVCs of trees can be found in polynomial time for every fixed $k\geq 3$.

For $k=2$, Demaine et al.~\cite{Demaine2015} proved that a TS-sequence of length $O(n^2)$ can be found in $O(n^2)$ time whenever the target is reachable, and they gave paths on which every such sequence has length $\Omega(n^2)$.
Their result transfers from independent sets to $2$-PVCs by taking complements.
The bound proved here for the direct pair construction does not recover that quadratic guarantee.
It remains open whether $O(n^2)$ slides suffice for every fixed $k\geq 3$.
Another question is whether the $O(kn^2)$ decision bound can be improved to linear time for each fixed $k\geq 3$, as is known for $k=2$.
\relax\space 

\FloatBarrier

\section*{Acknowledgements}
Duc A.\ Hoang's research was supported by the Vietnam National University, Hanoi under the project QG.25.07 ``A study on reconfiguration problems from algorithmic and graph-theoretic perspectives''.
\relax\space 

\section*{AI Declaration}
ChatGPT 6 Pro and Claude Fable 5.1 were used to assist with developing and reviewing mathematical arguments and drafting and revising the paper.
The author takes full responsibility for the correctness and content of the paper.
\relax\space 

\printbibliography

\end{document}